\documentclass[10pt,twocolumn,letterpaper]{article}

\usepackage[pagenumbers]{cvpr} % To force page numbers, e.g. for an arXiv version
\usepackage[accsupp]{axessibility}

\newif\ifarxiv
\arxivtrue
\makeatletter
\@namedef{ver@everyshi.sty}{}
\makeatother

\usepackage{colortbl}
\usepackage{currfile}
\usepackage{etoolbox}
\usepackage{multirow}
\usepackage{tikz}
\usetikzlibrary{positioning,arrows.meta, calc, fit, positioning, shapes.geometric, spy}
\usepackage{pgfplots}
\pgfplotsset{compat=1.18}
\usepackage{xfrac}
\usepackage{adjustbox}  % trim 에 상대치 사용
\usepackage{algorithm}
\usepackage{algpseudocode}
\usepackage{svg}
\usepackage{todonotes}
\usepackage{ragged2e}

\usepackage{tabularx}
\usepackage[dvipsnames,table, svgnames]{xcolor}

\usepackage{makecell}
\usepackage{nicefrac}
\usepackage{subcaption}
\usepackage[normalem]{ulem}

\definecolor{turquoise}{cmyk}{0.65,0,0.1,0.3}
\definecolor{purple}{rgb}{0.65,0,0.65}
\definecolor{dark_green}{rgb}{0, 0.5, 0}
\definecolor{orange}{rgb}{0.8, 0.6, 0.2}
\definecolor{red}{rgb}{0.8, 0.2, 0.2}
\definecolor{darkgray}{rgb}{0.5, 0.5, 0.5}
\definecolor{darkred}{rgb}{0.6, 0.1, 0.05}
\definecolor{blueish}{rgb}{0.0, 0.3, .6}
\definecolor{light_gray}{rgb}{0.7, 0.7, .7}
\definecolor{pink}{rgb}{1, 0, 1}
\definecolor{greyblue}{rgb}{0.15, 0.25, 0.65}

\newcommand{\copydanger}[1]{\textbf{OMITTED COPY/PASTE TEXT}}

\newcommand{\ignore}[1]{}

\usepackage{bbm}
\usepackage{lipsum}
\usepackage{blindtext}

\renewcommand{\paragraph}[1]{\vspace{.5em}\noindent\textbf{#1}.}

\newcommand{\OurMethod}{DiffBMP\xspace}

\newcommand{\SupplementaryMaterial}[1]{{supplementary material}}

\usepackage{enumitem}
\setlist[itemize]{noitemsep,leftmargin=*,topsep=0em}
\setlist[enumerate]{noitemsep,leftmargin=*,topsep=0em}

\newcommand{\height}{\text{H}}

\newcommand{\rgt}{\aftergroup\mathclose\aftergroup{\aftergroup}\right}

\newcommand\commentout[1]{}

\definecolor{tabfirst}{rgb}{1, 0.7, 0.7} %
\definecolor{tabsecond}{rgb}{1, 0.85, 0.7} %
\definecolor{tabthird}{rgb}{1, 1, 0.7} %

\tikzset{
  outline text/.style={
    text=white,
    preaction={draw=black, line width=1.5pt},
  }
}

\tikzset{
    graybox/.style={
        fill=black!80, fill opacity=0.35,
        text=white, text opacity=1,
        inner sep=3pt,
        font=\small,
        align=left
    },
    lightblack_box_white_text/.style={
        fill=black, fill opacity=0.7,
        text=white, text opacity=1,
        inner sep=3pt,
        font=\small,
        align=right
    },
    white_text_nobox/.style={
        fill=none, 
        text=white, text opacity=1,
        inner sep=3pt,
        font=\small,
        align=left
    },
    black_box_white_text/.style={
        fill=black, 
        text=white, 
        inner sep=1pt,
        font=\small,
        align=right
    },
    lightblue_box_black_text/.style={
        fill=blue!20, fill opacity=0.8,
        text=black, text opacity=1,
        inner sep=3pt,
        font=\small,
        align=left
    },
    yellow_box_black_text/.style={
        fill=yellow!80, fill opacity=0.9,
        text=black, text opacity=1,
        inner sep=1pt,
        font=\small,
        align=left
    },
    white_box_black_text/.style={
        fill=white, fill opacity=1,
        text=black, text opacity=1,
        inner sep=3pt,
        font=\small,
        align=left
    },
}

\tikzset{
    overlay text/.style={
        rectangle,
        fill=black!85, % 배경색 흰색 85% 투명도
        text=white,
        font=\small,
        inner sep=1pt,
        anchor=north east % 우상단 정렬
    },
    primitive label/.style={
        rectangle,
        fill=white!85,
        opacity=0.9,
        text=black,
        font=\small,
        inner sep=1pt,
        anchor=north west % 좌상단 정렬
    }
}

\newcommand{\boxtext}[1]{%
    \tikz[baseline=-0.0pt]{
        \node[
            fill=black, 
            fill opacity=0.7, 
            text=white, 
            text opacity=1, 
            font=\small, % 텍스트 크기를 \small로 설정
            anchor=base, 
            inner sep=1pt, 
            outer sep=0pt % 노드 주변의 바깥쪽 여백 (margin) 설정
        ] {#1};
    }%
}

\usepackage{contour}     
\contourlength{0.05em}   % 윤곽선 두께 조절 (취향대로)
\usepackage{fontawesome} 
\usepackage{scalerel}

\newcommand{\condensedSB}[1]{\scalebox{0.6}[1.0]{\sffamily \textbf{#1}}%
}

\usepackage{pifont}
\definecolor{darkgreen}{RGB}{0, 150, 0} 
\newcommand{\cmark}{\textcolor{darkgreen}{\ding{51}}}%
\newcommand{\xmark}{\textcolor{red}{\ding{55}}}%

\newcommand{\vc}{\boldsymbol{c}}

\usepackage{kotex}

\counterwithin*{subfigure}{figure} 
\usepackage{comment}

\tikzset{bbox1/.style={draw=red,line width=1pt},
bbox2/.style={draw=red,line width=1.2pt},
linkline/.style={line width=1pt, draw=red, 
dash pattern=on 3pt off 1.5pt}}

\newcommand{\relbox}[5]{%
  \begin{pgfonlayer}{fg}%
  \path
    let \p1 = (#1.south west),  % (x1,y1)
        \p2 = (#1.north east),  % (x2,y2)
        \n1 = {\x1 + (#2)*(\x2-\x1)}, % x0
        \n2 = {\y1 + (#3)*(\y2-\y1)}, % y0
        \n3 = {\x1 + (#4)*(\x2-\x1)}, % x1
        \n4 = {\y1 + (#5)*(\y2-\y1)}  % y1
    in
      coordinate (ll) at (\n1,\n2)
      coordinate (ur) at (\n3,\n4);
  \draw[bbox2] (ll) rectangle (ur);
  \end{pgfonlayer}%
}

\newcommand{\relboxlinkto}[6]{%
  \begin{pgfonlayer}{fg}%
    \path
      let \p1 = (#1.south west),
          \p2 = (#1.north east),
          \n1 = {\x1 + (#3)*(\x2-\x1)}, % x0
          \n2 = {\y1 + (#4)*(\y2-\y1)}, % y0
          \n3 = {\x1 + (#5)*(\x2-\x1)}, % x1
          \n4 = {\y1 + (#6)*(\y2-\y1)}  % y1
      in
        coordinate (ll) at (\n1,\n2)   % SW of bbox on SRC
        coordinate (ur) at (\n3,\n4)   % UR of bbox on SRC
        coordinate (ul) at (\n1,\n4);  % SE of bbox on SRC
    \draw[bbox1] (ll) rectangle (ur);

    \path
      let \p3 = (#2.south west),
          \p4 = (#2.north east)
      in
        coordinate (NW) at (\x3+4pt,\y4-4pt);   % NW of DST
    \draw[linkline] (ul) -- (NW);
  \end{pgfonlayer}%
}

\newlength\imgSmall   
\newlength\imgLarge   
\newlength\xGap       
\newlength\yDrop      

\newlength\xGapLarge  
\newlength\Lshift     

\tikzset{
  imgSmall/.style={inner sep=0,anchor=west},
  imgLarge/.style={inner sep=0,anchor=west},
  toplabel/.style={inner sep=0},
  arrowlab/.style={fill=red, inner sep=2pt, text=black},
  redbox/.style={
    fill=red,
    fill opacity=0.8,
    inner sep=3pt,
    text=white,
    text opacity=1,
    font=\small,
    align=left
  }
}

\pgfdeclarelayer{bg}
\pgfdeclarelayer{fg}
\pgfsetlayers{bg,main,fg}

\usepackage{listings}

\newcommand\blfootnote[1]{%
  \begingroup
  \renewcommand\thefootnote{}\footnote{#1}%
  \addtocounter{footnote}{-1}%
  \endgroup
}
\makeatletter
\def\blfootnote{\gdef\@thefnmark{}\@footnotetext}
\makeatother

\usepackage[symbol]{footmisc}

\usepackage{cuted}
\usepackage{capt-of} % \captionof 사용을 위해 필요

\usepackage[scaled=0.85]{beramono}
\usepackage{amsmath,amsfonts,bm}

\def\eqref#1{equation~\ref{#1}}
\def\1{\bm{1}}

\def\vb{{\bm{b}}}
\def\vc{{\bm{c}}}

\DeclareMathAlphabet{\mathsfit}{\encodingdefault}{\sfdefault}{m}{sl}
\SetMathAlphabet{\mathsfit}{bold}{\encodingdefault}{\sfdefault}{bx}{n}

\newcommand{\R}{\mathbb{R}}

\definecolor{cvprblue}{rgb}{0.21,0.49,0.74}
\usepackage[pagebackref,breaklinks,colorlinks,allcolors=cvprblue]{hyperref}
\usepackage{graphicx}

\def\paperID{7325} % *** Enter the Paper ID here
\def\confName{CVPR}
\def\confYear{2026}

\title{\OurMethod{}: Differentiable Rendering with Bitmap Primitives}
\author{Seongmin Hong$^{1,*}$, \quad \quad \quad Junghun James Kim$^{2,*}$, \quad \quad \quad Daehyeop Kim$^{3}$, \vspace{3pt}\\ \vspace{3pt} 
Insoo Chung$^{3}$, \quad \quad \quad Se Young Chun$^{1,2,3,\dagger}$\\
$^1$INMC, $^2$IPAI, $^3$Dept. of ECE, \ Seoul National University, \ Republic of Korea\\
{\tt\small \{smhongok, jonghean12, 2012abcd, insoo\_chung, sychun\}@snu.ac.kr}\vspace{2pt}
\\ 
\href{https://diffbmp.com/}{\uline{\tt diffbmp.com}}
}

\begin{document}

\maketitle
\begin{strip}

% 1. 모든 티저 이미지를 위한 Savebox 정의
\newsavebox{\boxA}
\newsavebox{\boxB}
\newsavebox{\boxCone}
\newsavebox{\boxCtwo}
\newsavebox{\boxCthree}
\newsavebox{\boxCfour}

% 2. 크기 계산 및 이미지 사전 렌더링 (단위 pt 명시)
\pgfmathsetmacro{\commonH}{0.35*\linewidth}
\pgfmathsetmacro{\teaserImagew}{0.185*\linewidth}

% (a), (b)는 높이 기준
\sbox{\boxA}{\includegraphics[height=\commonH pt]{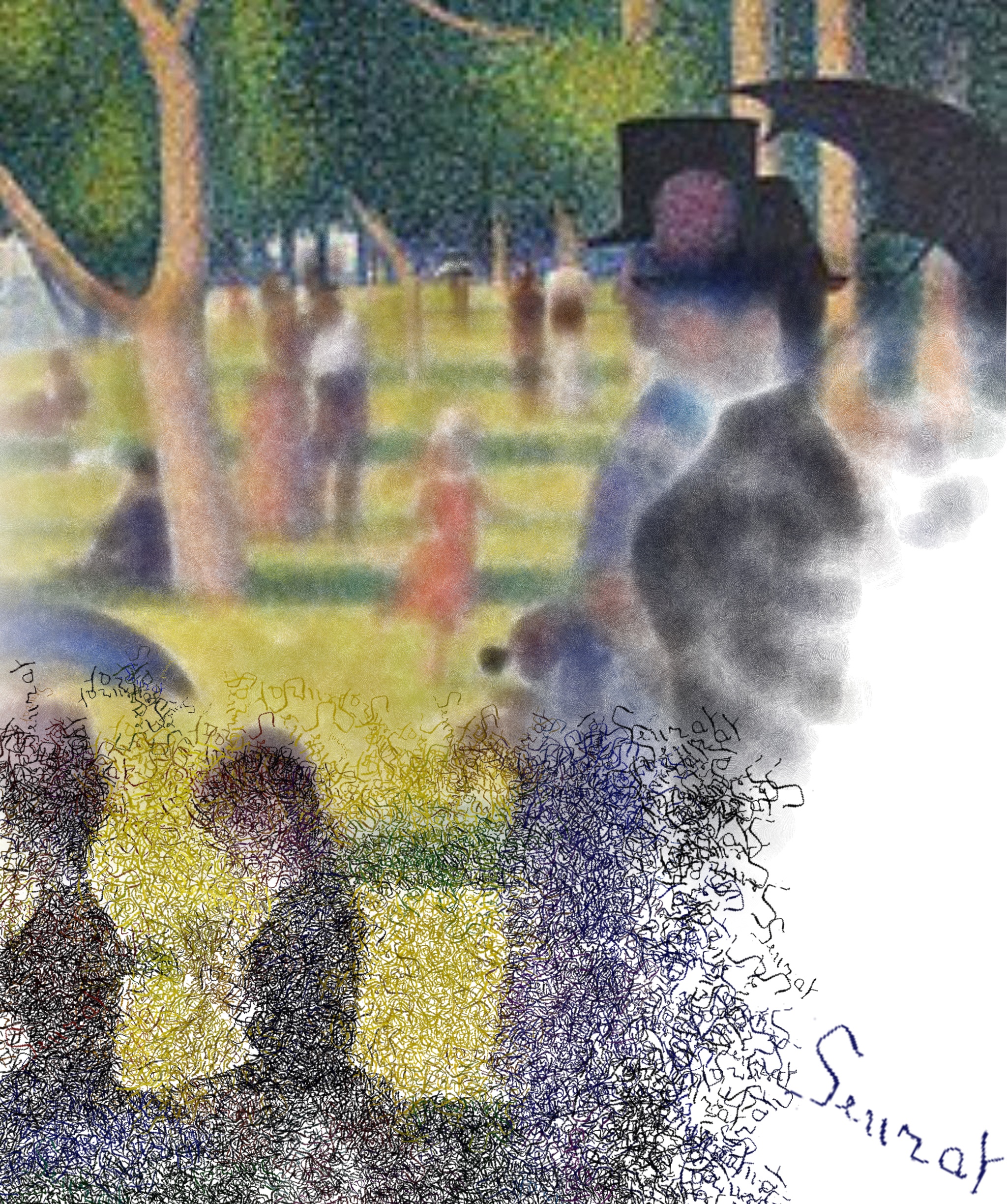}}
\sbox{\boxB}{\includegraphics[height=\commonH pt]{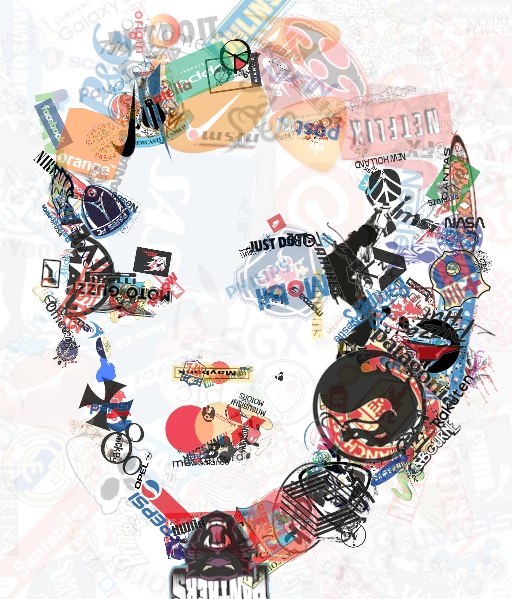}}

% (c)는 너비 기준 및 adjincludegraphics 활용
\sbox{\boxCone}{\adjincludegraphics[width=\teaserImagew pt, trim={0pt} {0.14\height} {0pt} {0.14\height}, clip]{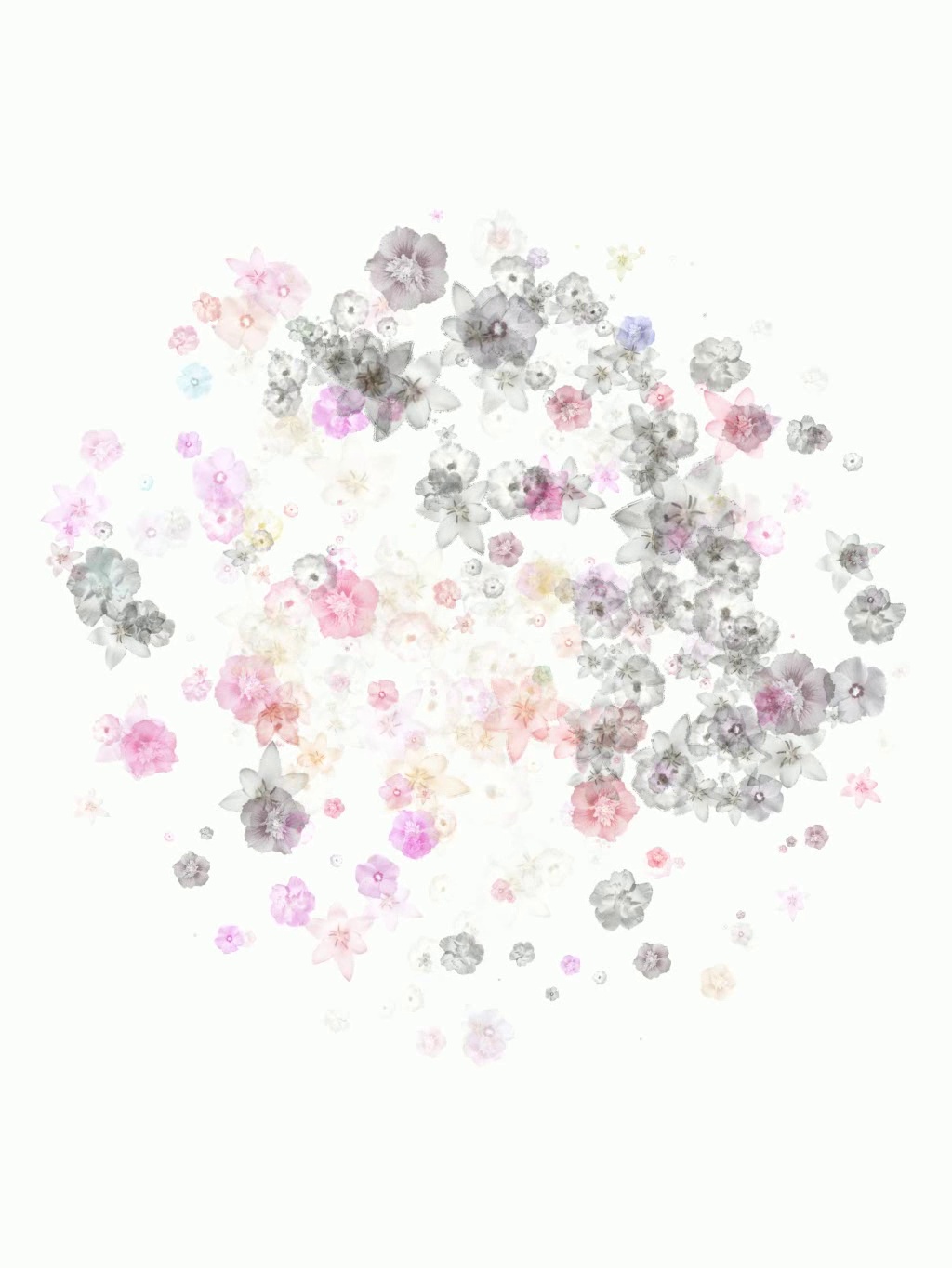}}
\sbox{\boxCtwo}{\adjincludegraphics[width=\teaserImagew pt, trim={0pt} {0.14\height} {0pt} {0.14\height}, clip]{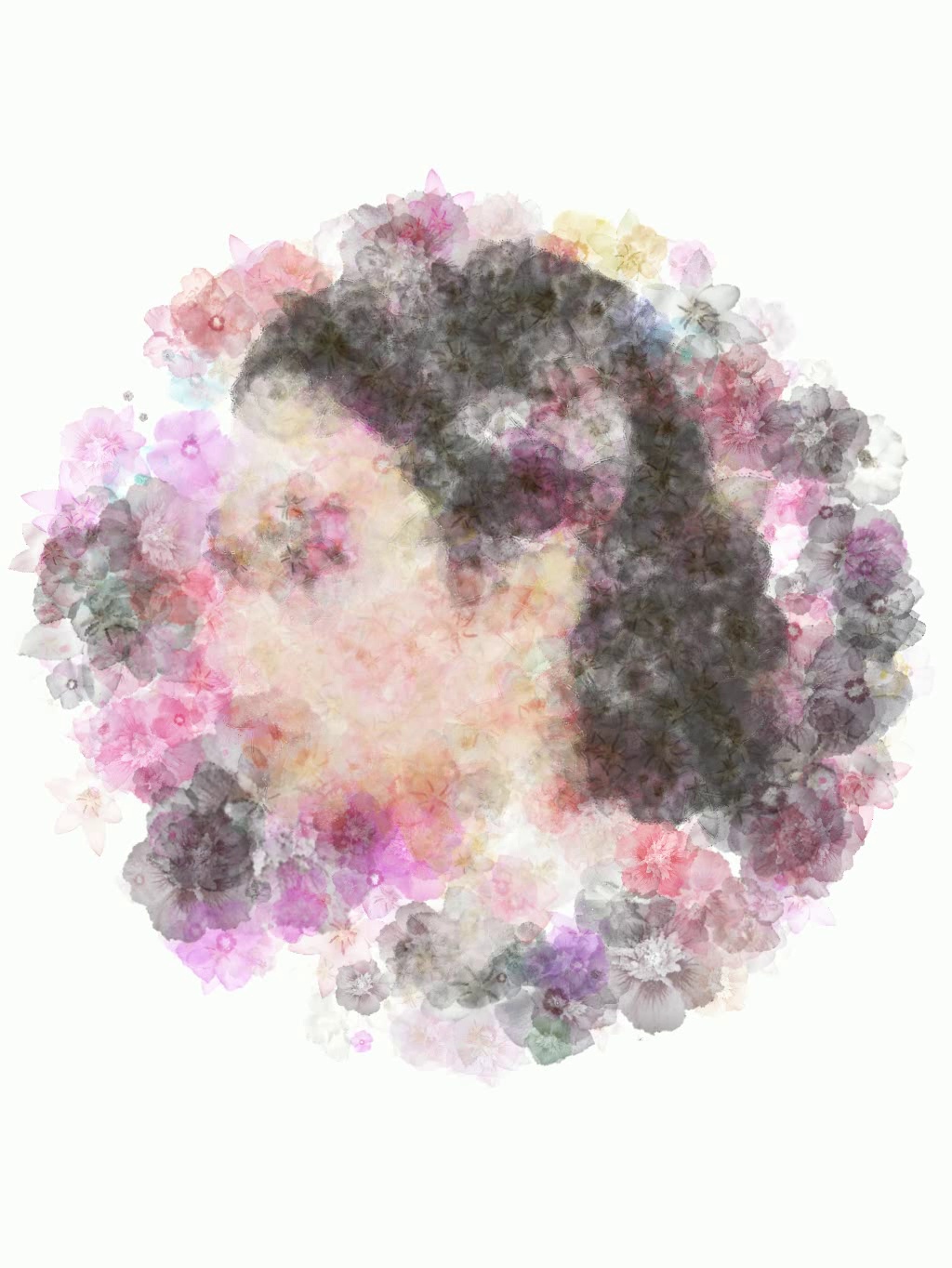}}
\sbox{\boxCthree}{\adjincludegraphics[width=\teaserImagew pt, trim={0pt} {0.145\height} {0pt} {0.145\height}, clip]{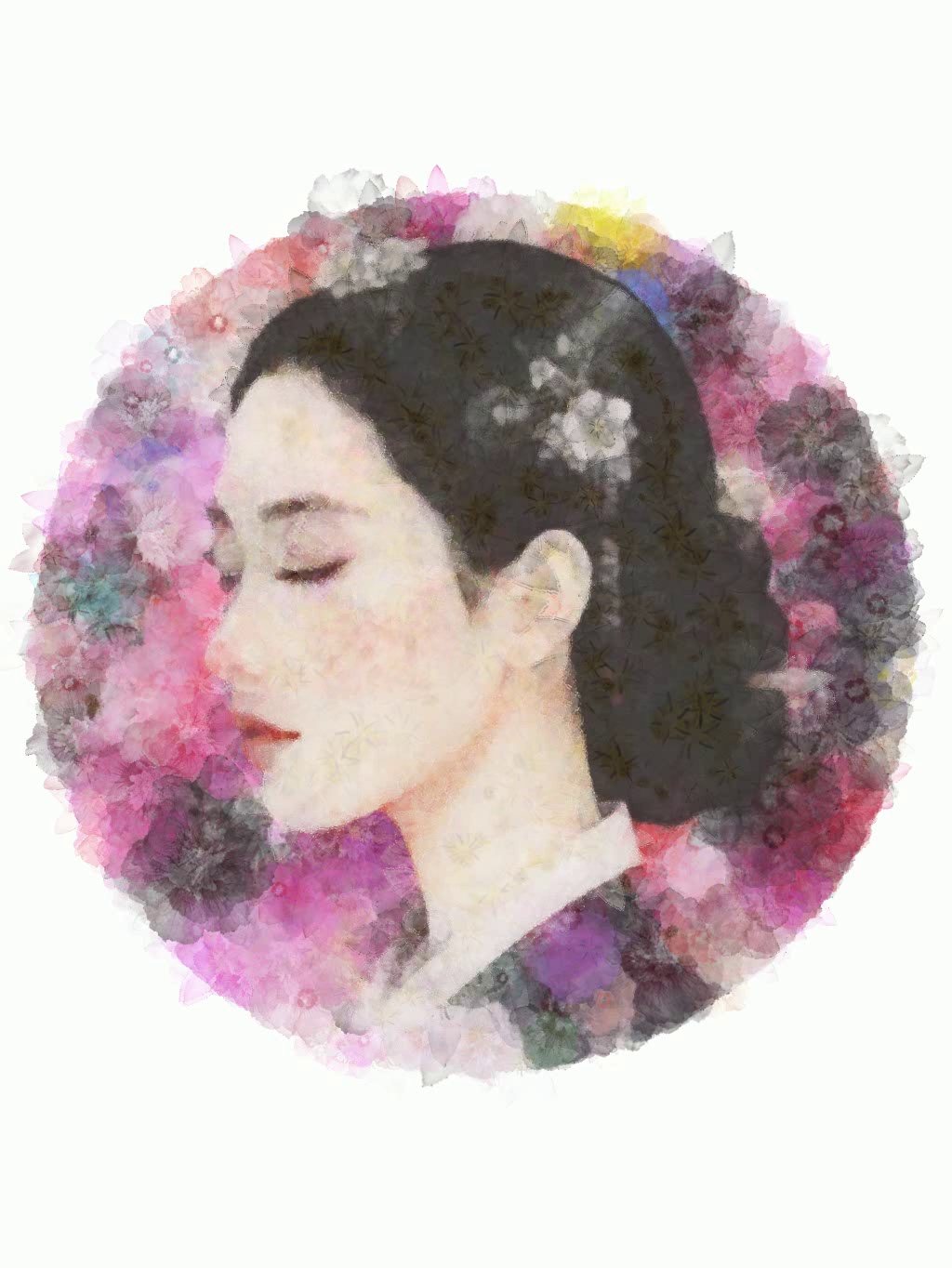}}
\sbox{\boxCfour}{\adjincludegraphics[width=\teaserImagew pt, trim={0pt} {0.145\height} {0pt} {0.145\height}, clip]{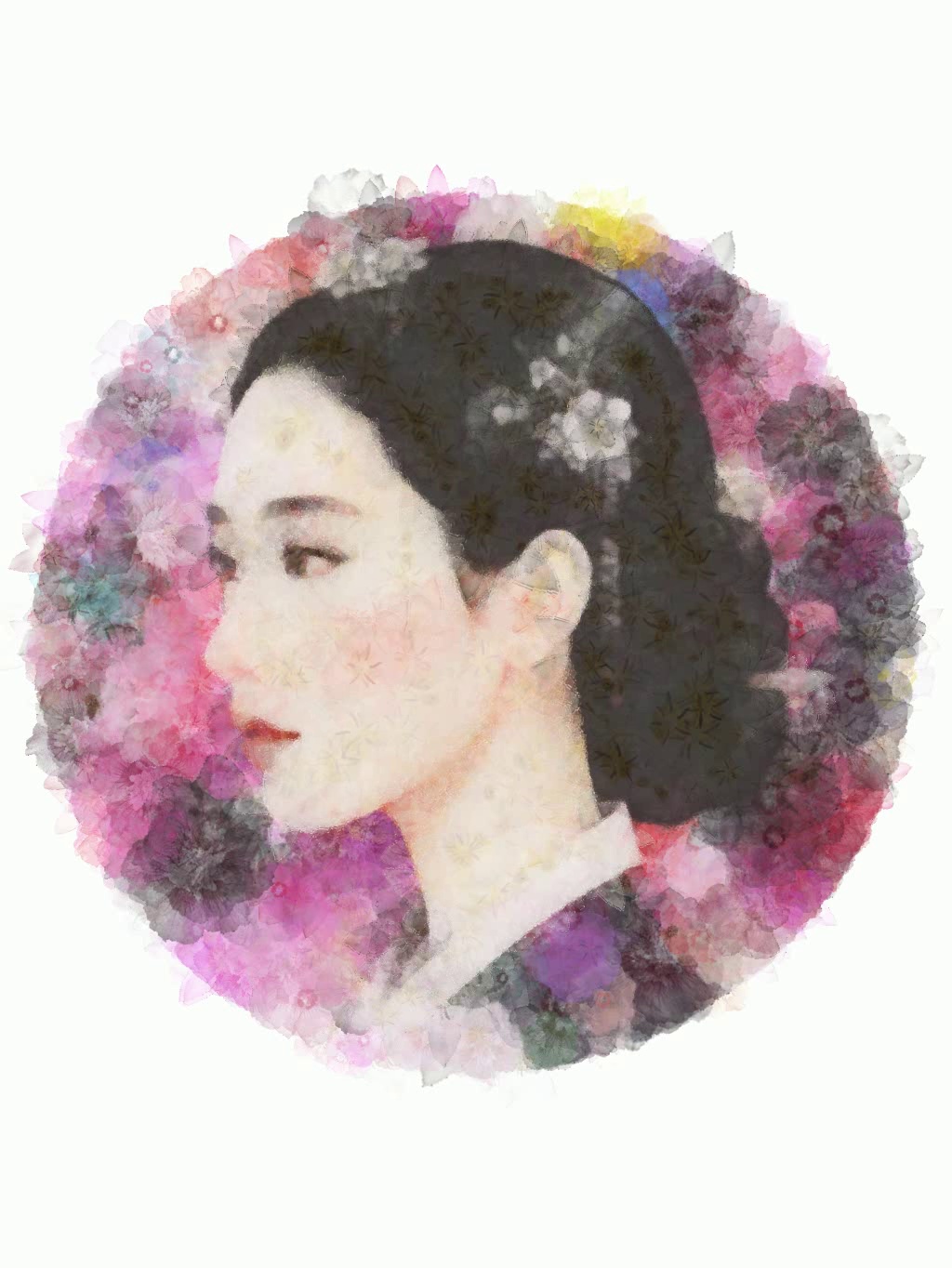}}

    \centering
    \vspace{-40pt} % 타이틀과의 간격 조절 (필요시 음수 값으로 조정)
    \makebox[\textwidth][c]{
\begin{tabular}[t]{c@{\hspace{5pt}}c@{\hspace{5pt}}c}      
    % (a) TikZ 이미지
    \begin{tikzpicture}[spy using outlines={rectangle, draw=red, line width=4pt, magnification=4}]
        \node[anchor=south west, inner sep=0pt] (img1) at (0,0) {\usebox{\boxA}};
        
        \node[lightblack_box_white_text, anchor=north east, font=\footnotesize, align=right] at (img1.north east) {\faBullseye~Painting of \textit{G. Seurat}};    
        \node[lightblack_box_white_text, anchor=north east, font=\scriptsize, align=right] at ([yshift=-0.27*\commonH pt]img1.north east) {
            \scalebox{0.88}[0.95]{\faExpand~1024$\times$809} \\ \condensedSB{VRAM}~2.0 GB\\ \faClockO~70s 
        }; 
        \node[lightblack_box_white_text, anchor=north east, font=\scriptsize, align=right] at ([yshift=-0.60*\commonH pt]img1.north east) {
            \scalebox{0.88}[0.95]{\faExpand~1024$\times$609} \\ \condensedSB{VRAM}~1.5 GB\\ \faClockO~53s
        };
        
        % Spy overlays
        \spy[magnification=6, size=1.2cm] on ($ (img1.center) !0.65! (img1.east) $) in node [anchor=west, draw=green, line width=1.5pt, fill=white, xshift=1pt] at (img1.west);
        \spy[magnification=3.0, size=1.2cm] on ([shift={(0.18cm, -0.25cm)}]$ (img1.center) !0.5! (img1.south east) $) in node [anchor=south west, draw=green, line width=1.5pt, fill=white, xshift=1pt, yshift=1pt] at (img1.south west);
    \end{tikzpicture}
    &
    % (b) TikZ 이미지
    \begin{tikzpicture}
        \node[anchor=south west, inner sep=0pt] (img2) at (0,0) {\usebox{\boxB}};
        
        \node[lightblack_box_white_text, anchor=north east, font=\footnotesize, align=right] at (img2.north east) {\faBullseye~Marilyn Monroe\\ \faExpand~512$\times$599}; 
        \node[lightblack_box_white_text, anchor=south east, font=\footnotesize, align=right] at (img2.south east) {\condensedSB{VRAM}~1.0 GB\\ \faClockO~22s}; 
    \end{tikzpicture}
    &
    % (c) TikZ 이미지
    \begin{tikzpicture}[spy using outlines={rectangle, draw=red, line width=1pt, magnification=4}]
        \pgfmathsetmacro{\nodew}{0.37*\linewidth}
        \node[anchor=south west, inner sep=0pt, minimum height=\commonH pt, minimum width=\nodew pt] (img3) at (0,0) {};      
        
        % 4분할 이미지 배치
        \node[anchor=north west, inner sep=0pt] (img3-1) at (img3.north west) {\usebox{\boxCone}};
        \node[anchor=north east, inner sep=0pt] (img3-2) at (img3.north east) {\usebox{\boxCtwo}};
        \node[anchor=south west, inner sep=0pt] (img3-3) at (img3.south west) {\usebox{\boxCfour}};
        \node[anchor=south east, inner sep=0pt] (img3-4) at (img3.south east) {\usebox{\boxCthree}};

        % 라벨 및 Spy
        \node[lightblack_box_white_text, anchor=south east, font=\footnotesize] at (img3-1.south east) {1}; 
        \spy[magnification=3, size=1.4cm] on ($ (img3-1.center) !0.5! (img3-1.north east) $) in node [anchor=south west, draw=green, line width=1.5pt, fill=white, xshift=-1pt, yshift=2pt] at (img3-1.south west);
        
        \node[lightblack_box_white_text, anchor=south west, font=\footnotesize] at (img3-2.south west) {2}; 
        \node[lightblack_box_white_text, anchor=north east, font=\footnotesize] at (img3-3.north east) {3}; 
        \node[lightblack_box_white_text, anchor=north west, font=\footnotesize] at (img3-4.north west) {4};
        
        \node[lightblack_box_white_text, anchor=north east, font=\footnotesize, align=right] at (img3-2.north east) {\scalebox{0.8}[0.9]{\faBullseye~A girl \& Flowers}\\ \faExpand~512$\times$512}; 
        \node[lightblack_box_white_text, anchor=south east, font=\footnotesize, align=right] at (img3-4.south east) {\condensedSB{VRAM}~1.8 GB\\ \faClockO~67s+7s/f}; 
    \end{tikzpicture}          
    \\ \vspace{1pt}
    \makecell{\refstepcounter{subfigure}\label{fig:teaser:a}(a) Image composition with\\ 3000 fingerprints / 2000 autographs} &
    \makecell{\refstepcounter{subfigure}\label{fig:teaser:b}(b) Intrinsic-preserving graphic \\ assemblage with 300 logos} &
    \makecell{\refstepcounter{subfigure}\label{fig:teaser:c} (c) Video composition under spatial constraint \\ w/ 2000 flowers, followed by post-processing} 
  \end{tabular}
    }
    \vspace{7pt}
    % \captionof{figure}{\textbf{Explore the new creative horizons opened by \OurMethod{}.} Essentially, \OurMethod{} makes the position, rotation, scale, color, and opacity of arbitrary bitmap images \emph{differentiable} (\eg, a fingerprint or the \textit{`Seurat'} autograph in \textcolor{cvprblue}{a}). This capability extends operations previously limited to vector graphics to any bitmap primitive (\eg, \textcolor{cvprblue}{b}: intrinsic-preserving graphic assemblage of Marilyn Monroe using numerous brand logos). \OurMethod{} is highly flexible and expandable, so that it can process videos, spatially constrained images, or both simultaneously, as shown in \textcolor{cvprblue}{c}(\boxtext{3}, \boxtext{4}).  Moreover, \OurMethod{} outputs 2D layers in an editable format ($\mathtt{.psd}$), enabling diverse post-processing creation, as the blooming effect in \textcolor{cvprblue}{c}(\boxtext{1}-\boxtext{3}). The various image/video creations, including \textcolor{cvprblue}{c}, can be viewed in the supplementary video. GPU VRAM usage and runtimes were measured on an NVIDIA RTX 3090. Note that \textcolor{cvprblue}{c}'s runtime excludes post-processing.}
    % --- 에러 방지용 수동 캡션 시작 ---
    \refstepcounter{figure} % Figure 카운터 강제 증가
    \label{fig:teaser}      % 라벨 지정 (\ref{fig:teaser} 작동함)
    \begin{minipage}{\textwidth} % 텍스트 너비 조절
        \small % 캡션 폰트 사이즈
        \textbf{Figure \thefigure.} % "Figure 1." 출력
        \textbf{Explore the new creative horizons opened by \OurMethod{}.} 
        Essentially, \OurMethod{} makes the position, rotation, scale, color, and opacity of arbitrary bitmap images \emph{differentiable} (\eg, a fingerprint or the \textit{`Seurat'} autograph in \textcolor{cvprblue}{a}). This capability extends operations previously limited to vector graphics to any bitmap primitive (\eg, \textcolor{cvprblue}{b}: intrinsic-preserving graphic assemblage of Marilyn Monroe using numerous brand logos). \OurMethod{} is highly flexible and expandable, so that it can process videos, spatially constrained images, or both simultaneously, as shown in \textcolor{cvprblue}{c}(\boxtext{3}, \boxtext{4}). Moreover, \OurMethod{} outputs 2D layers in an editable format ($\mathtt{.psd}$), enabling diverse post-processing creation, as the blooming effect in \textcolor{cvprblue}{c}(\boxtext{1}-\boxtext{3}). The various image/video creations, including \textcolor{cvprblue}{c}, can be viewed in the supplementary video. GPU VRAM usage and runtimes were measured on an NVIDIA RTX 3090. Note that \textcolor{cvprblue}{c}'s runtime excludes post-processing.
    \end{minipage}
    %\label{fig:teaser}
    \vspace{8pt}
\end{strip}

\begin{abstract} 
We introduce \textbf{\OurMethod{}}, a scalable and efficient differentiable rendering engine for a collection of bitmap images. Our work addresses a limitation that traditional differentiable renderers are constrained to vector graphics, given that most images in the world are bitmaps. Our core contribution is a highly parallelized rendering pipeline, featuring a custom CUDA implementation for calculating gradients. This system can, for example, optimize the position, rotation, scale, color, and opacity of thousands of bitmap primitives all in under 1 min using a consumer GPU. We employ and validate several techniques to facilitate the optimization: soft rasterization via Gaussian blur, structure-aware initialization, noisy canvas, and specialized losses/heuristics for videos or spatially constrained images. We demonstrate \OurMethod{} is not just an isolated tool, but a practical one designed to integrate into creative workflows. It supports exporting compositions to a native, layered file format, and the entire framework is publicly accessible via an easy-to-hack Python package.$^1$
\end{abstract}
\blfootnote{\noindent \hspace{-18pt} $^*$Equal contribution.\; $^\dagger$Corresponding author.\; $^1$\texttt{pip install pydiffbmp}}
%\blfootnote{\noindent $^1$Installation via \texttt{ pip install pydiffbmp } is available.}

% ================================================================= % 1. INTRODUCTION % ================================================================= 
\section{Introduction}
Many problems in the real world can be formulated as \emph{optimization}: $\underset{\Theta}{\min} \, \mathcal{L}(f(\Theta))$, where $\Theta\in\R^p$ is the parameter, $f$ is a system, and $\mathcal{L}$ is the loss function. For large-scale optimization problems (\ie, $p \gg 1$), first-order methods~\cite{boyd2004convex} have become the standard, due to their computational efficiency and effectiveness. However, this imposes a critical prerequisite: the gradient $\frac{\partial f}{\partial \Theta}$ must be well-defined. Ideally, it should also be computationally efficient.

In computer vision and graphics, this same principle has given rise to the field of differentiable rendering~\cite{kato2020differentiable}. To this end, significant research efforts have focused on enabling informative gradients to flow from the rendered output $f(\Theta)$ back to the 2D or 3D scene parameters $\Theta$. This has allowed a vast array of inverse graphics tasks to be efficiently and effectively solved. To obtain spatial gradients from meshes where they are not naturally defined, researchers have either approximated the backward pass~\cite{loper2014opendr, kato2018neural} or reformulated the forward pass to be differentiation-friendly~\cite{rhodin2015versatile, liu2019soft}. Similar solutions have been employed for other representations like voxels~\cite{yan2016perspective} and point clouds~\cite{roveri2018pointpronets,yifan2019differentiable} to achieve end-to-end differentiability. While the fundamental frameworks of modern volume rendering methods like NeRF~\cite{mildenhall2021nerf} and 3DGS~\cite{kerbl20233d} are inherently differentiable, considerable effort has been invested in modifying the forward pass~\cite{fridovich2022plenoxels, sun2022direct, muller2022instant} or creating custom CUDA kernels~\cite{muller2022instant, kerbl20233d, ye2025gsplat} for the efficient collection and accumulation of gradients.

Despite this progress, a critical gap remains in the 2D domain. Existing widely used frameworks are constrained to vector primitives. For example, DiffVG~\cite{li2020differentiable} and its follow-up works~\cite{li2020differentiable, ma2022towards, chen2023editable, chen2024towards} masterfully handle vector paths, which are very memory-efficient while flexible. But the vast majority of real-world 2D assets are not vector graphics; they are bitmap images. While the foundational mechanism for differentiably transforming raster images was introduced~\cite{jaderberg2015spatial}, its adoption for creation has been limited to non-general tasks like pattern composition~\cite{reddy2020discovering}. Due to the inherent nature of bitmap images—being discrete, high-dimensional arrays of pixel values—they impose a substantial memory and computational burden. Consequently, a general and scalable engine for differentiably optimizing the transforms of thousands of bitmap images has not existed. This has made it impossible to automatically create compelling compositions, such as those in \cref{fig:teaser}, using first-order optimization.

To open the door for solving a new class of problems using first-order optimization on bitmap images, we introduce \OurMethod{}: a highly parallelized differentiable rendering engine designed specifically for bitmap primitives. By leveraging custom CUDA kernels for both forward and backward passes, \OurMethod{} enables arbitrary raster images to become full participants in the gradient-based optimization ecosystem. 
Our main contributions are as follows:
\begin{itemize}
\item \textbf{Scalable Differentiable Framework:} A novel differentiable renderer that overcomes the memory and computational burdens of raster-based optimization, scaling to thousands of bitmap primitives.
\item \textbf{Research Insights \& Optimization Aids:} Analysis of gradient sparsity and convergence dynamics in bitmap optimization, addressed via soft rasterization, structure-aware initialization, and noisy canvas techniques.
\item \textbf{Algorithmic Innovations \& Extensions:} Specialized solutions for video modeling and spatially constrained rendering, integrated into artist-friendly workflows through a Python interface and $\mathtt{.psd}$ exports for post-processing.
\end{itemize}

\begin{figure*}[t]
    \centering
    \includegraphics[width=\linewidth]{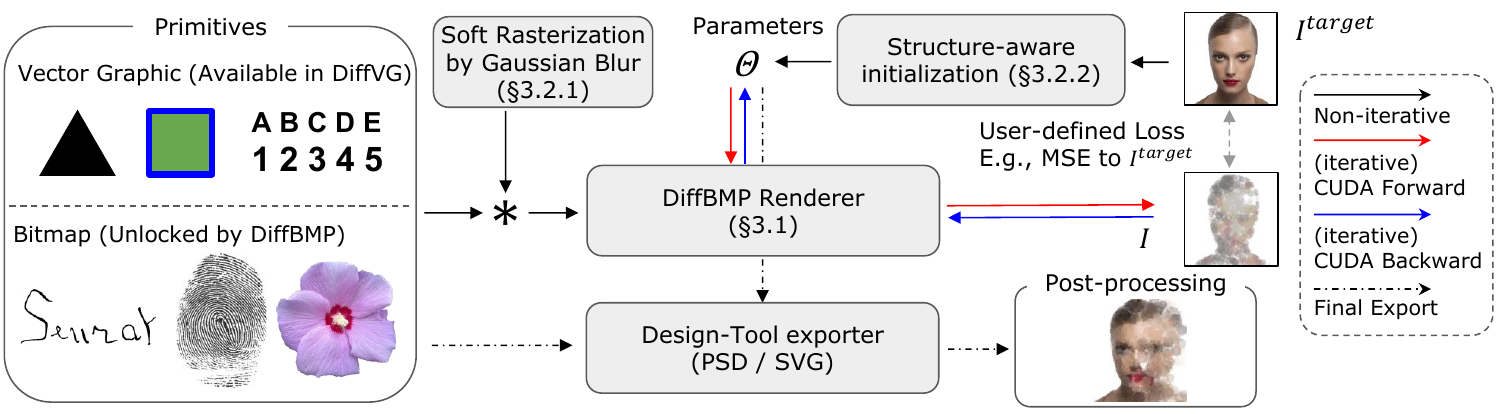}
    \caption{
    \textbf{An illustration of the algorithm flow of \OurMethod{}.} 
    Our core contribution is the renderer for bitmap images, introduced in \cref{sec:method:1}, which enables scalable and efficient differentiable rendering using a custom CUDA kernel. Additionally, we employ soft rasterization (\cref{sec:method:2:blur}) for improved gradient propagation and an initialization technique (\cref{sec:method:2:init}) for better target approximation. After optimization, a native, layered file is generated, which enables diverse post-processing creation.
    }
    \label{fig:framework}
\end{figure*}

% ================================================================= 
% 2. RELATED WORK (REVISED)
% ================================================================= 
\section{Related Work}

\begin{table*}[t]
\caption{\textbf{Comparison with Prior Work.} Prior work can be broadly classified into two main categories. DiffVG~\cite{li2020differentiable} is high-performing and broadly applicable, but it is limited to vector primitives. Due to the difficulty of matching DiffVG's performance using bitmap primitives, only a few studies, such as \citet{reddy2020discovering}, have attempted implementations for specialized tasks, but their efficiency and performance fall short of DiffVG. To put simply, \OurMethod{} aims to be the bitmap primitive counterpart to DiffVG.}
\label{tab:comparison_final}
\centering
\begin{tabular}{p{0.24\linewidth}  p{0.09\linewidth}  p{0.27\linewidth} p{0.09\linewidth} p{0.07\linewidth} p{0.09\linewidth}}
\toprule
\textbf{2D Differentiable} \newline \textbf{rendering method} & \textbf{Primitive} \newline \textbf{type}  & \textbf{Applications} &  \textbf{Opacity} \newline \textbf{Support} & \textbf{Para-}\newline\textbf{llelism} & \textbf{Engine /} \newline \textbf{Interface}  \\
\midrule
Differentiable Composition \newline by \citet{reddy2020discovering} & Bitmap & Specialized for \emph{patterns} \newline \eg, manipulation, tiling & \xmark & \xmark & Pytorch /\newline Python \\
\addlinespace
\textbf{DiffVG}~\cite{li2020differentiable} and its extensions~\cite{reddy2021im2vec, ma2022towards, chen2023editable, chen2024towards,zhou2024segmentation, zhao2025less} & 
Vector &  
Very diverse \newline \eg, vectorization, painterly, etc. & \cmark & \cmark & C++ /\newline Python \\
\addlinespace
\textbf{\OurMethod{} (Ours)} & 
Bitmap &
Very diverse. \eg, painterly, \newline assemblage, video, etc. & \cmark & \cmark & CUDA /\newline Python \\
\bottomrule
\end{tabular}
\end{table*}

\paragraph{Differentiable Rendering for Vector Graphics}\label{sec:diffvg}
Differentiable rendering first emerged in the 3D domain as a technique to compute the gradient of a rendering output (typically a mesh) with respect to its parameters \cite{de2011model, loper2014opendr, kato2018neural, chen2019learning, liu2019soft, li2018differentiable}. The first practical system to differentiably connect the parameters of vector graphics with a 2D raster image was DiffVG \cite{li2020differentiable}. Its high quality, computational efficiency, and user-friendly Python interface enabled various applications, including image vectorization \cite{reddy2021im2vec, ma2022towards, chen2023editable, chen2024towards,zhou2024segmentation, zhao2025less} and text-based vector graphics generation \cite{frans2022clipdraw, song2023clipvg, jain2023vectorfusion,xing2024svgdreamer}. Although non-DiffVG-based method \cite{liu2025bezier} has recently been proposed, but also is confined to vector graphics. One might wonder if the function of \OurMethod{} could be achieved by first vectorizing complex bitmap primitives (albeit with significant complexity), applying an existing method like DiffVG, and then simply exporting the result back to a bitmap. However, as demonstrated in $\text{\cref{sec:experiment:eval}}$, we experimentally show that DiffVG struggles significantly with bitmap-level complexity or equivalent complex SVG primitives. This difficulty underscores the necessity of the proposed \OurMethod{}.

\paragraph{Differentiable Rendering for Bitmap Images}
Differentiable rendering for bitmap images, while theoretically possible, lags significantly behind the vector graphics domain in computational efficiency and scalability. The foundational method for making raster image geometry differentiable is to apply differentiable operations, such as bilinear interpolation, to the image grid. This was introduced by Spatial Transformer Networks~\cite{jaderberg2015spatial}, originally for learning spatial invariance for feature maps within neural networks. Building on this mechanism, \citet{reddy2020discovering} proposed a method for creating patterns from bitmap images. However, this approach was implemented without specialized parallelism and was limited to a relatively narrow task: composing repetitive, opaque (opacity=1) patterns. \Cref{tab:comparison_final} compares \OurMethod{} with existing differentiable rendering methods for both vector graphics and bitmap images.

% ================================================================= 
% 3. THE \OurMethod{} FRAMEWORK 
% ================================================================= 
\section{The \OurMethod{} Framework}\label{sec:method} 
\subsection{Differentiable Forward / Backward}\label{sec:method:1} 
The core of our method is an end-to-end differentiable module that can render a collection of raster primitives with spatial and color transformations. 
To achieve practical speed/memory performance, we implement a custom CUDA kernel for the forward/backward pass of the renderer, and made available to developers through a Python interface.

\subsubsection{Forward Rendering Process Basic}\label{sec:method:1:1} 
\paragraph{Coordinate Transformation and Primitive Sampling}
Here, we present how spatial transformations are performed to bitmap primitives differentiably.
Let $N$ be the number of (bitmap) primitives. Each $i$-th primitive has parameters of position $x_i, y_i$, scale $s_i$, rotation $\theta_i$, opacity logit $\nu_i$, and RGB color logit $\vc_i \in \R^3$. Let $(x, y)$ be a coordinate on the canvas. We first transform $(x, y)$ to normalized primitive coordinates:
\begin{equation}\label{eqn:affine}
\begin{bmatrix} u \\ v \end{bmatrix} = \begin{bmatrix} \cos \theta_i & \sin \theta_i \\ -\sin \theta_i & \cos \theta_i \end{bmatrix} \begin{bmatrix} (x-x_i)/s_i \\ (y-y_i)/s_i \end{bmatrix} \in [-1,1]^2 
\end{equation}
These normalized coordinates are then mapped to discrete bitmap coordinates $(U,V)$:
\begin{equation}\label{eqn:bitmap_coords}
\begin{bmatrix} U \\ V \end{bmatrix} = \begin{bmatrix} (u\!+\!1)/2 \cdot (W_i\!-\!1) \\ (v\!+\!1)/2 \cdot (H_i\!-\!1) \end{bmatrix} \in [0, W_i\!-\!1] \times [0, H_i\!-\!1], 
\end{equation}
where $(H_i,W_i)$ is the size of the $i$-th bitmap primitive. 
Note that $U$ and $V$ are almost surely non-integer. As in STN~\cite{jaderberg2015spatial}'s differentiable image sampling, which is also used in \cite{reddy2020discovering}, we compute the primitive contribution $M_i(x,y)$ via bilinear interpolation, using four nearest integer grid points (\ie, $\{\lfloor U \rfloor, \lfloor U \rfloor\!+\!1\} \times \{\lfloor V \rfloor, \lfloor V \rfloor\!+\!1\}$), to get the gradient on spatial transformation.

\paragraph{Alpha Compositing}
The alpha value for primitive $i$ at pixel $(x,y)$ is:
\begin{equation}\label{eqn:apply_alpha} \alpha_i(x,y) = \alpha_{\max} \cdot \sigma(\nu_i) \cdot M_i(x,y),
\end{equation}
where $\sigma$ is the sigmoid function and $\alpha_{\max} \in(0,1]$ is a hyperparameter.
Using Porter-Duff over compositing~\cite{porter1984compositing}, the transmittance and final color are:
\begin{align} T_k(x,y) &= \prod_{j=0}^{k-1}(1 - \alpha_j(x,y)) \label{eqn:T_k} \\ I(x,y) &= \sum_{k=0}^{N-1} T_k(x,y) \alpha_k(x,y) \sigma(\vc_k) \in [0,1]^3 \label{eqn:I} \end{align}

\paragraph{Parallelization} 
To compute Eqs. (\ref{eqn:affine}-\ref{eqn:I}) practically, we adopt a tile-and-bin CUDA pipeline, following tile-based differentiable splatting practices as in \cite{ye2025gsplat}, adapted here for 2D bitmap primitives. On the CPU, we first partition the image plane into \(T\times T\) tiles (default \(T{=}32\)), and each primitive is assigned to every tile whose bounding box on the image plane overlaps the tile region with a small padding margin. On the GPU, we launch one CUDA thread block per tile with \(T\times T\) threads, for achieving complete pixel-level parallelism. Threads in a block cooperatively stage per-primitive parameters in shared memory and composite the tile-local list in front-to-back order for computing \cref{eqn:T_k,eqn:I}.

\paragraph{Optional constraint on $\Theta$} To provide finer user control, we optionally apply a constraint on $\Theta$. Especially, we preserve each primitive’s original color $\vc_i^{\text{org}}$ by blending it with a learnable color $\vc^\text{var}_i$ , enabling a unique application. To this end, we fix $\vc^\text{org}_i$ by a ratio defined by the hyperparameter $\mu_\text{blend} \in [0,1]$ within the set of $\vc_i$'s, and only update the remaining $(1-\mu_\text{blend})$ as follows:
\begin{equation}
    \vc_i = \mu_\text{blend} \vc^\text{org}_i + (1 - \mu_\text{blend}) \vc^\text{var}_i
\end{equation}
For example, $\mu_\text{blend} = 1$ was used in \cref{fig:teaser:b} to use the original colors in the output; otherwise, $\mu_\text{blend} = 0$ is default.

\subsubsection{Backward Pass in Half Precision}\label{sec:backward}
\paragraph{Gradient calculation} The gradients with respect to position, scale, and rotation are computed via the chain rule through the coordinate transformation:
\begin{equation}\label{eqn:backward}
    \small
    \frac{\partial I(x,y)}{\partial x_i} = \frac{\partial I(x,y)}{\partial M_i(x,y)} \left ( \frac{\partial M_i(x,y)}{\partial u} \frac{\partial u}{\partial x_i} + \frac{\partial M_i(x,y)}{\partial v} \frac{\partial v}{\partial x_i} \right )
\end{equation}
For simplicity, we only show $\frac{\partial I(x,y)}{\partial u}$ in \cref{eqn:backward}; gradients with respect to $\{y_i, s_i, \theta_i\}$ are similar with $x_i$ case. We can exactly calculate (\ie, do not need any approximations) $\frac{\partial I(x,y)}{\partial M_i(x,y)}$ using Eqs. (\ref{eqn:apply_alpha}, \ref{eqn:T_k}, \ref{eqn:I}). We can also easily obtain $\frac{\partial M_i(x,y)}{\partial u}$ using \cref{eqn:bitmap_coords} and regarded bilinear interpolation~\cite{jaderberg2015spatial}. At last, we get $\frac{\partial u}{\partial x_i}, \frac{\partial u}{\partial y_i}, \frac{\partial u}{\partial s_i}$ and $\frac{\partial u}{\partial \theta_i}$ from \cref{eqn:affine}. Gradients with respect to color and alpha are more straightforward than the geometric parameters, so we provide explanations in the supplementary material. 

\paragraph{Half-precision for efficiency} For precision and performance, texture fetches and per-pixel temporaries use FP16 (packed as $\mathtt{\_\_half2}$ where profitable), while accumulation of the displayed color/transmittance is evaluated in FP16 arithmetic tailored to our packed data path. This choice follows established mixed-precision practice that reduces bandwidth and VRAM usage~\cite{micikevicius2017mixed}. In the backward pass, per-parameter gradients are accumulated \emph{per pixel} directly into packed $\mathtt{\_\_half2}$ buffers via $\mathtt{\_\_half2}$ $\mathtt{atomicAdd}$ over two-channel gradient groups, followed by a lightweight post pass that unpacks to legacy FP16 arrays. This design omits block-local reductions and aligns with our accuracy–throughput goal.

\subsubsection{Dedicated CUDA Kernel for Export}
Our optimization kernels (\cref{sec:method:1:1,sec:backward}) use atomic operations, which are unsuitable for generating layered PSD files. We therefore implement a dedicated export kernel that uses primitive-level parallelism to efficiently render isolated, editable layers without backward pass overhead. This architecture provides a key advantage: optimization can run at a low resolution while the final, high-quality PSD is exported at a much higher resolution (\eg, 2$\times$, 4$\times$). Full implementation details are provided in the supplementary material.

\subsection{Techniques for Improved Optimization}\label{sec:method:2}
\OurMethod{} provides significant flexibility throughout the algorithm, including its hyperparameters, to reflect the diverse intentions of a creator. Nevertheless, just as meaningful artistic expression relies on foundational skill to precisely realize a vision, this creative flexibility is only powerful if it can achieve high fidelity to the target image. We therefore introduce three techniques to achieve this: (\cref{sec:method:2:blur}) gradient enriching via soft rasterization, (\cref{sec:method:2:init}) a method for achieving good initialization based on the target image and (\cref{sec:method:3:uniform}) blending uniform noise canvas to encourage primitives to fill all image regions up.

\begin{figure}[t]
    \centering
    \begin{subfigure}[b]{0.30\linewidth}
        \centering
        \includegraphics[width=\linewidth]{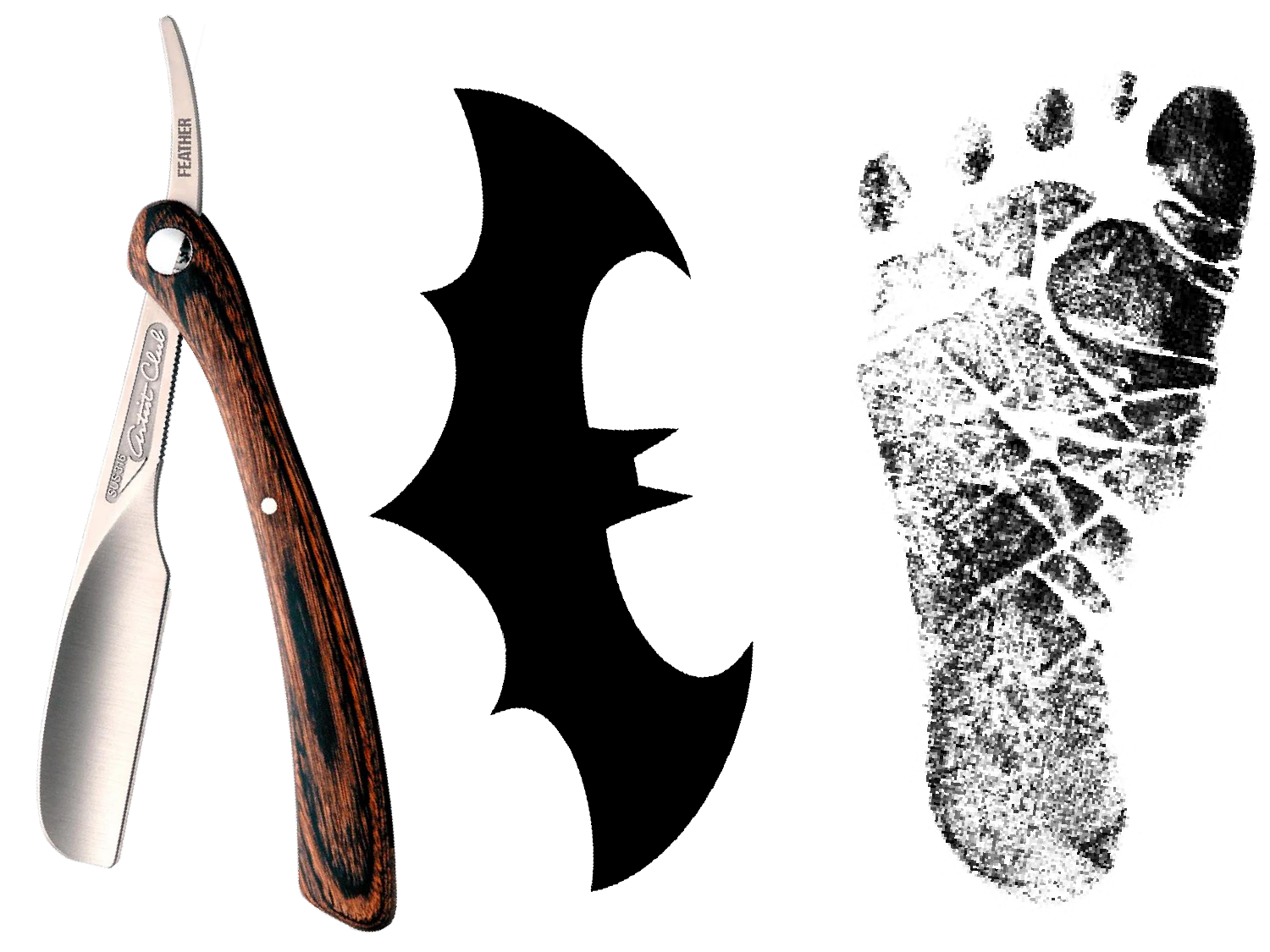}
        \caption{Primitives}
        \label{fig:primitive}
    \end{subfigure}
    \hfill
    \begin{subfigure}[b]{0.30\linewidth}
        \centering
        \includegraphics[width=\linewidth]{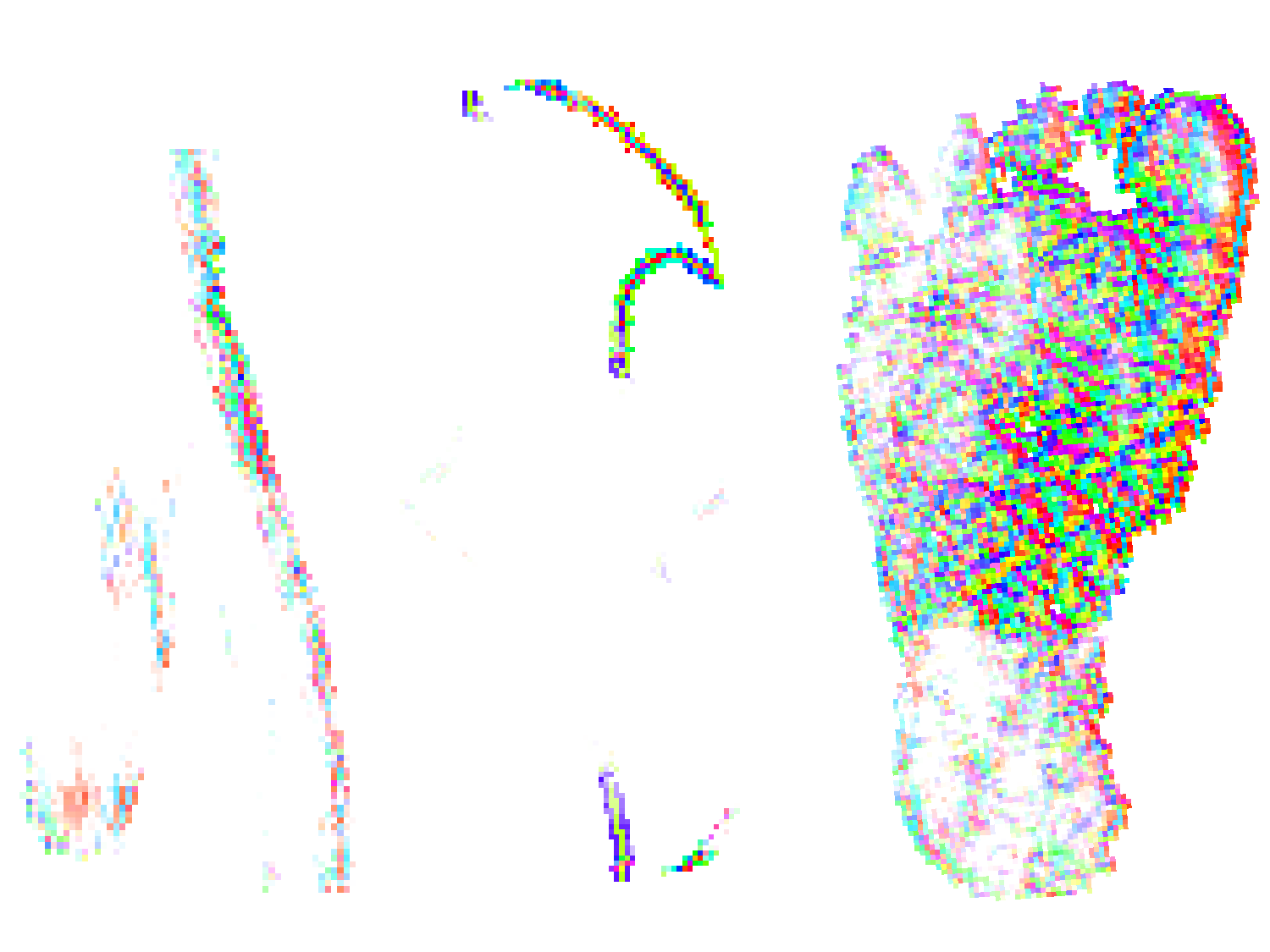}
        \caption{w/o blur}
        \label{fig:blur:no}
    \end{subfigure}
    \hfill
    \begin{subfigure}[b]{0.30\linewidth}
        \centering
        \includegraphics[width=\linewidth]{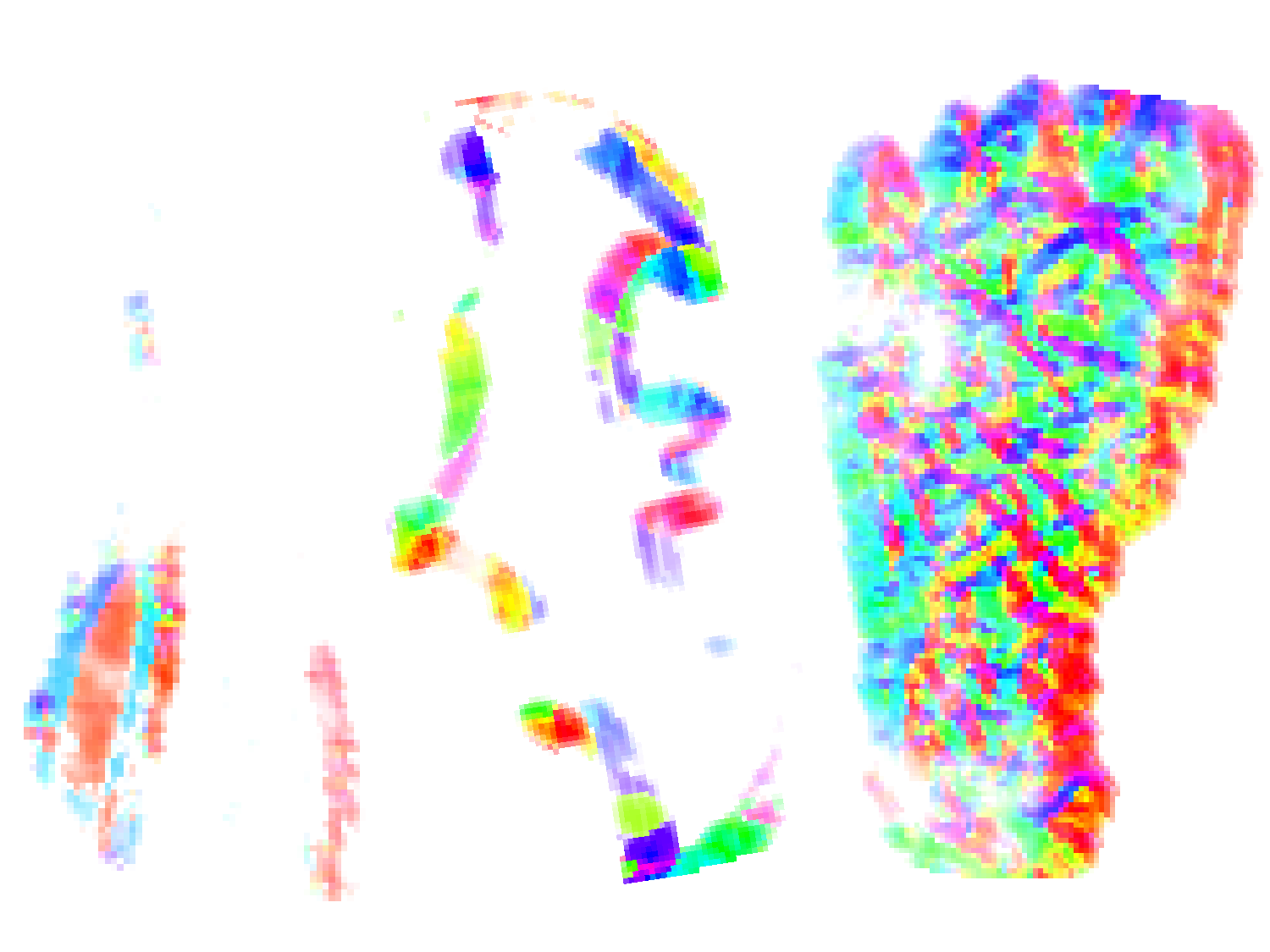}
        \caption{w/ blur ($\S\ref{sec:method:2:blur}$)}
        \label{fig:blur:yes}
    \end{subfigure}
    \caption{Blur enriches the spatial gradients of primitives. The figure displays the per-pixel gradients with respect to $x_i$ and $y_i$ for three different primitives (a), extracted during the experiments of \cref{tab:ablation_result}, both (b) with and (c) without applying blur (\cref{sec:method:2:blur}). Color indicates the direction of the gradient, and opacity represents its magnitude on a log scale (from $10^{-6}$ to $10^{-3}$). Applying blur results in richer and more coherent gradients. This, in turn, leads to the better optimization results shown in \cref{tab:ablation_result}.}
    \label{fig:three_subfigures}
\end{figure}
\subsubsection{Blurring Primitives for Enriching Gradients}\label{sec:method:2:blur}
In \cref{sec:backward}, we established a well-defined gradient path. However, its quantity and extent could be improved for better optimization. In \cref{fig:blur:no}, the vanilla implementation (\ie, without blurring) shows that $\frac{\partial |I(\cdot,\cdot)-I^{\text{target}}(\cdot,\cdot)|^2}{\partial (x_i,y_i)}$ occurs in relatively fewer pixels. This sparsity arises because the gradient is generated by the non-uniformity of the four operands of bilinear interpolation, which typically have similar values everywhere except at the object boundaries. To enrich (or align) such sparse gradients, soft rasterization (\ie, smoothing the edge of the primitives) is a well-known and prevailing solution, introduced in many prior works~\cite{rhodin2015versatile, liu2019soft, petersen2019pix2vex, fischer2023plateau, reddy2020discovering}. Following these approaches, we simply apply Gaussian blur to each primitives before optimization. As seen in \cref{fig:blur:yes}, this resulted in enriched and aligned gradients with no significant computational cost, consequently leading to better optimization results, which will be shown in \cref{sec:exp:ablation}.

\subsubsection{Structure-aware Initialization}\label{sec:method:2:init}
Effective initialization is critical for gradient-based optimization. We design a structure-aware initialization strategy that adapts primitive placement and scale to local image complexity. We compute local variance (of $I$) across RGB channels using a 7$\times$7 sliding window, which identifies regions requiring fine detail (high variance) versus smooth areas (low variance). Initial $(x_i,y_i)$ are sampled among lattice points with probability $\propto 0.1 + 0.9 × \mathrm{NLV}(x,y)$, where $\mathrm{NLV}:\mathrm{canvas} \rightarrow [0,1]$ is the normalized local variance. $s_i$ was initialized with $s_i = s_{\max} - (s_{\max} - s_{\min}) \cdot\mathrm{NLV}(x_i,y_i) $. This ensures dense and fine coverage in complex regions while maintaining sparse and course coverage in flat areas. We set $c_i \sim \mathcal{N}(I(x_i,y_i),\sigma_c^2)$ for warm start, while we fixed
$\nu_i=-2.0$ ($\approx$12\% opacity) to ensure gradient flow through all layers. $\theta_i$ is sampled uniformly in $[0, 2\pi)$.

\subsubsection{Canvas with Uniform Noise to Impose Primitives}\label{sec:method:3:uniform}
A problem arises when a target region shares the same color as the canvas background, as this can prevent primitives from properly splatting onto such regions. We provide an optional mechanism to enforce the placement of primitives in these areas (especially for \cref{sec:3.3.2}): \emph{setting the canvas background} $\vb(x,y) \sim \mathcal{U}[0,1]^3$. This optional technique modifies the forward function (\cref{eqn:I}) as follows:
\begin{equation}\label{eqn:uniform noise background blending}    
    I_{\text{FG+BG}} = I_{\text{FG}} + T_N\odot \vb.
\end{equation} 
This idea is modified from \cite{tian2022noisebg}'s work for triangle mesh rendering~\cite{laine2020modular}. Unlike \cite{tian2022noisebg}, which samples $\vb$ five times to average the gradients, we sample $\vb$ only once per iteration.

% ================================================================= 
% 4. APPLICATIONS 
% ================================================================= 
\subsection{Heuristics and Losses for Dynamic and Spatially Constrained Rendering}\label{sec:applications} 
%The \OurMethod{} engine is highly flexible and can be applied to various tasks. In this section, we present some example applications.
While \OurMethod{} defaults to ordinary images on a rectangular canvas as the target, it should be capable of handling videos and spatially constrained images for wider applicability. This section details the heuristics and losses designed to achieve this.

\subsubsection{Dynamic \OurMethod{} for Videos}\label{sec:4.1} 
\OurMethod{} can be extended to model dynamic content. A scalable and straightforward way to achieve both \emph{frame-wise fidelity} and \emph{anti-flicker} is to optimize sequentially, as in \cite{luiten2024dynamic} for 3DGS. \cite{luiten2024dynamic} initializes the current frame’s parameters $\Theta^{f}$ from the \textit{converged} parameters of the previous frame $\Theta^{f-1\star}$. 
However, we observed that some highly visible primitives (with large $s_i$ and $\alpha_i$, and ordered at front) hinder the sequential optimization and \emph{get stuck} at the previous frame, perhaps because they block gradients to other primitives. As shown in \cref{Dynamic:RemStk}, simply removing them improved frame-wise fidelity.

\cite{luiten2024dynamic} optimizes background primitives only in the first frame and uses the same primitives in the following frames. However, since \OurMethod{} does not explicitly distinguish the background, we heuristically (using frame-difference or flow-consistency mask) identify \emph{unchanged regions} and reuse their parameters without re-optimization, which improved anti-flicker. See \cref{sec:exp:ablation} for the results.

\begin{figure}
\centering
\begin{tikzpicture}
  % ===== Row 1: four small images =====
  \begin{pgfonlayer}{bg}

    % Triplet (a,b, c) to the right; aligned and evenly spaced
    \node[imgSmall] (S2) at (0,0)
      {\includegraphics[width=\imgSmall]{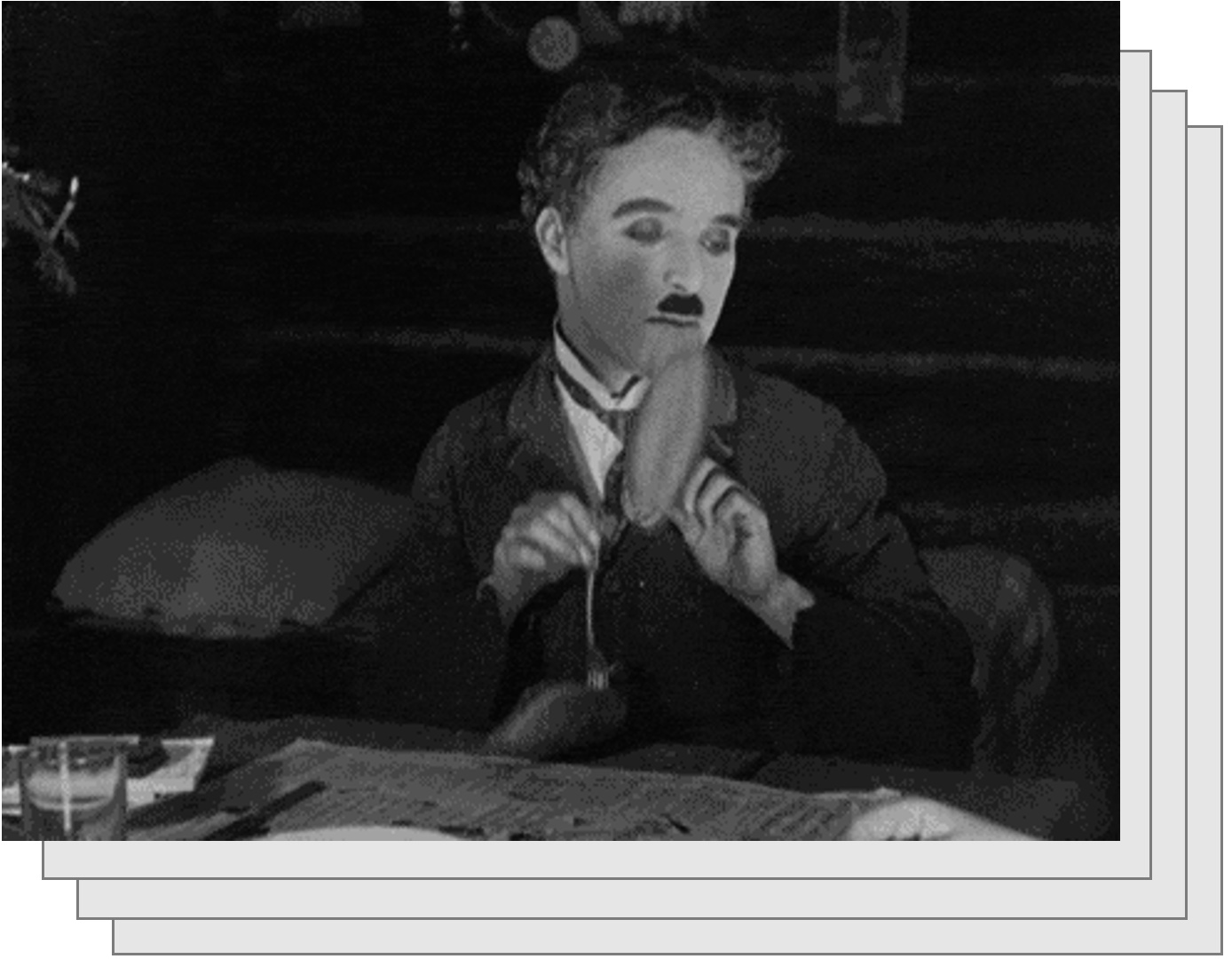}};
    \node[imgSmall] (S3) at ($(S2.east)+(\xGap,0)$)
      {\includegraphics[width=\imgSmall]{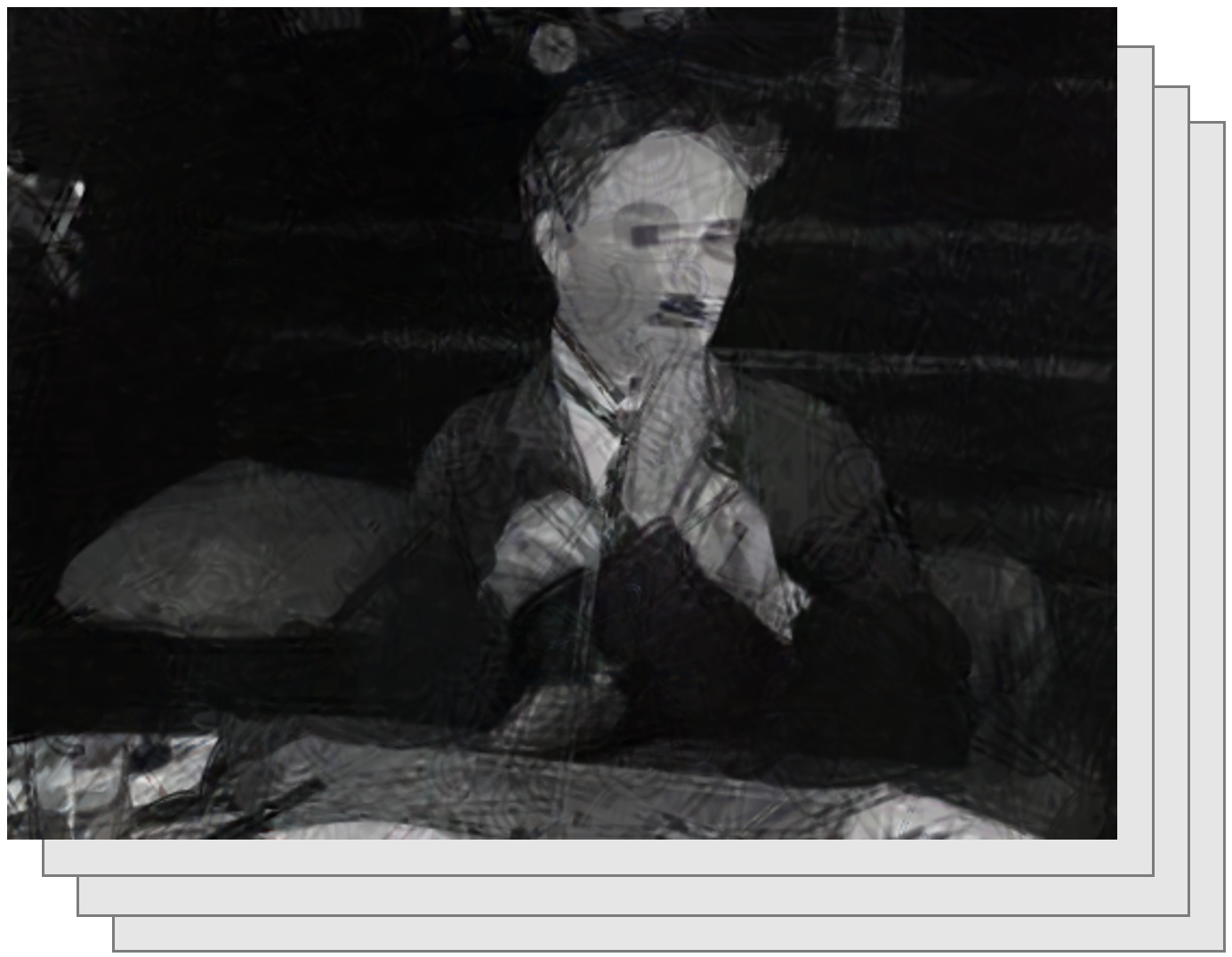}};
    \node[imgSmall] (S4) at ($(S3.east)+(\xGap,0)$)
      {\includegraphics[width=\imgSmall]{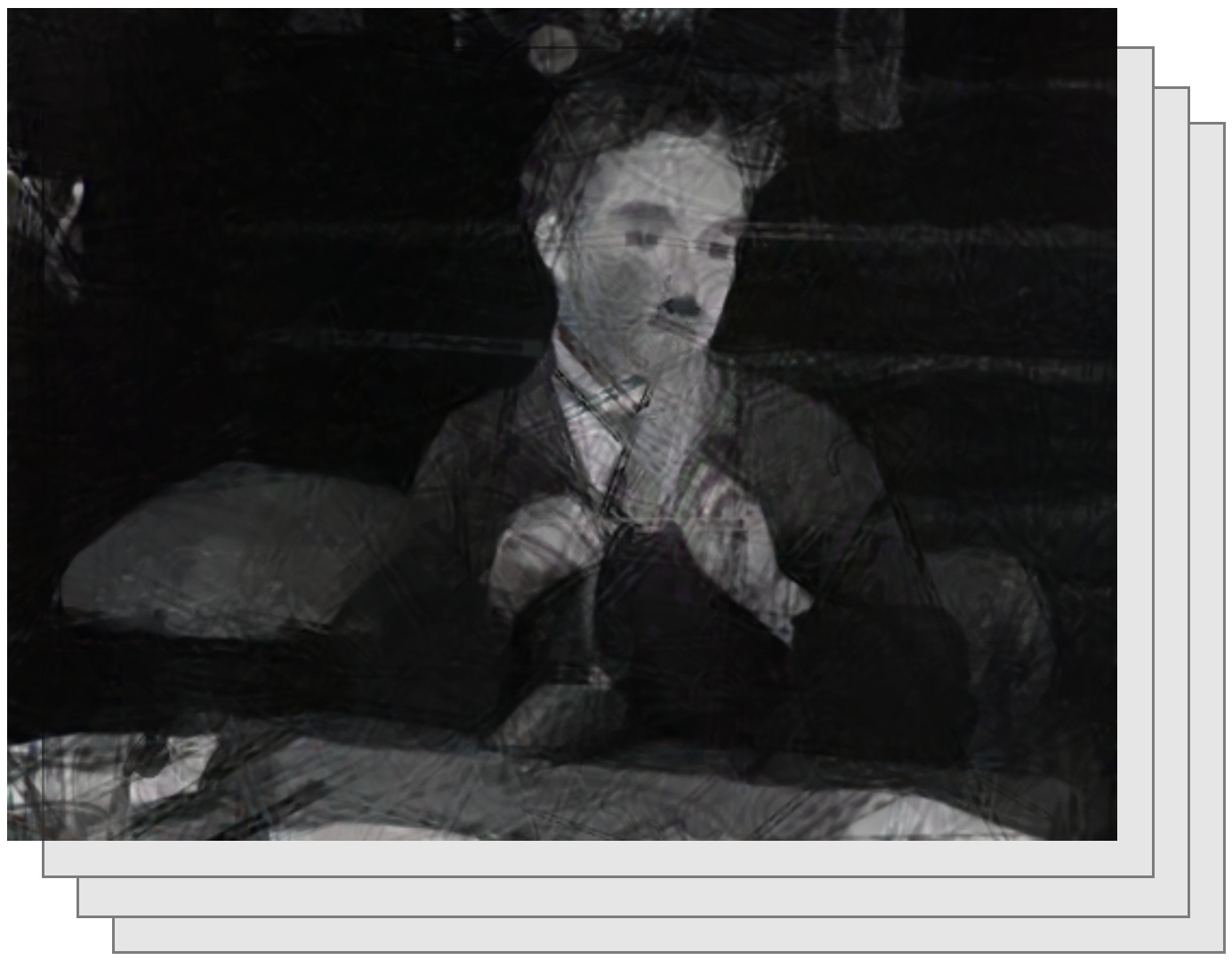}};
  \end{pgfonlayer}

% ===== Row 2: three large images (non-overlapping, L2 shifted left of S2) =====
\begin{pgfonlayer}{bg}
  % L2: shifted left by \Lshift relative to S2
  \node[imgLarge] (L2) at ($(S2.south west)+(0,-\yDrop)$)
    {\includegraphics[width=\imgLarge]{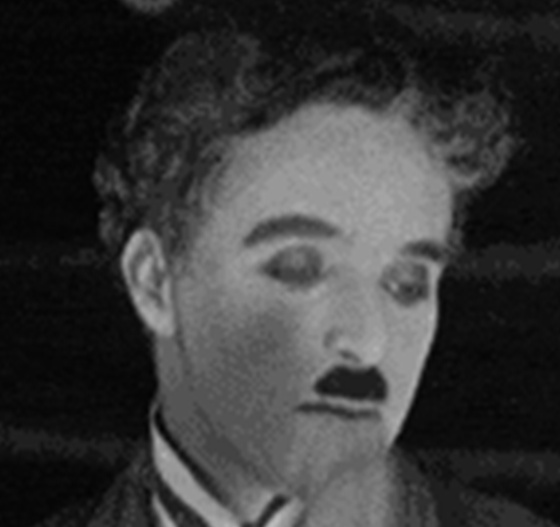}};

  % L3: placed to the right of L2 by a fixed gap; guarantees no overlap
  \node[imgLarge] (L3) at ($(L2.east)+(\xGapLarge,0)$)
    {\includegraphics[width=\imgLarge]{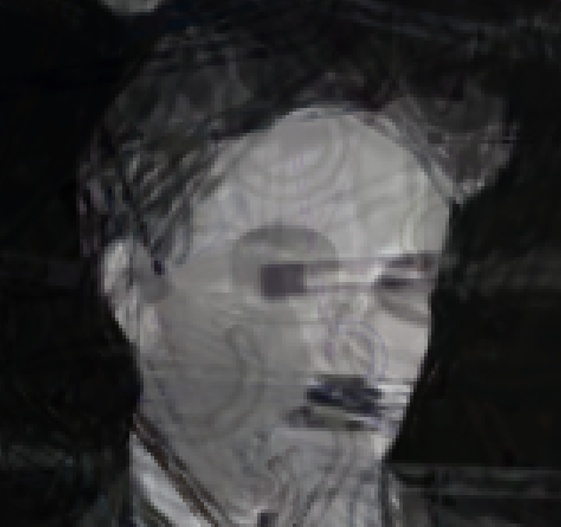}};

  % L4: same strategy (to the right of L3)
  \node[imgLarge] (L4) at ($(L3.east)+(\xGapLarge,0)$)
    {\includegraphics[width=\imgLarge]{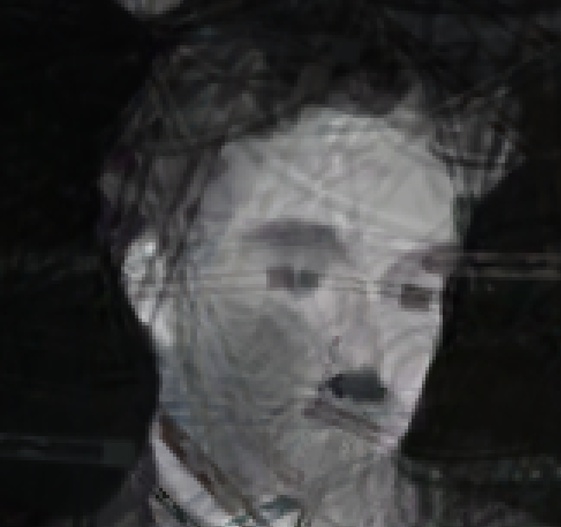}};
\end{pgfonlayer}

  % Headings above row-1 images
  \node[toplabel, anchor=north, font=\footnotesize] at ($(L2.south)+(0,-7pt)$) {(a) Target Frame};

\node[toplabel, anchor=north, font=\footnotesize] at ($(L3.south)+(0,-4pt)$) {%
  {%
    \shortstack{(b) Initialized \\ from $\Theta^{f-1\star}$}%
  }%
};

\node[toplabel, anchor=north, font=\footnotesize] at ($(L4.south)+(0,-4pt)$) {%
  {%
    \shortstack{(c) (b) + Removing \\ Stuck primitives}%
  }%
};

% Example fractions (tune these)
\def\xlo{0.38}\def\ylo{0.56}\def\xhi{0.65}\def\yhi{0.92}

% In your current figure (since you have S2,S3,S4 and L2,L3,L4):
\relboxlinkto{S2}{L2}{\xlo}{\ylo}{\xhi}{\yhi}
\relboxlinkto{S3}{L3}{\xlo}{\ylo}{\xhi}{\yhi}
\relboxlinkto{S4}{L4}{\xlo}{\ylo}{\xhi}{\yhi}

% large row
\def\xlo{0.05}\def\ylo{0.02}\def\xhi{0.95}\def\yhi{0.95}
\relbox{L2}{\xlo}{\ylo}{\xhi}{\yhi}
\relbox{L3}{\xlo}{\ylo}{\xhi}{\yhi}
\relbox{L4}{\xlo}{\ylo}{\xhi}{\yhi}

\begin{pgfonlayer}{fg}
  \node[lightblack_box_white_text, anchor=north east, font=\scriptsize, align=right, inner sep=1pt] at ($(L2.north east)+(-3pt,3pt)$) {\faFilm~8 frames\\ \faExpand~256$\times$192};
  \node[lightblack_box_white_text, anchor=north east, font=\scriptsize, align=right, inner sep=1pt] at ($(L3.north east)+(-3pt,3pt)$) {29.26 dB \\ \faClockO~4.7s, \condensedSB{VRAM}~\scalebox{0.88}[0.95]{0.8GB}};
  \node[lightblack_box_white_text, anchor=north east, font=\scriptsize, align=right, inner sep=1pt] at ($(L4.north east)+(-3pt,3pt)$) {29.58 dB \\ \faClockO~4.8s, \condensedSB{VRAM}~\scalebox{0.88}[0.95]{0.8GB}};
\end{pgfonlayer}

% ---- images (drawn earlier)
\begin{pgfonlayer}{fg}
  \node[lightblack_box_white_text, anchor=south east, font=\scriptsize, align=right] (prim_label) at ($(S2.south east)+(-5pt,0.22\imgSmall)$) {Primitives};
\end{pgfonlayer}

\begin{pgfonlayer}{main}
  \node[imgLarge, anchor=north east] (primitives) at ($(prim_label.south east)+(0,1pt)$)
    {\includegraphics[height=0.45\imgLarge, angle=90]{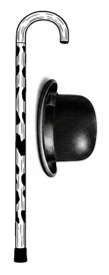}};
\end{pgfonlayer}

\end{tikzpicture}
\caption{\textbf{Heuristics for Dynamic \OurMethod{}.} Sequential optimization (\ie, initializing $\Theta^{f}$ with $\Theta^{f-1\star}$) can lead to local minima, as in \textcolor{cvprblue}{b}, where a large primitive is stuck on the face. \textcolor{cvprblue}{c} shows our remedy: reducing the opacity of such overly dominant primitives, resulting in PSNR gain.}
\label{Dynamic:RemStk}
\end{figure}

\begin{table*}[t]
\caption{\textbf{Per-iteration performance across implementations and resolutions.} Each cell reports a triple ``forward / backward / VRAM'' in the order of forward runtime, backward runtime, and total VRAM usage per iteration. Rows list the compared implementations (\eg, PyTorch baseline and our custom CUDA kernels in FP32/FP16), and columns specify the target image resolution and tile size.
}
\label{tab:evaluation}
\centering
\setlength{\tabcolsep}{3pt}
\resizebox{\linewidth}{!}{%
\begin{tabular}{@{}llcccc@{}}
\toprule
   \multicolumn{2}{l}{Image res \;/\; Tile size}  & 512$\times$512 \;/\; 16$\times$16 & 1024$\times$1024 \;/\; 16$\times$16 & 1024$\times$1024 \;/\; 32$\times$32 & 2048$\times$2048 \;/\; 32$\times$32  \\ \midrule
   \multirow{2}{*}{\shortstack[l]{PyTorch}}
   & RTX 3090 & 1360ms / 2337ms / 6.4 GB & 5514ms / 9811ms / 12.0 GB & 1393ms / 2477ms / 5.0 GB & 5405ms / 9483ms / 9.0 GB \\ 
   & L40S & 1342ms / 2413ms / 6.6 GB & 4570ms / 8151ms / 12.2 GB & 1423ms / 2507ms / 5.2 GB & 4914ms / 8707ms / 9.2 GB \\ 
   \multirow{2}{*}{CUDA-32bit} 
   & RTX 3090 & 3.9ms / 11.6ms / 1.0 GB & 7.2ms / 9.7ms / 2.1 GB & 7.6ms / 9.3ms / 2.0 GB & 16.1ms / 10.0ms / 6.1 GB \\
   & L40S & 4.7ms / 2.9ms / 1.2 GB & 10.5ms / 2.8ms / 2.3 GB & 8.5ms / 2.9ms / 2.2 GB & 23.3ms / 3.8ms / 6.3 GB \\ 
   \multirow{2}{*}{CUDA-16bit} 
   & RTX 3090 & 2.3ms / 6.2ms / 1.1 GB & 4.2ms / 5.8ms / 1.6 GB & 4.3ms / 5.5ms / 1.6 GB & 9.0ms / 6.4ms / 3.8 GB \\ 
   & L40S & 2.0ms / 2.1 ms / 1.2 GB & 5.4ms / 2.3ms / 1.8 GB & 4.5ms / 2.3ms / 1.8 GB & 12.8ms / 3.0ms / 4.0 GB \\ 
   \bottomrule
\end{tabular}
}
\end{table*}

\subsubsection{Rendering with Spatial Constraint}\label{sec:3.3.2}
Our model supports rendering only the foreground, enabling downstream applications such as appearance editing. So our goal is:  $\forall i, I^\text{target}_\alpha(x,y) = 0 \Rightarrow M_i(x,y) = 0$, where $I^\text{target}_\alpha$ is the target image's alpha.
We optimize $\Theta$ by applying the following loss to reduce the opacity of background primitives:
\begin{equation} \label{eqn:obj_loss}
    \mathcal{L} = \| (I^\text{target}_\alpha>0) \odot (I- I^\text{target})|_2^2 + \lambda_{\alpha} \| I_\alpha - I^\text{target}_\alpha \|_2^2,
\end{equation}
where $I_\alpha= 1- T_{N}$ from \cref{eqn:T_k}. Instead of pruning the transparent primitives as in \cite{rogge2025OC2DGS}, we re-initialize these primitives for further applications. See the results in \cref{fig:no_bg_result}.

% ================================================================= 
% 5. EXPERIMENTS AND RESULTS 
% ================================================================= 
\section{Evaluations and Applications} We conduct a series of experiments to validate the effectiveness and versatility of our proposed method. 

\subsection{Evaluation}\label{sec:experiment:eval}
We implemented our highly optimized renderer in CUDA and wrapped with Python interface for usability. To quantify computational benefits of our CUDA implementation (our main contribution), we also implemented a na\"ive PyTorch baseline~\cite{paszke2019pytorch}. We report per-iteration forward/backward runtime and peak VRAM on the same GPUs (RTX 3090, L40S) with identical inputs and hyperparameters. Since all remaining modules (partitioning, loss computation, optimizer, and other Python-side code) share the same PyTorch implementation across all three variants, we restrict the breakdown to the renderer’s forward/backward passes. We compare three implementations (PyTorch, CUDA-32bit, and CUDA-16bit) evaluated at 512\textsuperscript{2}, 1K\textsuperscript{2}, and 2K\textsuperscript{2} resolutions with tile sizes 16 or 32. \Cref{tab:evaluation} summarizes these measurements. This setup isolates implementation effects from workload variance and shows that our custom CUDA parallelization, especially CUDA-16bit, is critical for reducing runtime and memory footprint, enabling scalable \OurMethod{} at higher resolutions and primitive counts.

\begin{figure}[t]
    \centering
    \begin{tabular}{@{}p{0.20\linewidth} @{ } p{0.38\linewidth} @{ } p{0.38\linewidth}@{}}
        \multicolumn{1}{c}{\textbf{Primitive}} & \multicolumn{1}{c}{\textbf{DiffVG~\cite{li2020differentiable}}} & \multicolumn{1}{@{}c@{}}{\textbf{\OurMethod{} (Ours)}} \\
        % --- Row 1 ---
        \begin{tikzpicture}
            \node[inner sep=0pt] (img) {\includegraphics[width=\linewidth]{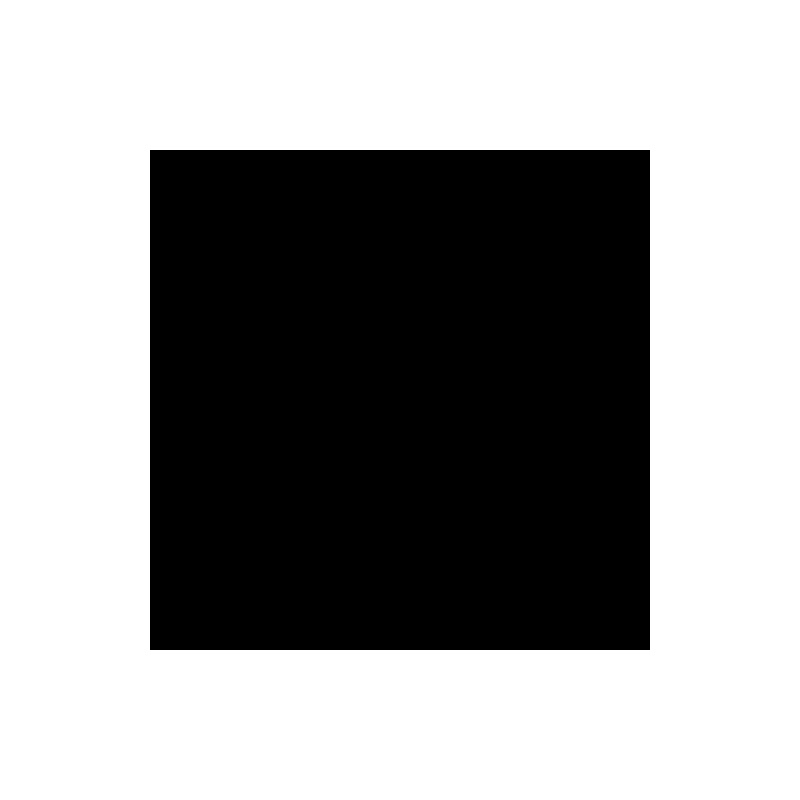}};
            \node[primitive label, shift={(0, 2mm)}] at (img.north west) {\small Simple SVG};
        \end{tikzpicture}
        &
        \begin{tikzpicture}
            \node[inner sep=0pt] (img) {\includegraphics[width=\linewidth]{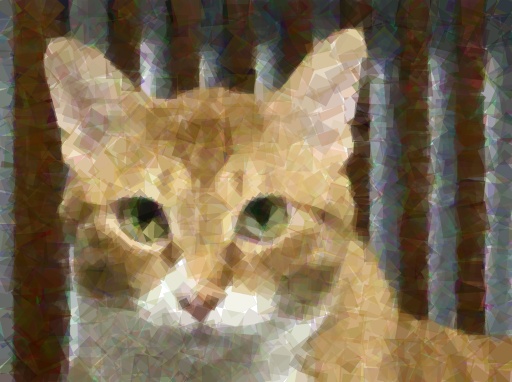}};
            \node[lightblack_box_white_text, anchor=north east] at (img.north east) {%
                \parbox{1.2cm}{\raggedleft 24.26 dB \\ \faClockO~67s}%
            };
        \end{tikzpicture}
        &
        \begin{tikzpicture}
            \node[inner sep=0pt] (img) {\includegraphics[width=\linewidth]{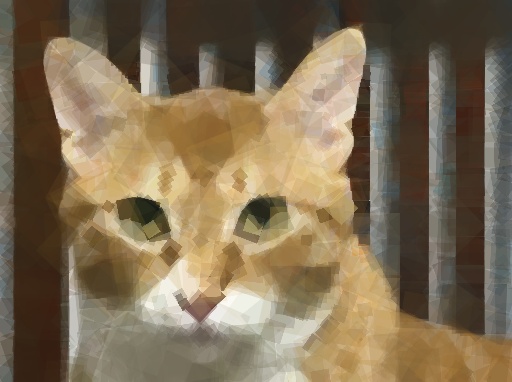}};
            \node[lightblack_box_white_text, anchor=north east] at (img.north east) {%
                \parbox{1.2cm}{\raggedleft 25.04 dB \\ \faClockO~51s}%
            };
        \end{tikzpicture}
        \\ % 줄 바꿈
        % --- Row 2 ---
        \begin{tikzpicture}
            \node[inner sep=0pt] (img) {\includegraphics[width=\linewidth]{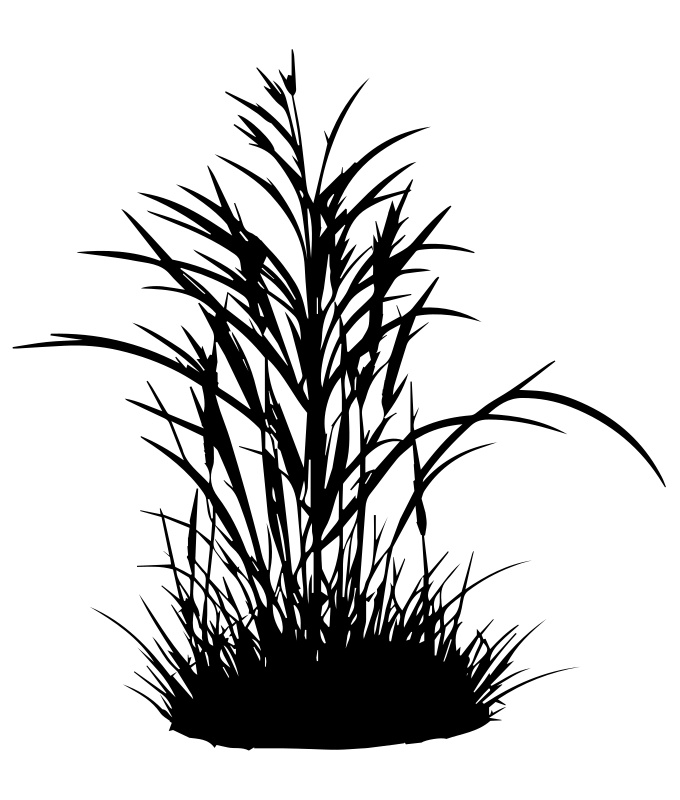}};
            \node[primitive label, shift={(0, 3mm)}] at (img.north west) {\scalebox{0.95}[1.0]{\footnotesize Complex SVG}};
        \end{tikzpicture}
        &
        \begin{tikzpicture}
            \node[inner sep=0pt] (img) {\includegraphics[width=\linewidth]{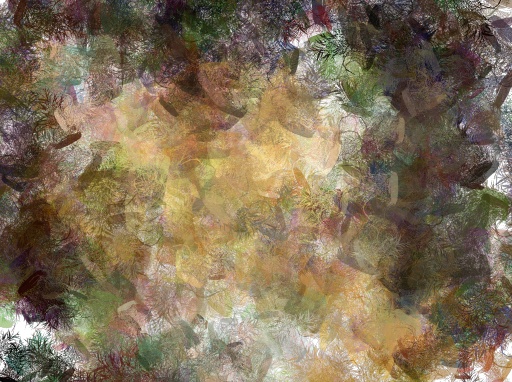}};
            \node[lightblack_box_white_text, anchor=north east] at (img.north east) {%
                \parbox{1.2cm}{\raggedleft 14.35 dB \\ \faClockO~477s}%
            };
        \end{tikzpicture}
        &
        \begin{tikzpicture}
            \node[inner sep=0pt] (img) {\includegraphics[width=\linewidth]{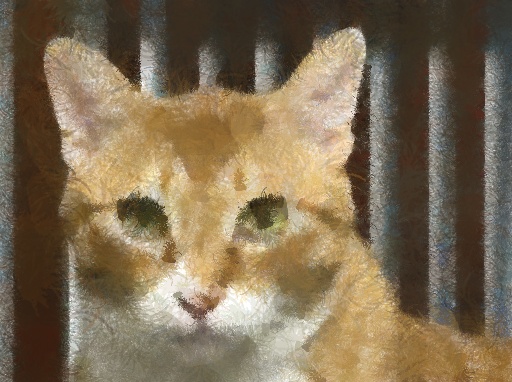}};
            \node[lightblack_box_white_text, anchor=north east] at (img.north east) {%
                \parbox{1.2cm}{\raggedleft 23.35 dB \\ \faClockO~115s}%
            };
        \end{tikzpicture}
        \\ % 줄 바꿈
        % --- Row 3 ---
        \begin{tikzpicture}
            \node[inner sep=0pt] (img) {\includegraphics[width=\linewidth]{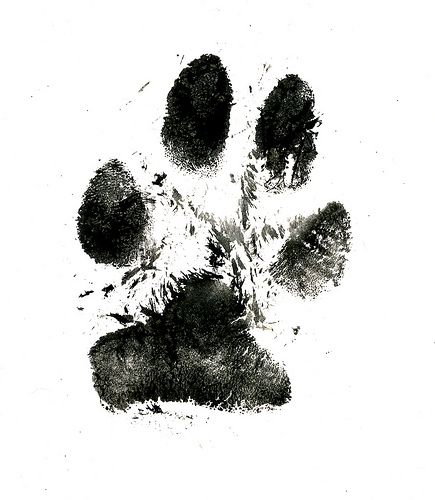}};
            \node[primitive label, shift={(3mm, 2mm)}] at (img.north west) {Bitmap};
        \end{tikzpicture}
        &
        \multicolumn{1}{@{}c@{}}{%
            \begin{tikzpicture}
                % Create an invisible node with the same dimensions as the other images
                \node[inner sep=0pt] (img) {\phantom{\includegraphics[width=0.35\linewidth]{figs/diffvg_vs_ours/cat_square_diffvg.jpg}}};
                % Place the N/A text in the absolute center of this invisible node
                \node[anchor=center, font=\Large] at (img.center) {N/A};
            \end{tikzpicture}
        }
        &
        \begin{tikzpicture}
            \node[inner sep=0pt] (img) {\includegraphics[width=\linewidth]{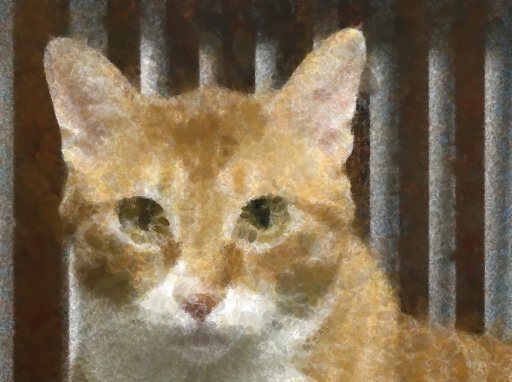}};
            \node[lightblack_box_white_text, anchor=north east] at (img.north east) {%
                \parbox{1.2cm}{\raggedleft 24.26 dB \\ \faClockO~36s}%
            };
        \end{tikzpicture}
    \end{tabular}
    \caption{Our \OurMethod{} works well for complex SVG primitives (second row) and bitmap primitives (third row), whereas existing DiffVG~\cite{li2020differentiable} fails or is not available, even though it works well for simple SVG primitives (first row). Note that SVG exportation is included in runtimes for row 1-2, while bitmap exports only PNG, so the third row takes less time (36s).}
    \label{fig:diffvg_comparison}
\end{figure}

\paragraph{DiffVG's struggle with complex SVG limits its extension to bitmaps}
\OurMethod{} addresses key limitations of existing vector-graphic differentiable renderers like DiffVG~\cite{li2020differentiable}. While DiffVG remains effective for standard SVG-based vector graphics (as shown in the first row of \cref{fig:diffvg_comparison}), its reliance on an analytic calculation leads to low fidelity (poor PSNR) and drastically increased runtime when processing complex SVG curves, as shown in the second row of \cref{fig:diffvg_comparison}. This suggests that DiffVG may struggle with many bitmap images even if image vectorization methods are applied. In contrast, \OurMethod{} does not suffer from it by dealing them as raster, which maintain consistent performance regardless of primitive complexity. Above all, the critical distinction is that, as illustrated in the third row, \OurMethod{} is the first model capable of performing differentiable rendering using arbitrary raster 2D primitives.

\subsection{Ablation study}\label{sec:exp:ablation}
\paragraph{Soft Rasterization and Initialization}
We ablate the two optimization helpers introduced in \cref{sec:method:2}—soft rasterization (\cref{sec:method:2:blur}) and structure-aware initialization (\cref{sec:method:2:init})—as in the unified protocol of \cref{sec:experiment:eval}. As summarized in \cref{tab:ablation_result}, applying Gaussian blur alone yields consistent PSNR gains and stabilizes optimization by providing rich, non-vanishing gradients near primitive boundaries; enabling structure-aware initialization alone accelerates convergence and improves final fidelity, particularly at higher primitive counts. Using both (the last row in \cref{tab:ablation_result}) produces the best results across all targets, with sharper edges, fewer spurious fragments, and reduced seed-to-seed variance, whereas disabling both often leads to destabilization of the fit and missing fine details. These helpers operate only before the optimization stage and do not alter the objective; their runtime overhead is almost the same as in \cref{tab:evaluation}.
\begin{table}
\caption{\textbf{Ablation of our optimization helpers: soft rasterization (\cref{sec:method:2:blur}) and structure-aware initialization (\cref{sec:method:2:init}).} We report PSNR on three targets with indicating whether each component is enabled. Soft rasterization enriches gradients and yields consistent gains, while structure-aware initialization further boosts fidelity; enabling both gives the best PSNR across all scenarios.
}
\label{tab:ablation_result}
\centering
\small
\setlength{\tabcolsep}{3.2pt}
\begin{tabular}{ccccc}
\toprule
\makecell{SoftRas\\ \cref{sec:method:2:blur}}&\makecell{SA-Init. \\ \cref{sec:method:2:init}} & \makecell{Scenario 1 \\ (512$\times$512)} & \makecell{Scenario 2 \\ (512$\times$512)} & \makecell{Scenario 3 \\ (1024$\times$1024)} \\
\midrule
\xmark  &\xmark  & 24.4 & 20.6 & 25.9 \\
\cmark  &\xmark &\cellcolor{tabthird}24.7 & \cellcolor{tabsecond}21.5 & \cellcolor{tabthird}26.5 \\
\xmark  & \cmark &\cellcolor{tabsecond}25.5 & \cellcolor{tabthird}21.0 & \cellcolor{tabsecond}27.1 \\
 \cmark &\cmark &\cellcolor{tabfirst}25.7 & \cellcolor{tabfirst}21.7 & \cellcolor{tabfirst}27.4 \\ \bottomrule
\end{tabular}
\end{table}

\begin{figure}[t]
    \centering
    % target image
\begin{subfigure}[b]{0.3\linewidth}
    \centering
    \begin{tikzpicture}
        % Rendered Image
        \node[inner sep=0pt, anchor=north west, xshift=0.05\linewidth] (default_bg) at (0,0) {\includegraphics[width=0.9\linewidth]{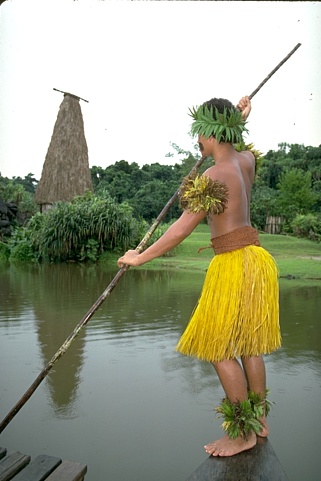}};
        
        % Primitive (작은 이미지)
        \node[inner sep=0pt, anchor=south east] (small_img) at (default_bg.south east) {\includegraphics[width=0.35\linewidth]{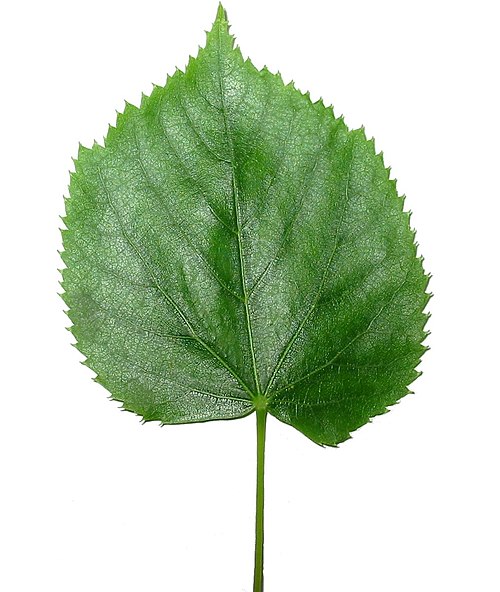}};
        
        % 작은 이미지 위에 스타일 적용 텍스트
\node[lightblack_box_white_text, anchor=south east, font=\scriptsize]
      at ([xshift=0pt,yshift=0pt]small_img.north east) {Primitive};
    \end{tikzpicture}
    \caption{Target Image}\label{fig:noisycanvas:1}
\end{subfigure}
    \hfill
    % white background
    \begin{subfigure}[b]{0.3\linewidth}
        \centering
        \begin{tikzpicture}
            % Rendered Image
            \node[inner sep=0pt, anchor=north west, xshift=0.05\linewidth] (default_bg) at (0,0) {\includegraphics[width=0.9\linewidth]{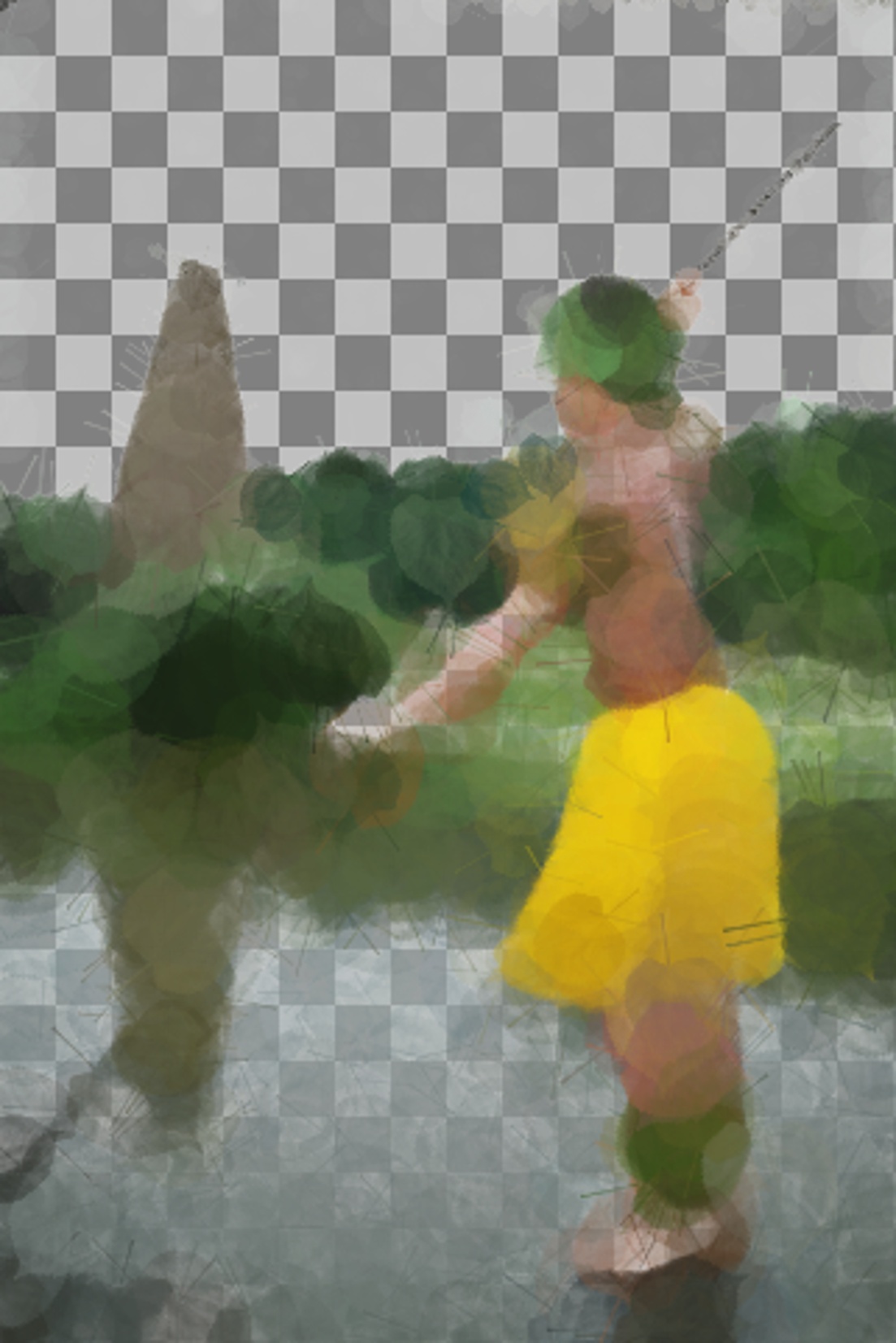}};
        \end{tikzpicture}
        \caption{w/ white canvas}\label{fig:noisycanvas:2}
    \end{subfigure}
    \hfill
    % uniform noise background
    \begin{subfigure}[b]{0.3\linewidth}
        \centering
        \begin{tikzpicture}
            % Rendered Image
            \node[inner sep=0pt, anchor=north west, xshift=0.05\linewidth] (default_bg) at (0,0) {\includegraphics[width=0.9\linewidth]{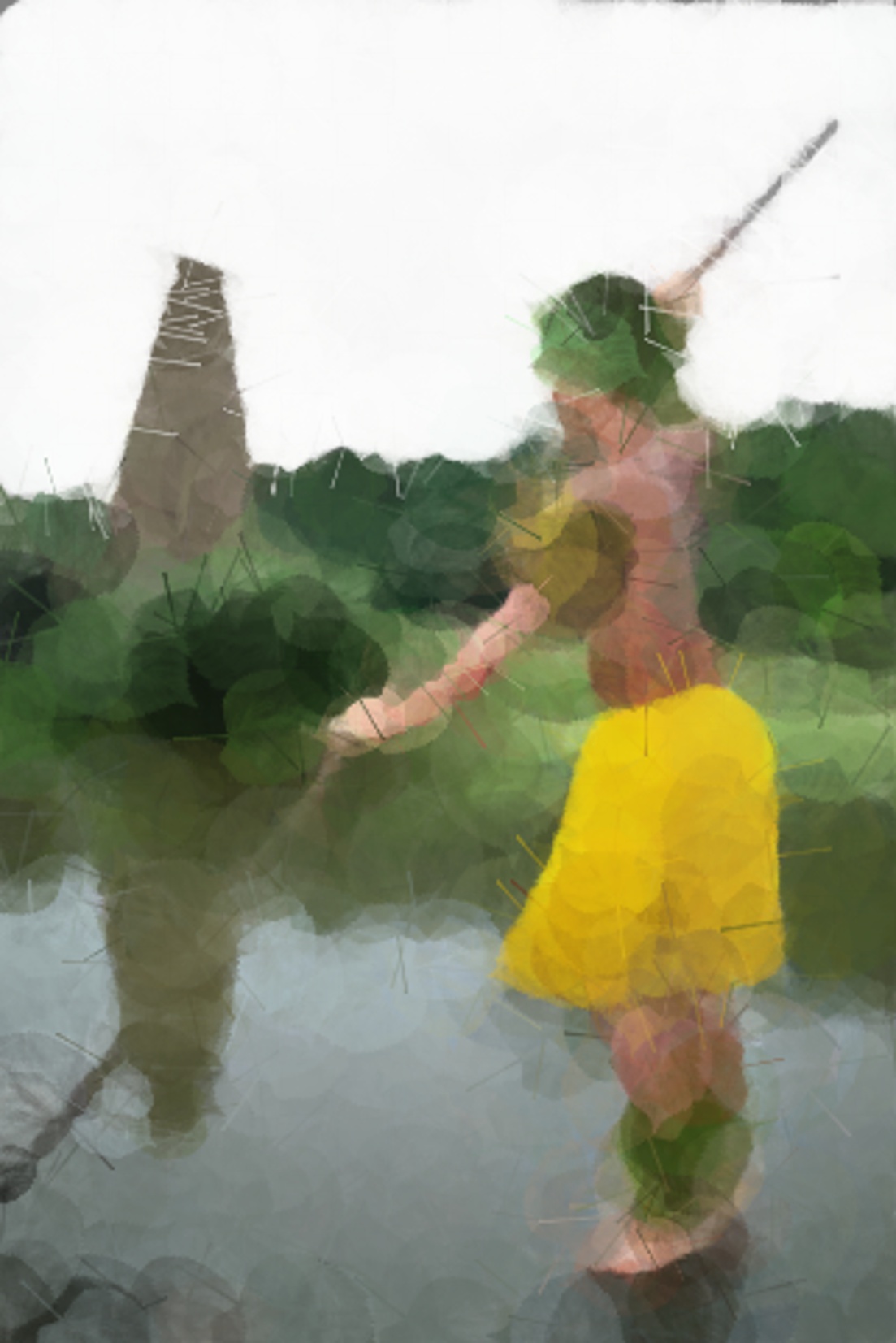}};
        \end{tikzpicture}
        \caption{w/ random canvas}\label{fig:noisycanvas:3}
    \end{subfigure}

    \caption{Uniform noise (\cref{sec:method:3:uniform}) makes canvas fully covered by primitives, as shown in (c). (b) Conversely, using the canvas as white removes the need for white primitives to be positioned.}
    \label{fig:ablation_random_bg}
\end{figure}

\paragraph{Canvas with Uniform Noise to Impose Primitives}
As explained in \cref{sec:method:3:uniform}, primitives tend to prioritize covering canvas areas with colors distinct from a designated blending color. This causes some part of the canvas to remain uncovered by primitives, as shown in \cref{fig:noisycanvas:2}. Optimizing with a uniformly noisy canvas resolved this, as in \cref{fig:noisycanvas:3}.

\paragraph{Dynamic \OurMethod{}}
We compare configurations by toggling three lightweight heuristics from \cref{sec:4.1} and report per-frame fidelity (PSNR, SSIM) and anti-flicker (tOF, tLP; \cite{chu2018tecoGAN}). Metrics are computed per frame, averaged within each video, then reported as mean$\pm$std over 17 videos. As summarized in \cref{tab:dynamic_result}, initializing from $\Theta^{f-1\star}$ already reduces flicker and modestly improves fidelity. On top of this initialization, removing stuck primitives yields the \emph{highest fidelity}, while freezing primitives in unchanged region delivers a \emph{large flicker reduction}. Combining all three achieves the best temporal stability overall while maintaining competitive fidelity.

\begin{table}[t]
\caption{\textbf{Dynamic \OurMethod{} ablations.} We observed that improvements in flicker suppression can trade off with per-frame fidelity; using all three of our heuristics in \cref{sec:4.1} targets a balanced solution, having competitive fidelity and the best anti-flicker.}
\label{tab:dynamic_result}
\centering
\small
\setlength{\tabcolsep}{3pt}
\resizebox{\linewidth}{!}{
\begin{tabular}{cccrrrr} 
\toprule
\multirow{2}{*}{\makecell{Init.\\ $\Theta^{f-1\star}$}} &
\multirow{2}{*}{\makecell{Rem.\\ Stk.}}&
\multirow{2}{*}{\makecell{Freez.\\ Unch.}} &
\multicolumn{2}{c}{Frame-wise Fidelity} &
\multicolumn{2}{c}{\makecell{Anti-flicker}}\\
\cmidrule(lr){4-5}\cmidrule(lr){6-7}
& & &
\makecell{PSNR$\uparrow$} &
\makecell{SSIM$\uparrow$} &
\makecell{tLP$\downarrow$} & 
\makecell{tOF$\downarrow$}\\ 
\midrule
\xmark & \xmark & \xmark & 24.19\textsubscript{$\pm$2.36} & 0.616\textsubscript{$\pm$0.106} & 7.41\textsubscript{$\pm$8.89} & 2.23\textsubscript{$\pm$1.10}\\
\cmark & \xmark & \xmark & \cellcolor{tabthird} 24.26\textsubscript{$\pm$1.73} & \cellcolor{tabthird} 0.629\textsubscript{$\pm$0.073} & 5.39\textsubscript{$\pm$4.64} & \cellcolor{tabsecond} 1.88\textsubscript{$\pm$1.13}\\
\cmark & \cmark & \xmark & \cellcolor{tabfirst} 24.66\textsubscript{$\pm$2.03} & \cellcolor{tabfirst} 0.647\textsubscript{$\pm$0.088} & \cellcolor{tabthird} 4.98\textsubscript{$\pm$4.93} &  \cellcolor{tabthird} 1.89\textsubscript{$\pm$1.11}\\
\cmark & \xmark & \cmark & 24.23\textsubscript{$\pm$2.13} & 0.617\textsubscript{$\pm$0.099} & \cellcolor{tabsecond} 3.50\textsubscript{$\pm$2.52} & 1.91\textsubscript{$\pm$1.32}\\
\cmark & \cmark & \cmark & \cellcolor{tabsecond} 24.38\textsubscript{$\pm$2.23} & \cellcolor{tabsecond} 0.630\textsubscript{$\pm$0.069} & \cellcolor{tabfirst} 3.49\textsubscript{$\pm$2.26} & \cellcolor{tabfirst} 1.84\textsubscript{$\pm$1.19} \\
\bottomrule
\end{tabular}
}
\end{table}

\begin{figure}[t]
    \centering
    % MSE
    \begin{minipage}[t]{0.32\linewidth}
        \centering
        \begin{tikzpicture}
            % Rendered Image
            \node[inner sep=0pt, anchor=north west, xshift=0.05\linewidth] (default_bg) at (0,0) {\includegraphics[width=0.9\linewidth]{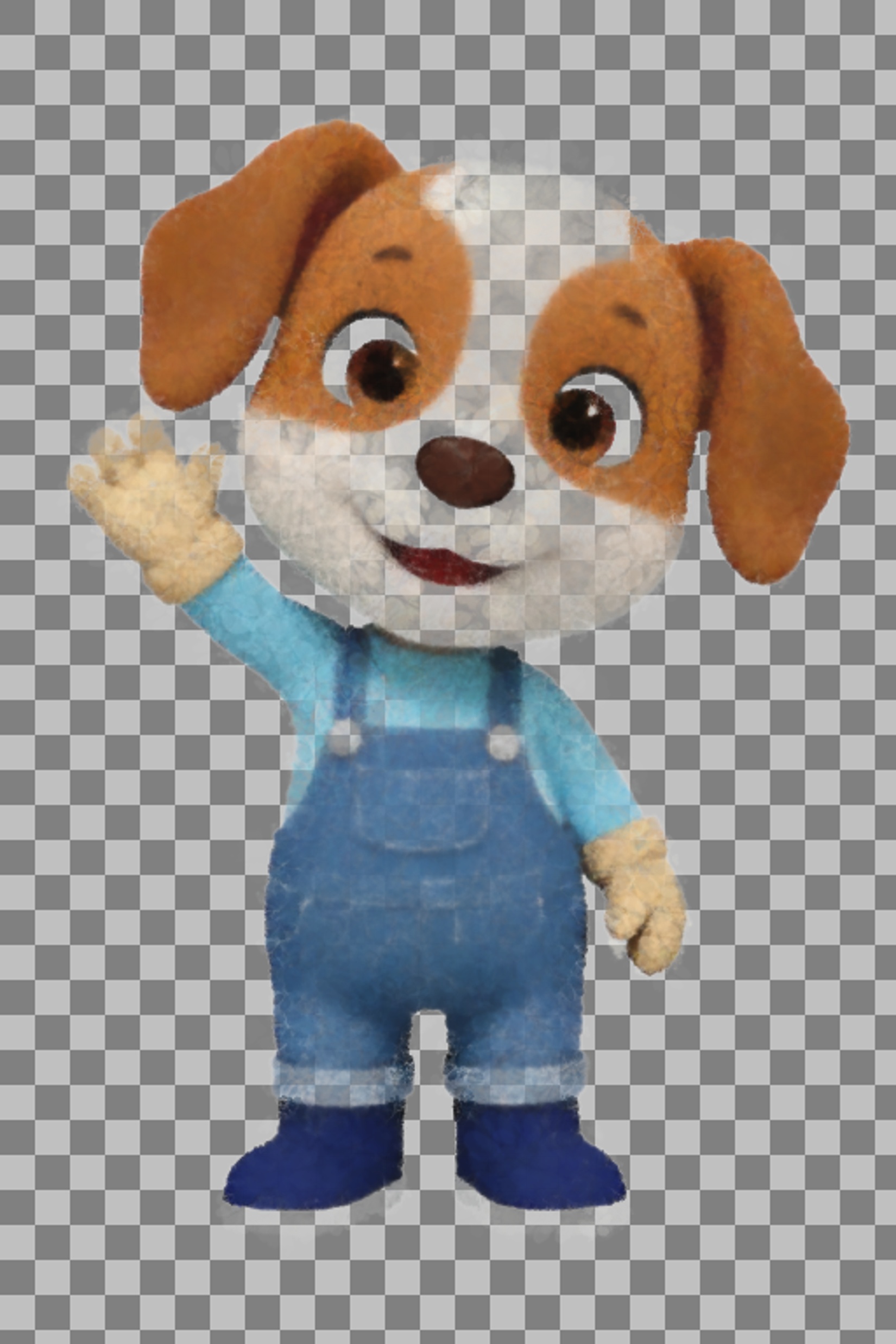}};
            
            % Spatial Constraint Check
            \node[anchor=south east, inner sep=0pt, xshift=0.15\linewidth, yshift=0.00\linewidth] at (default_bg.south east) {\includegraphics[width=0.4\linewidth]{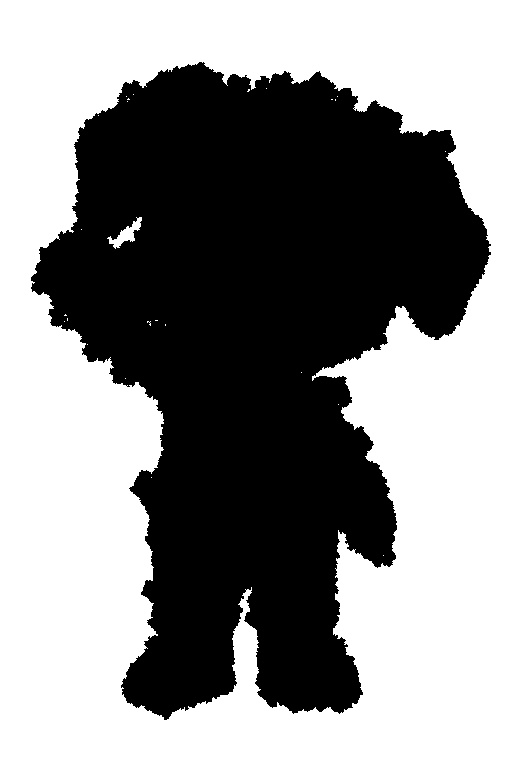}};
            % Comment
            \node[lightblack_box_white_text, anchor=south west, font=\footnotesize, align=left] at (default_bg.south west) {
            \condensedSB{VRAM}~\scalebox{0.8}[1]{1.6 GB} \\ \faClockO~40s}; 
        \end{tikzpicture}
        \caption*{(a) Ours (default)}
    \end{minipage}
    \hfill
    % MSE + Alpha
    \begin{minipage}[t]{0.32\linewidth}
        \centering
        \begin{tikzpicture}
            % Rendered Image
            \node[inner sep=0pt, anchor=north west, xshift=0.05\linewidth] (default_bg) at (0,0) {\includegraphics[width=0.9\linewidth]{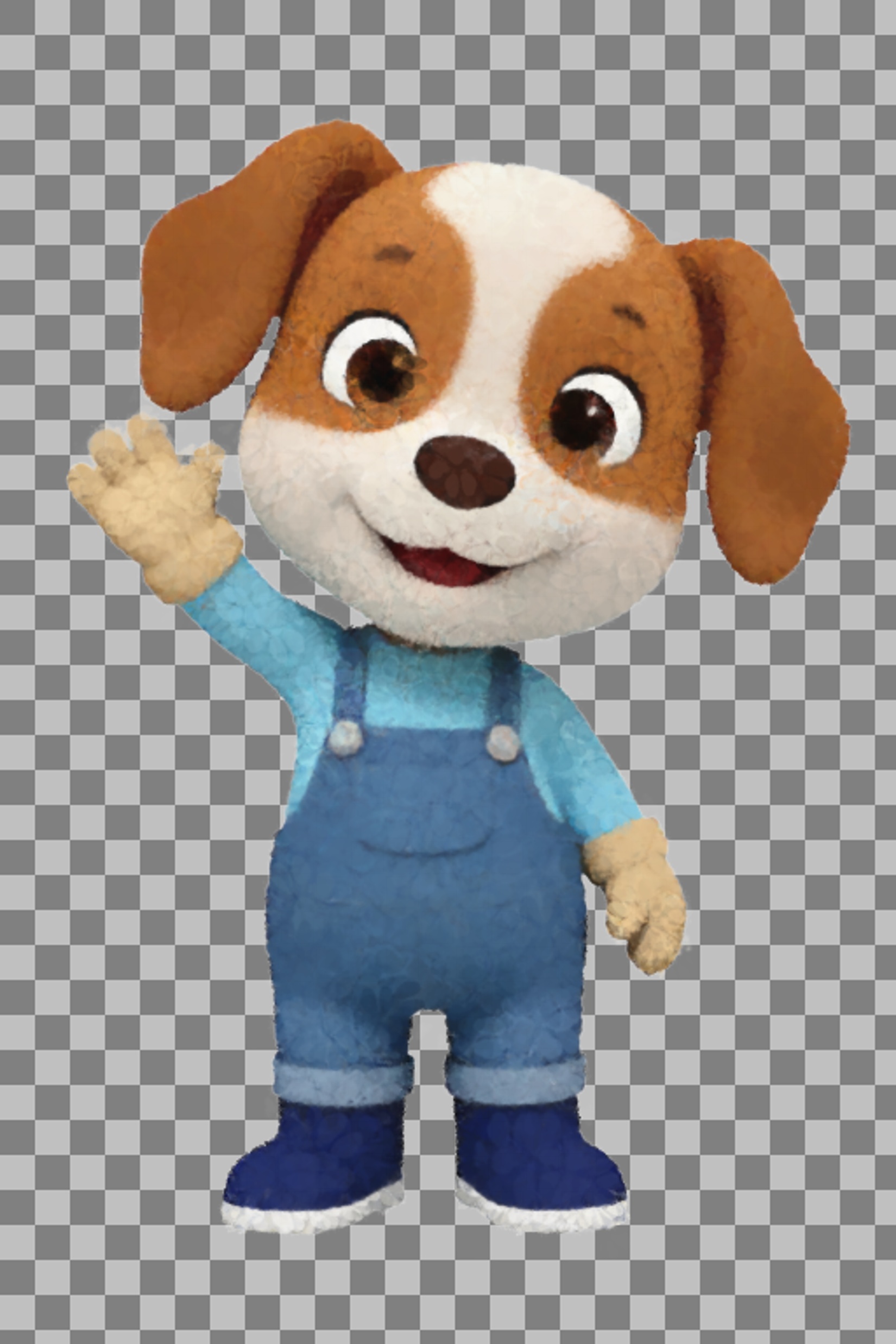}};
            
            % Spatial Constraint Check
            \node[anchor=south east, inner sep=0pt, xshift=0.15\linewidth, yshift=0.00\linewidth] at (default_bg.south east) {\includegraphics[width=0.4\linewidth]{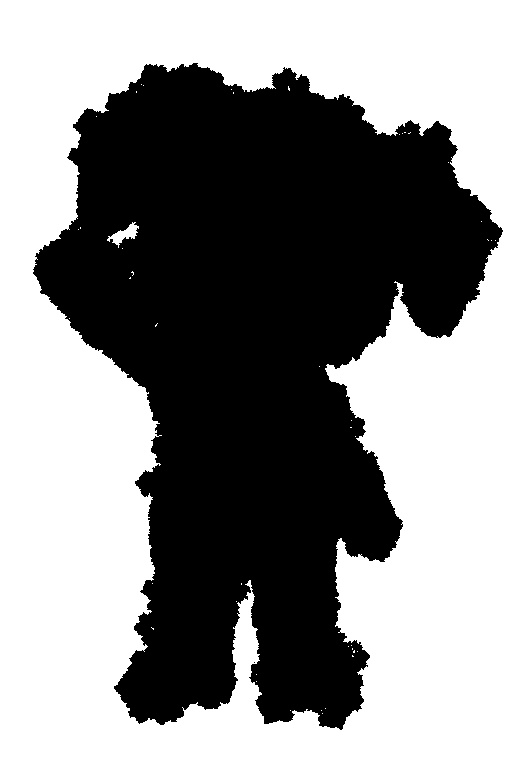}};
            % Comment
            \node[lightblack_box_white_text, anchor=south west, font=\footnotesize, align=left] at (default_bg.south west) {
            \condensedSB{VRAM}~\scalebox{0.8}[1]{1.3 GB} \\ \faClockO~35s}; 
        \end{tikzpicture}
        \caption*{(b) (a) + opacity loss}
    \end{minipage}
    \hfill
    % MSE + Alpha + Prune
    \begin{minipage}[t]{0.32\linewidth}
        \centering
        \begin{tikzpicture}
            % Rendered Image
            \node[inner sep=0pt, anchor=north west, xshift=0.05\linewidth] (no_bg_bg) at (0,0) {\includegraphics[width=0.9\linewidth]{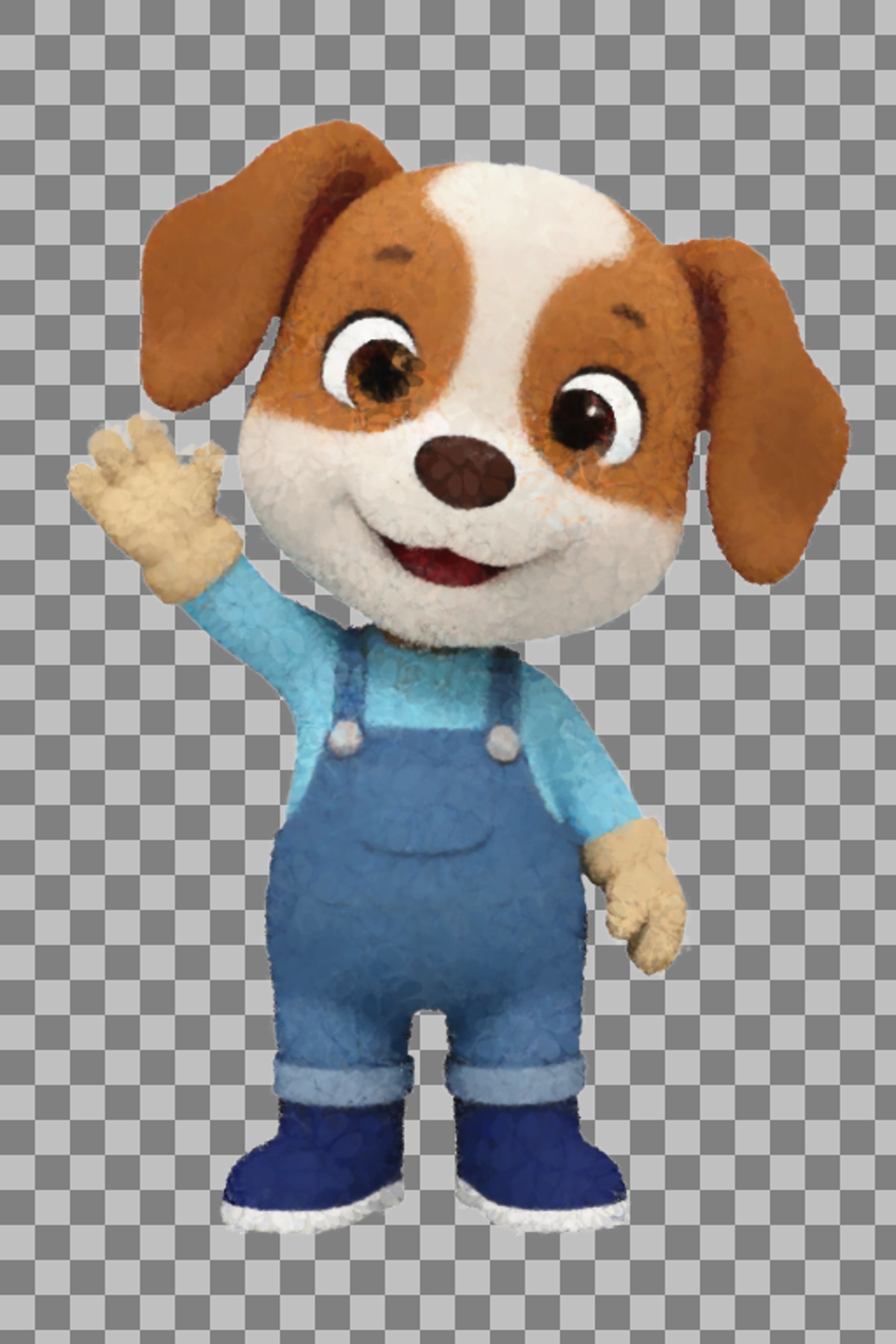}};
            
            % Spatial Constraint Check
            \node[anchor=south east, inner sep=0pt, xshift=+0.15\linewidth, yshift=0.00\linewidth] at (no_bg_bg.south east) {\includegraphics[width=0.4\linewidth]{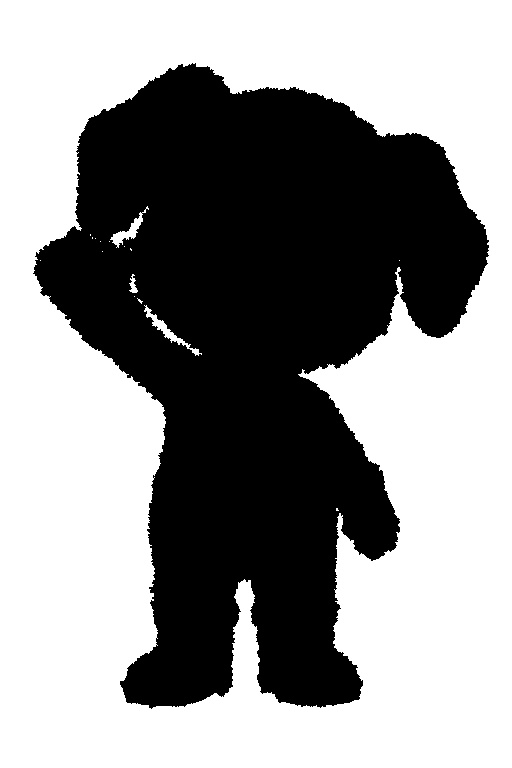}};
            % Comment
            \node[lightblack_box_white_text, anchor=south west, font=\footnotesize, align=left] at (default_bg.south west) {
            \condensedSB{VRAM}~\scalebox{0.8}[1]{1.3 GB} \\ \faClockO~35s}; 
        \end{tikzpicture}
        \caption*{(c) (b)+re-init.}
    \end{minipage}

    \caption{Comparison of rendered output $I$
  and primitives with spatial constraint under different settings: (a) Ours (default), (b) Ours with opacity loss, and (c) Ours with both opacity loss and re-initialization. Applying both opacity loss and re-initialization produces the best results.}
    \label{fig:no_bg_result}
\end{figure}

\paragraph{Spatially Constrained}
In \cref{fig:no_bg_result}, we present a qualitative comparison of the methods discussed in \cref{sec:3.3.2}. The target object is rendered with full opacity by applying opacity loss, which is crucial in post-processing. However, when using this loss, some primitives remain on the boundary of the object with low opacity. To address this issue, we re-initialize such low-opacity primitives by reusing them for further optimization.

\subsection{Creative Workflow Examples}\label{sec:creative}
A core objective of \OurMethod{} is to serve not just as an optimization algorithm, but as a practical tool that integrates into existing creative pipelines. The ability to export compositions into native, layered PSD files is central to this goal. While previous sections focus on quantitative metrics, this section demonstrates the qualitative benefits of our artist-friendly exports through several workflow examples, showcasing how \OurMethod{} can act as a collaborative tool.

\paragraph{Intrinsic-Preserving Graphic Assemblage}\label{sec:creative:assemblage}
\Cref{fig:teaser:b}, which we term an `Intrinsic-Preserving Graphic Assemblage', expands upon traditional photo mosaics~\cite{finkelstein1998image}. Conventional mosaics often alter the color tint of source images to match a target palette and are typically restricted to a rigid, grid-based layout. In contrast, \OurMethod{} preserves the original color of each primitive (\eg, various brand logos~\cite{koustubhk_popular_2025}) and arranges them in a grid-free manner, successfully forming a recognizable final shape. This result suggests that \OurMethod{} can serve as a powerful tool for pioneering new forms of computational art.

\begin{figure}
    \centering
\begin{tabularx}{\linewidth}{>{\centering\arraybackslash}X >{\centering\arraybackslash}X} 
        \toprule
        \textbf{Primitive / Prompt} & \textbf{Output} \\
        \midrule
    \end{tabularx}   
    \begin{tikzpicture}
        \node[anchor=south east, inner sep=0] (img1) at (0,0) {
        \includegraphics[width=\linewidth]{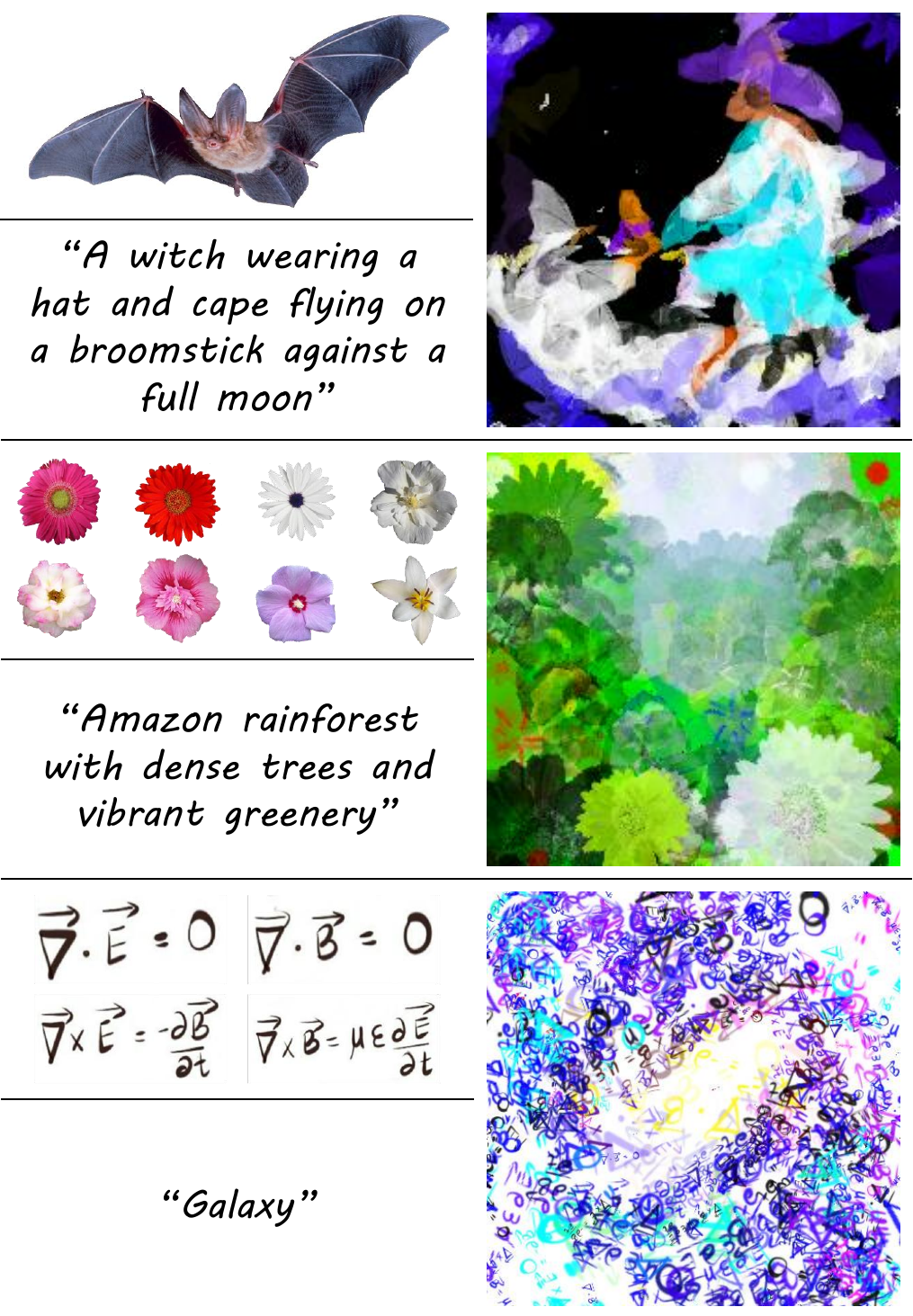}
        };

        \node[lightblack_box_white_text, anchor=south east, font=\footnotesize, align=right] at ([yshift=2.095cm,xshift=-2.8pt]img1.east) {\condensedSB{VRAM}~1.5 GB\\ \faClockO~55s};

        \node[lightblack_box_white_text, anchor=south east, font=\footnotesize, align=right] at ([yshift=-1.9cm,xshift=-2.8pt]img1.east) {\condensedSB{VRAM}~1.5 GB\\ \faClockO~46s};
        
        \node[lightblack_box_white_text, anchor=east, font=\footnotesize, align=right] at ([yshift=-5.53cm,xshift=-2.8pt]img1.east) {\condensedSB{VRAM}~1.8 GB\\ \faClockO~56s};
    \end{tikzpicture}
    \caption{\OurMethod{} can also be combined with CLIP for creation. Each caption shows the text prompt and the primitive used.}
    \label{fig:CLIP_examples}
    \label{fig:example}
\end{figure}

\paragraph{Creation from text} 
\OurMethod{} is capable of optimizing an arbitrary loss function on the output raster. Since CLIP~\cite{radford2021learning} is a neural network trained to learn the similarity between an image and text, combining it with \OurMethod{} enables us to optimize the parameters $\Theta$ to increase the similarity with a given text prompt. Following the approach of CLIPDraw~\cite{frans2022clipdraw}, which combines DiffVG and CLIP, we duplicate the output of \OurMethod{} and perform data augmentation through perspective transformation, cropping, and resizing. We minimize the negative of the cosine similarity between the augmented \OurMethod{} output images and the input text, adding the cosine similarity with some negative prompts. See \cref{fig:CLIP_examples} for the results.
% ================================================================= 
% 6. DISCUSSION AND FUTURE WORK 
% ================================================================= 
\section{Limitations and Future Work} 
\paragraph{\OurMethod{} requires a GPU}
While DiffVG~\cite{li2020differentiable} can be executed on a CPU, \OurMethod{} fundamentally requires a GPU, and our renderer was specifically implemented in CUDA based on this necessity. \OurMethod{} operates using bitmap primitives, which are data structures rather than purely mathematical representations. Since bitmap primitives require significant memory usage, GPU operation is essential for \OurMethod{}.

\paragraph{\OurMethod{} is sensitive to hyperparameters}
The generality of \OurMethod{} increases its sensitivity to hyperparameter selection and initialization, which can make some tasks susceptible to local minima. Developing methods to automatically determine optimal hyperparameters based on the user's specific task, primitives, and target remains a significant area for future work.

\paragraph{Extending to Autoregressive / RL Drawing}\label{sec:SBR}
Graphical composition and drawing problems were previously addressed without a differentiable renderer, relying on rule-based~\cite{hertzmann1998painterly}, reinforcement learning (RL)~\cite{schaldenbrand2021content, huang2025attention, ganin2018synthesizing, mellor2019unsupervised, jia2019paintbot}, or neural network-based methods~\cite{liu2021paint, wang2023stroke, tang2024attentionpainter, song2024processpainter}. Since most of these methods do not utilize gradient-based $\emph{optimization}$, their performance has inherent limitations. We believe \OurMethod{} provides a critical foundation for applying first-order optimization to these types of autoregressive and RL drawing problems in future work.

% ================================================================= 
% 7. CONCLUSION 
% ================================================================= 
\section{Conclusion} 
In this paper, we presented \OurMethod{}, the first general-purpose differentiable rendering engine designed to utilize arbitrary 2D bitmap images as primitives. \OurMethod{} demonstrates a powerful ability to scalably and efficiently approximate target designs under various constraints. Crucially, this core capability not only enables the creation of highly artistic compositions but also ensures seamless integration into the professional designer's workflow. Furthermore, by packaging \OurMethod{} with a user-friendly interface, we anticipate that its public release will significantly broaden the creative horizons of computational art and design.

\section*{Acknowledgements}
This work was supported in part by Institute of Information \& communications Technology Planning \& Evaluation (IITP) grant funded by the Korea government(MSIT) [NO.RS-2021-II211343, Artificial Intelligence Graduate School Program (Seoul National University)], the National Research Foundation of Korea(NRF) grants funded by the Korea government(MSIT) (Nos. RS-2025-02263628, RS-2022-NR067592) and the BK21 FOUR program of the Education and Research Program for Future ICT Pioneers, Seoul National University.

{
    \small
    \bibliographystyle{ieeenat_fullname}
    \bibliography{refs}
}
\ifarxiv
\clearpage
\newpage
\begin{center}
    \Large \textbf{Appendix}
\end{center}

\crefname{section}{Sec.}{Secs.}
\Crefname{section}{Section}{Sections}
\Crefname{table}{Table}{Tables}
\crefname{table}{Tab.}{Tabs.}
\renewcommand{\thesection}{S\arabic{section}}
\renewcommand{\thefigure}{S\arabic{figure}}
\renewcommand{\thetable}{S\arabic{table}}
\renewcommand{\theequation}{S\arabic{equation}}
\setcounter{section}{0}
\setcounter{figure}{0}
\setcounter{table}{0}
\setcounter{equation}{0}

\section{Method Detail}

\subsection{Bilinear interpolation}
Here, we provide more detail on bilinear interpolation~\cite{jaderberg2015spatial}, where the spatial gradients occur. 
Let $\lfloor U \rfloor = u_0$, $\lfloor V \rfloor = v_0$, and define the fractional parts $w_u = U - u_0$, $w_v = V - v_0$. The bilinear weights form:
\begin{equation} \mathbf{w} = \begin{bmatrix} (1-w_u)(1-w_v) \\ w_u(1-w_v) \\ (1-w_u)w_v \\ w_u w_v \end{bmatrix} \end{equation}
The primitive values at the four neighboring pixels are:
\begin{equation} \mathbf{p} = \begin{bmatrix} P_i[v_0, u_0] \\ P_i[v_0, u_0+1] \\ P_i[v_0+1, u_0] \\ P_i[v_0+1, u_0+1] \end{bmatrix} \end{equation}
Those four points are depicted as green dots in \cref{fig:grid_sample}.
The interpolated value is then:
\begin{equation} M_i(x,y) = \mathbf{w}^T \mathbf{p} = \sum_{j=0}^{3} w_j p_j \end{equation}

\begin{figure}[b]
    \centering
    \begin{tikzpicture}
        \node[inner sep=0pt] (img) {\includegraphics[width=\linewidth]{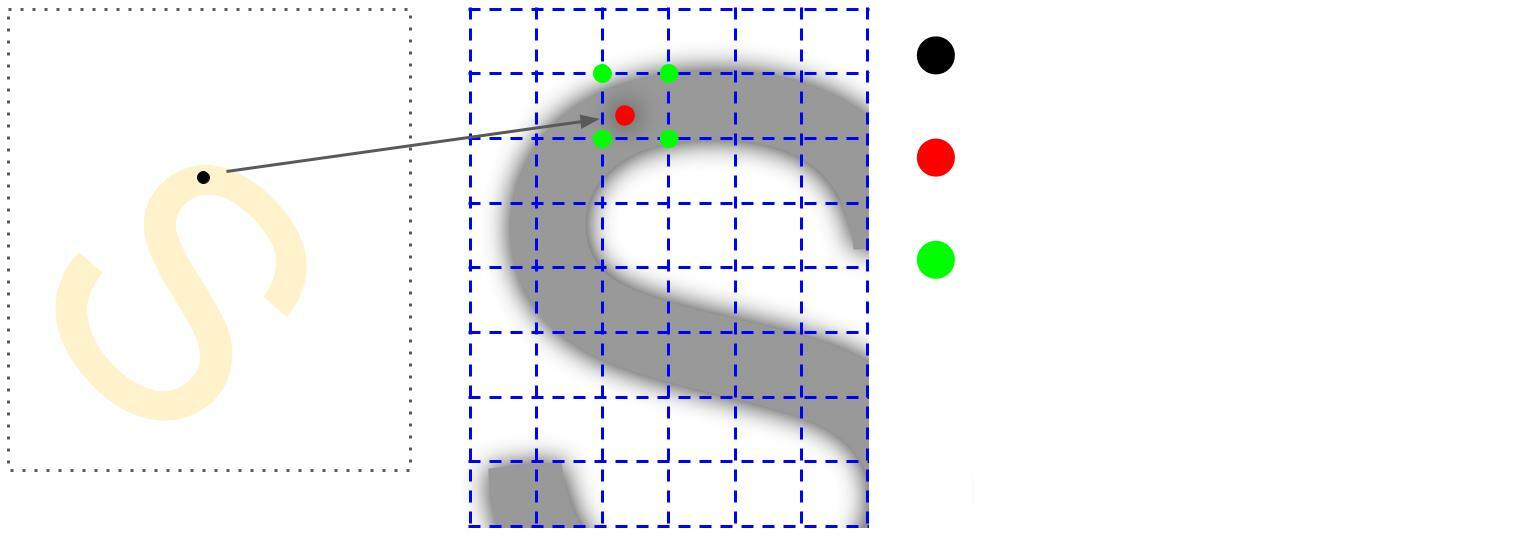}};

        \node[anchor=west] at (1.2cm, 1.15cm) {$(x,y)$};
        \node[anchor=west] at (1.2cm, 0.55cm) {$(U,V)$};
        \node[font=\small, anchor=west] at (1.2cm, -0.55cm) {\parbox{2.5cm}{$(\lfloor U \rfloor, \lfloor V \rfloor)$ \\ 
        $(\lfloor U \rfloor + 1, \lfloor V \rfloor)$\\
        $(\lfloor U \rfloor, \lfloor V \rfloor + 1)$\\
        $(\lfloor U \rfloor + 1, \lfloor V \rfloor + 1)$}};
    \end{tikzpicture}
    \caption{Coordinate transformation and primitive sampling. Color and opacity from $i$-th primitive at $(x,y)$ in canvas (black dot) is sampled from $(U,V)$ in the primitive's coordinate (red dot). Since $U$ and $V$ are not integers, bilinear interpolation from four nearest lattice points (green dots). }
    \label{fig:grid_sample}
\end{figure}

\subsection{Backward Pass}
Here, we state how the color gradients and alpha gradients are calculated. Entire forward-backward computation diagrams are shown in \cref{fig:supp-comp-graph}.

\begin{figure*}[h]
\centering
% everything is resized to fit the page width/height automatically
\begin{adjustbox}{max width=\textwidth, max totalheight=\textheight-5em, center}
\begin{tikzpicture}[node distance=15mm and 8mm,
  font=\sffamily,
  scale=1.0, every node/.style={transform shape},
  background rectangle/.style={fill=blue!5, rounded corners=10pt, inner sep=6pt}
]
\tikzset{
  box/.style   ={draw, fill=white, rounded corners=2pt, minimum width=10mm, minimum height=8mm, align=center},
  flow/.style  ={-{Latex[length=2.3mm,width=1.4mm]}, line width=0.7pt},
  deriv/.style ={red, densely dashed, -{Latex[length=2.3mm,width=1.4mm]}, line width=0.7pt},
  note_r/.style ={midway, inner sep=1pt, xshift=15pt},
  note_rb/.style ={midway, inner sep=1pt, xshift=25pt},
  note_rt/.style ={midway, inner sep=1pt, xshift=23pt},
  leaf/.style = {box, fill=gray!24, font=\bfseries}
}
%============================= TOP =============================%
\node[box] (L) {$\mathcal{L}$};
\node[box, below=of L, yshift=-1mm] (I) {
$\displaystyle
I(x,y)\;=\;\sum_{k=1}^{K} T_k\,\alpha_k\,\vc_k,\quad
T_{k+1}=T_k\,(1-\alpha_k)
\;\;\;
$};
\draw[flow] (I)  -- (L);
\draw[deriv] (L.south) ++(0.2,0) -- ++(0,-1.5)
  node[note_r]
  {$\dfrac{\partial \mathcal{L}}{\partial I}$};
%=========================== LEVEL 1 ===========================%
\node[box, leaf, below left=of I, xshift=15mm] (cn)  {$\vc_i$};
\node[box, below right=of I, xshift=-15mm] (alphaeq) {$\alpha_i=\alpha_{\max}\tilde{\sigma}_iM_i(x,y)$};
\draw[flow] ([xshift=-10pt]cn.north east) -- ([xshift=-15pt]I.south);
\draw[flow] (alphaeq.north west) -- ([xshift=15pt]I.south);
\draw[deriv] ([xshift=-5pt]I.south) -- (cn.north east) node[note_rb] {$\dfrac{\partial I}{\partial c_i}$};
\draw[deriv] ([xshift=25pt]I.south) -- ([xshift=10pt]alphaeq.north west) node[note_rt] {$\dfrac{\partial I}{\partial \alpha_i}$};

%==================== RIGHT BRANCH: ALPHA =====================%
% sigmoid param (opacity)
\node[box, leaf, below left=of alphaeq, xshift=5mm] (opa)  {$\nu_i$};
% compact Gaussian notation to save width
\node[box, below right=of alphaeq, xshift=-10mm] (Mn) {$M_i(x,y)=\mathrm{bilinear}\!\bigl(T_i,\;\mathrm{tex}(u,v)\bigr)$};
\node[box, below =of Mn] (tex) {$\mathrm{tex}(u,v)=\bigl(0.5(u{+}1)(W{-}1),\;0.5(v{+}1)(H{-}1)\bigr)$};
\node[box, below =of tex] (uv) {$
\begin{bmatrix}u\\ v\end{bmatrix}
=\frac{1}{s_i}
\begin{bmatrix}
\cos\theta_i & \sin\theta_i\\
-\sin\theta_i& \cos\theta_i
\end{bmatrix}
\begin{bmatrix}x-x_i\\[1pt] y-y_i\end{bmatrix}$};
\draw[flow]  ([xshift=-10pt]opa.north east) -- ([xshift=-15pt]alphaeq.south);
\draw[flow]  ([xshift=20pt]Mn.north west)  -- ([xshift=5pt]alphaeq.south);
\draw[deriv] ([xshift=-5pt]alphaeq.south) -- (opa.north east) node[note_rb] {$\dfrac{\partial \alpha_i}{\partial \nu_i}$};
\draw[deriv] ([xshift=15pt]alphaeq.south) -- ([xshift=30pt]Mn.north west) node[note_rt] {$\dfrac{\partial \alpha_i}{\partial M_i}$};
\draw[flow]  (tex.north) -- (Mn.south);
\draw[flow]  (uv.north)  -- (tex.south);
\draw[deriv] (Mn.south) ++(0.2,0) -- ++(0,-1.5) node[note_r] {$\dfrac{\partial M_i}{\partial \mathrm{tex}}$};
\draw[deriv] (tex.south) ++(0.2,0) -- ++(0,-1.5) node[note_r, xshift=2pt]  {$\dfrac{\partial \mathrm{tex}}{\partial (u,v)}$};
% coverage params
\node[box, leaf, below left=of uv, xshift=5mm] (xn) {$x_i$};
\node[box, leaf, below left=of uv, xshift=30mm] (yn) {$y_i$};
\node[box, leaf, below right=of uv, xshift=-30mm] (s) {$s_i$};
\node[box, leaf, below right=of uv, xshift=-5mm] (theta) {$\theta_i$};%{$\Sigma_n=R(\theta_n)\,s_n^2\,R(\theta_n)^\top$};
\draw[flow]  ([xshift=-10pt]xn.north east) -- ([xshift=-70pt]uv.south);
\draw[flow]  (yn.north) -- ([xshift=-20pt]uv.south);
\draw[flow]  (s.north) -- ([xshift=15pt]uv.south);
\draw[flow]  (theta.north west) -- ([xshift=60pt]uv.south);
\draw[deriv] ([xshift=-60pt]uv.south) -- (xn.north east) node[note_r, xshift=3pt] {$\dfrac{\partial M_i}{\partial x_i}$};
\draw[deriv] ([xshift=-15pt]uv.south) -- ([xshift=5pt]yn.north) node[note_r] {$\dfrac{\partial M_i}{\partial y_i}$};
\draw[deriv] ([xshift=20pt]uv.south) -- ([xshift=5pt]s.north) node[note_r] {$\dfrac{\partial M_i}{\partial s_i}$};
\draw[deriv] ([xshift=70pt]uv.south) -- ([xshift=10pt]theta.north west)  node[note_r, xshift=3pt] {$\dfrac{\partial M_i}{\partial \theta_i}$};

%============================= LEGEND =========================%
\node[draw, fill=white, align=left, anchor=north east, inner sep=3pt]
      at ($(I.north east)+(9,15mm)$) {
  \begin{tikzpicture}[baseline, x=10mm]
    \draw[flow]  (0,0) -- (0.7,0) node[right]{\scriptsize forward};
    \draw[deriv] (0,-0.32) -- (0.7,-0.32) node[right]{\scriptsize backward(gradient)};
  \end{tikzpicture}
};
\end{tikzpicture}
\end{adjustbox}
\caption{\textbf{\OurMethod{} computation tree (forward/backward).}
Forward uses premultiplied-alpha Over in front-to-back order; per-primitive opacity is a sigmoid-scaled value modulated by a template mask sampled at rotated/scaled coordinates. Learnable leaves (gray) are $c_i, \nu_i, x_i, y_i, s_i, \theta_i$. Red dashed arrows indicate gradient flow. See Sec.~3.1.1 for equations.}
\label{fig:supp-comp-graph}
\end{figure*}
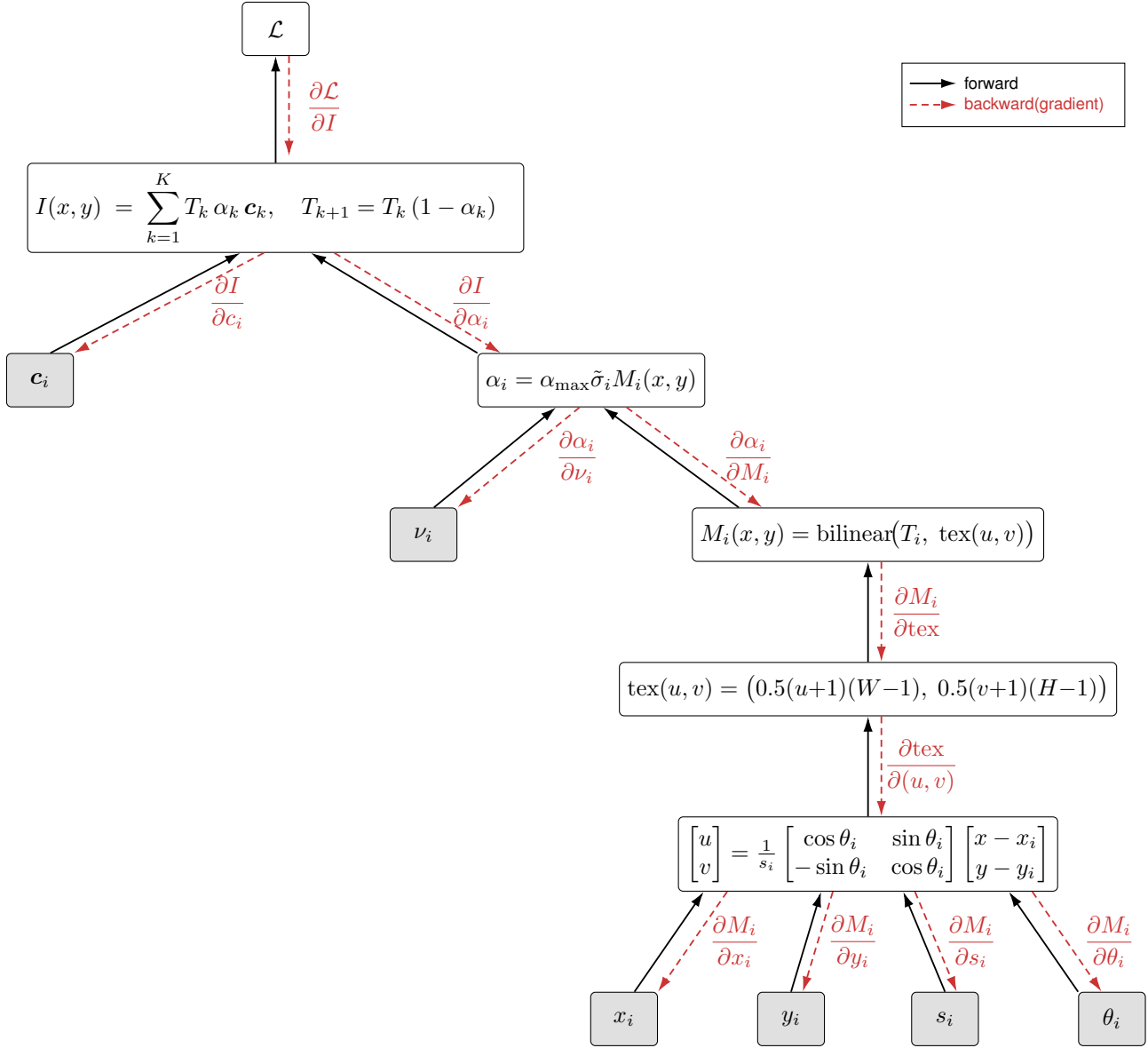

\paragraph{Color Gradients}
The gradient with respect to color logits follows the chain rule:

\begin{equation} \frac{\partial L}{\partial c_{k,j}} = \frac{\partial L}{\partial I_j} \cdot T_k \alpha_k \cdot \sigma(c_{k,j})(1 - \sigma(c_{k,j})) \end{equation}

\paragraph{Alpha Gradients}
For the Porter-Duff over operator~\cite{porter1984compositing}, the alpha gradient incorporates both direct color contribution and transmittance effects:

\begin{equation} \frac{\partial L}{\partial \alpha_k} = \sum_{j \in {r,g,b}} \frac{\partial L}{\partial I_j} \left( T_k \sigma(c_{k,j}) - \frac{S_j}{1-\alpha_k} \right) + \frac{\partial L}{\partial A} T_k B \end{equation}

where the suffix sum $S_j = \sum_{m>k} \sigma(c_{m,j}) \alpha_m T_m$ and back-product $B = \prod_{m>k}(1-\alpha_m)$ account for the interdependence of layered primitives.

\paragraph{Half2 Precision}
In the backward pass, for all gradient calculation, we use $\mathtt{\_\_half2}$ which packs two FP16, $\mathtt{\_\_half}$, in vector format. Per-parameter gradients are accumulated \emph{per pixel} directly into packed $\mathtt{\_\_half2}$ buffers via $\mathtt{\_\_half2}$ $\mathtt{atomicAdd}$ for the pairs \((m_x,m_y)\), \((\alpha,\text{scale})\), \((\theta,c_r)\), and \((c_g,c_b)\), followed by a lightweight post pass that unpacks to legacy FP16 arrays. In our measurements, $\mathtt{\_\_half}$ $\mathtt{atomicAdd}$ matched kernel time but degraded numerical accuracy, whereas $\mathtt{\_\_half}$ $\mathtt{atomicCAS}$ recovered accuracy at prohibitive cost. By contrast, packed $\mathtt{\_\_half2}$ based atomics preserved accuracy, improved throughput, and reduced memory traffic. Accordingly, we adopt this path and dispense with block-local reductions, aligning with our accuracy–throughput goal.

\subsection{CUDA Implementation Details}
In $T=32$ settings, we partition the canvas into $32\times32$ tiles on the CPU and assign each primitive to all tiles its bounding box overlaps. Each CUDA thread block processes one tile with $32\times32$ threads (one thread per pixel), achieving full pixel-level parallelism. We adopt a tile-and-bin CUDA pipeline, following tile-based differentiable splatting practices as in \cite{ye2025gsplat}, adapted here for 2D bitmap primitives.

\subsubsection{Forward Pass}

\begin{algorithm}
\caption{Forward Pass Kernel (FP16)}
\label{alg:forward}
\small
\begin{algorithmic}[1]
\Require $N$ primitives $\{(x_i, y_i, s_i, \theta_i, \nu_i, \vc_i)\}_{i=1}^N$, tile-primitive mapping
\Ensure Output image $I$, alpha channel $I_\alpha$

\State \textbf{Grid:} $(\lceil W_\text{canvas}/32 \rceil, \lceil H_\text{canvas}/32 \rceil)$ blocks, each with $32\times32$ threads
\State Each thread handles pixel $(x,y)$ in canvas coordinates

\For{each primitive $i$ assigned to this tile}
    \State Transform $(x,y) \to (u,v)$ in primitive space via Eq. (1)
    \State Compute $M_i(x,y)$ via bilinear interpolation at $(u,v)$
    \If{$M_i(x,y) < \epsilon$} skip \EndIf
    \State Compute $\alpha_k = \alpha_{max} \cdot \sigma(\nu_i) \cdot M_i(x,y)$, $\vc_k = \sigma(\vc_i)$
    \State Cache $\alpha_k, \vc_k$ in global memory (FP16)
\EndFor

\State \textbf{Alpha composite:} $T \gets 1$, $C \gets (0,0,0)$
\For{each cached primitive $k$}
    \State Store $T_k$ for backward; $C \gets C + T \alpha_k \vc_k$; $T \gets T(1 - \alpha_k)$
\EndFor
\State Write $I[x,y] \gets C$, $I_\alpha[x,y] \gets 1 - T$
\end{algorithmic}
\end{algorithm}

Primitive templates are loaded into shared memory per block. FP16 halves bandwidth and enables tensor core acceleration via $\mathtt{\_\_hfma()}$ intrinsics.

\subsubsection{Backward Pass}

The backward kernel uses the same grid structure ($32\times32$ threads per tile-block). Each thread rebuilds its cached primitive list and computes gradients via the chain rule. For the Over operator, gradients depend on suffix sums as shown in Sec.~3.1.2. Geometric gradients ($x_i, y_i, s_i, \theta_i$) flow through bilinear sampling of $M_i(x,y)$~\cite{jaderberg2015spatial} as in Sec.~3.1.2, where $\frac{\partial M_i(x,y)}{\partial u}$ and $\frac{\partial M_i(x,y)}{\partial v}$ come from bilinear interpolation. We accumulate gradients using packed $\mathtt{\_\_half2}$ atomic operations for parameter pairs $(x_i,y_i)$, $(\nu_i, s_i)$, $(\theta_i, c_{r,i})$, and $(c_{g,i},c_{b,i})$, which halves atomic contention while maintaining accuracy compared to unpacked FP16 atomics.

\subsubsection{PSD Export}

For per-primitive layer generation, we avoid atomics by rendering each primitive to its own cropped buffer. The export uses higher resolution (\eg, $2\times$ or $4\times$) and proceeds in two stages: First, we compute the bounding box $\text{bbox}_i$ for each primitive at the export scale $\rho$ in parallel across $N$ threads. Second, we launch a 3D CUDA grid with dimensions (tiles in $x$, tiles in $y$, primitives), where each primitive $i$ is rendered in parallel to its own layer $L_i$ by computing $M_i(x,y)$, $\alpha$, and $\vc$ as in the forward pass, and writing to local coordinates within $\text{bbox}_i$. This primitive-level parallelism eliminates atomic operations, while memory scales with bounding box areas rather than full canvas, enabling 4K+ exports. Layers are editable in Photoshop/After Effects.

\subsection{Heuristics and Losses for Dynamic and Spatially Constrained Rendering}

\subsubsection{Dynamic \OurMethod{} for Videos}
\OurMethod{} can be easily extended to rendering sequential frames by warm starting from previous frames (initialize $\Theta^{f}$ from $\Theta^{f-1\star}$), as in \cite{luiten2024dynamic} for 3DGS. We add two lightweight controls here, targeting two specific problems: (i) over-dominant ``stuck'' primitives in changing regions and (ii) drift in static regions that causes flicker.

\paragraph{Primitives ``Stuck'' in regions of change}
\cref{Dynamic:RemStk} illustrates the ``stuck'' failure and why it occurs. After warm start, new foreground content in $I^{f}$ may appear where the previous frame $I^{f-1}$ contained background. Without rigidity constraints~\cite{luiten2024dynamic}, the optimizer takes the steepest path by \emph{recoloring} background primitives instead of \emph{relocating} the correct foreground ones. This is a situation we do not want. Large, opaque, front-ordered primitives that sit over high inter-frame change absorb the gradients the other primitives. They just get recolored, suppressing high-frequency detail and leaving the finer primitives behind them suboptimal.

\paragraph{Primitives modified in static region}
Updating a single primitive affects all pixels under its footprint—including regions with no inter-frame change. These unintended edits in static areas perturb the loss and trigger compensatory updates in neighboring primitives, creating a cascade that propagates across the frame (see \cref{Dynamic:FreezUnch}\textcolor{cvprblue}{b}). To arrest this drift, we compute at each step a difference mask $D:=\mathbbm{1}({I_\text{target}^{f-1} \neq I_\text{target}^{f})}$ and \emph{freeze} every primitive whose bounding box does not intersect $D$; only primitives overlapping $D$ are allowed to update. This localizes parameter changes to actually changing content, prevents the cascade and visible flicker.

\cref{alg:adaptive_control} formalizes the procedure: (1) compute per-primitive bounding boxes $\mathcal{B}_i$ and freeze flags from the inter-frame difference mask $D$; (2) partition the canvas into an $n_h{\times}n_w$ spatial grid to localize decisions; (3) within each region, rank non-frozen primitives by a visibility-weighted score and decay the opacity logit of the top-$K$ candidates by a factor $\eta\!\in\!(0,1)$. A primitive is considered stuck if it satisfies all three criteria simultaneously: \emph{large scale} ($s_i \ge \tau_{\text{scale}}\!\cdot\!W$), \emph{high opacity} ($\alpha_i \ge \tau_\alpha$, where $\alpha_i=\alpha_{\max}\sigma(\nu_i)$), and \emph{front z-order} (depth rank above the $\zeta$ percentile within its region). Here, $(n_h, n_w)$ controls spatial granularity, $K$ caps interventions per region, $\zeta$ targets front-most strokes, $\tau_{\text{scale}}$ and $\tau_\alpha$ gate eligibility, and $\eta$ controls decay strength. The procedure is lightweight, data-agnostic, and adds negligible overhead.

\begin{figure}
\centering
\begin{tikzpicture}
\pgfdeclarelayer{bg}
\pgfdeclarelayer{fg}
  \pgfsetlayers{bg,main,fg}

  % ===== Row 1: four small images =====
  \begin{pgfonlayer}{bg}

    % Triplet (a,b, c) to the right; aligned and evenly spaced
    \node[imgSmall] (S2) at (0,0)
      {\includegraphics[width=\imgSmall]{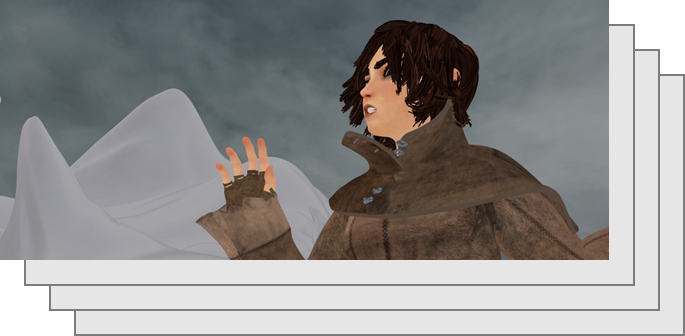}};
    \node[imgSmall] (S3) at ($(S2.east)+(\xGap,0)$)
      {\includegraphics[width=\imgSmall]{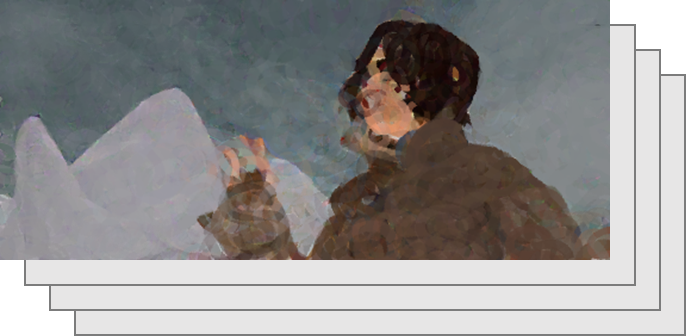}};
    \node[imgSmall] (S4) at ($(S3.east)+(\xGap,0)$)
      {\includegraphics[width=\imgSmall]{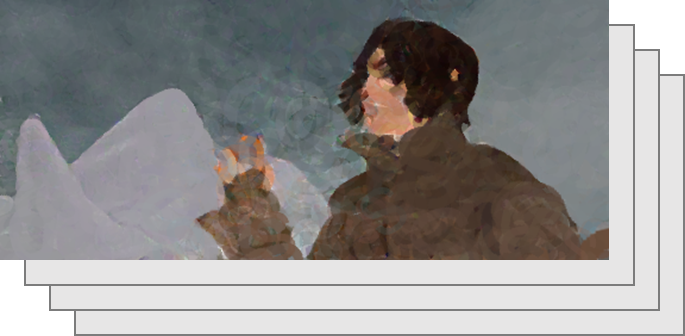}};
  \end{pgfonlayer}

% ===== Row 2: three large images (non-overlapping, L2 shifted left of S2) =====
\begin{pgfonlayer}{bg}
  % L2: shifted left by \Lshift relative to S2
  \node[imgLarge] (L2) at ($(S2.south west)+(0,-\yDrop)$)
    {\includegraphics[width=\imgLarge]{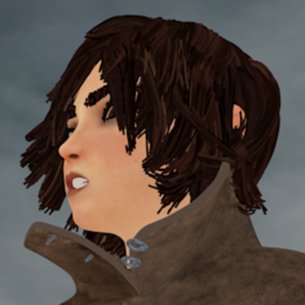}};

  % L3: placed to the right of L2 by a fixed gap; guarantees no overlap
  \node[imgLarge] (L3) at ($(L2.east)+(\xGapLarge,0)$)
    {\includegraphics[width=\imgLarge]{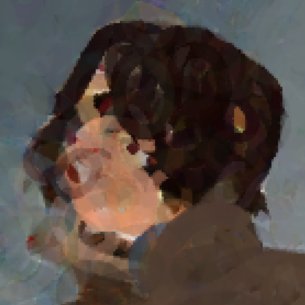}};

  % L4: same strategy (to the right of L3)
  \node[imgLarge] (L4) at ($(L3.east)+(\xGapLarge,0)$)
    {\includegraphics[width=\imgLarge]{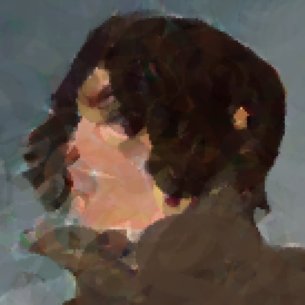}};
\end{pgfonlayer}

  % Headings above row-1 images
  \node[toplabel, anchor=north, font=\footnotesize] at ($(L2.south)+(0,-7pt)$) {(a) Target Frame};

\node[toplabel, anchor=north, font=\footnotesize] at ($(L3.south)+(0,-4pt)$) {%
  {%
    \shortstack{(b) Initialized \\ from $\Theta^{f-1\star}$}%
  }%
};

\node[toplabel, anchor=north, font=\footnotesize] at ($(L4.south)+(0,-4pt)$) {%
  {%
    \shortstack{(c) (b) + Removing \\ Stuck primitives}%
  }%
};

% Example fractions (tune these)
\def\xlo{0.47}\def\ylo{0.46}\def\xhi{0.72}\def\yhi{0.95}

% In your current figure (since you have S2,S3,S4 and L2,L3,L4):
\relboxlinkto{S2}{L2}{\xlo}{\ylo}{\xhi}{\yhi}
\relboxlinkto{S3}{L3}{\xlo}{\ylo}{\xhi}{\yhi}
\relboxlinkto{S4}{L4}{\xlo}{\ylo}{\xhi}{\yhi}

% large row
\def\xlo{0.05}\def\ylo{0.02}\def\xhi{0.95}\def\yhi{0.95}
\relbox{L2}{\xlo}{\ylo}{\xhi}{\yhi}
\relbox{L3}{\xlo}{\ylo}{\xhi}{\yhi}
\relbox{L4}{\xlo}{\ylo}{\xhi}{\yhi}

\begin{pgfonlayer}{fg}
  \node[lightblack_box_white_text, anchor=north east, font=\scriptsize, align=right, inner sep=1pt] at ($(L2.north east)+(-3pt,3pt)$) {\faFilm~3 frames\\ \faExpand~256$\times$109};
  \node[lightblack_box_white_text, anchor=north east, font=\scriptsize, align=right, inner sep=1pt] at ($(L3.north east)+(-3pt,3pt)$) {\faClockO~3.8s, \condensedSB{VRAM}~\scalebox{0.88}[0.95]{0.7GB} \\ 27.20 dB };
  \node[lightblack_box_white_text, anchor=north east, font=\scriptsize, align=right, inner sep=1pt] at ($(L4.north east)+(-3pt,3pt)$) {\faClockO~3.8s, \condensedSB{VRAM}~\scalebox{0.88}[0.95]{0.7GB} \\ 27.89 dB};
\end{pgfonlayer}

\end{tikzpicture}
\caption{\textbf{Frame-wise Fidelity Heuristic of Dynamic \OurMethod{}.} 
Warm-starting from $\Theta^{f-1\star}$ can leave an over-dominant primitive sitting over a high-change area; rather than relocating, it gets recolored and suppresses local detail, as in \textcolor{cvprblue}{b}. We apply \emph{Removing Stuck Primitives}: adaptively decaying the opacity of large, opaque, front-ordered strokes so finer primitives behind them take over. This restores facial detail and improves fidelity under the same budget.}
\label{Dynamic:RemStk}
\end{figure}

\begin{figure}
\centering
\begin{tikzpicture}
\pgfdeclarelayer{bg}
\pgfdeclarelayer{fg}
  \pgfsetlayers{bg,main,fg}

  % ===== Row 1: 6 small images =====
  \begin{pgfonlayer}{bg}
    \node[imgSmall] (S2) at (0,0)
      {\includegraphics[width=0.5\imgSmall]{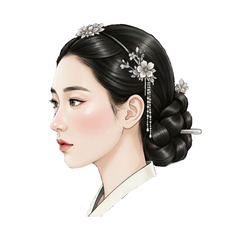}};
    \node[imgSmall] (S3) at ($(S2.east)+(0,0)$)
      {\includegraphics[width=0.5\imgSmall]{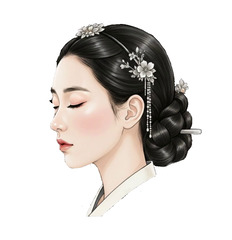}};
    \node[imgSmall] (S4) at ($(S3.east)+(\xGap,0)$)
          {\includegraphics[width=0.5\imgSmall]{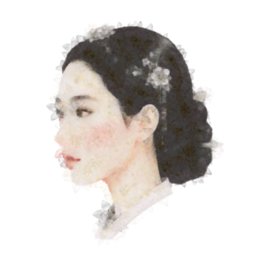}};
    \node[imgSmall] (S5) at ($(S4.east)+(0,0)$)
      {\includegraphics[width=0.5\imgSmall]{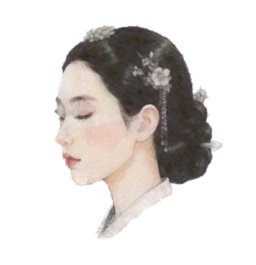}};
    \node[imgSmall] (S6) at ($(S5.east)+(\xGap,0)$)
          {\includegraphics[width=0.5\imgSmall]{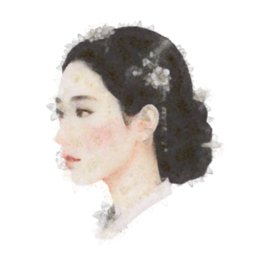}};
    \node[imgSmall] (S7) at ($(S6.east)+(0,0)$)
      {\includegraphics[width=0.5\imgSmall]{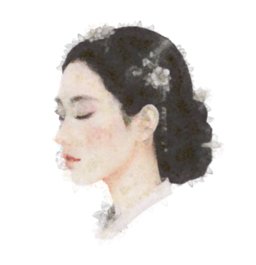}};
  \end{pgfonlayer}

% ===== Row 2: three large images (non-overlapping, L2 shifted left of S2) =====
\begin{pgfonlayer}{bg}
  % L2: shifted left by \Lshift relative to S2
  \node[imgLarge] (L2) at ($(S2.south west)+(0,-\yDrop)$)
    {\includegraphics[width=\imgLarge]{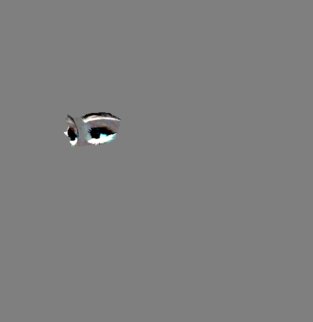}};

  % L3: placed to the right of L2 by a fixed gap; guarantees no overlap
  \node[imgLarge] (L3) at ($(L2.east)+(\xGapLarge,0)$)
    {\includegraphics[width=\imgLarge]{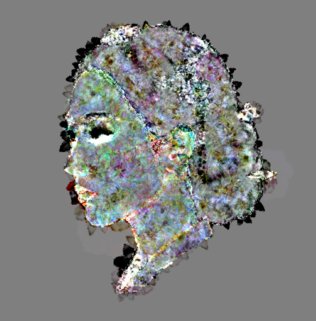}};

  % L4: same strategy (to the right of L3)
  \node[imgLarge] (L4) at ($(L3.east)+(\xGapLarge,0)$)
    {\includegraphics[width=\imgLarge]{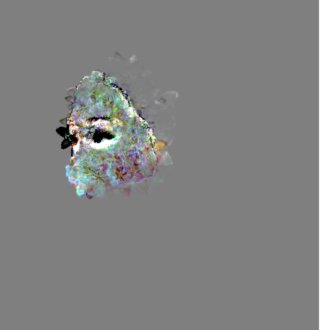}};
\end{pgfonlayer}

  % Headings above row-1 images
  \node[toplabel, anchor=north, font=\footnotesize] at ($(L2.south)+(0,-7pt)$) {(a) Ground Truth};

\node[toplabel, anchor=north, font=\footnotesize] at ($(L3.south)+(0,-4pt)$) {%
  {%
    \shortstack{(b) Initialized \\ from $\Theta^{f-1\star}$}%
  }%
};

\node[toplabel, anchor=north, font=\footnotesize] at ($(L4.south)+(0,-4pt)$) {%
  {%
    \shortstack{(c) (b) + Freezing \\ Unchanged Primitives}%
  }%
};

\begin{pgfonlayer}{fg}
  \node[lightblack_box_white_text, anchor=north east, font=\scriptsize, align=right, inner sep=1pt] at ($(L2.north east)+(-3pt,3pt)$) {\faFilm~17 frames\\ \faExpand~512$\times$529};
  \node[lightblack_box_white_text, anchor=north east, font=\scriptsize, align=right, inner sep=1pt] at ($(L3.north east)+(-3pt,3pt)$) {\faClockO~240s, \condensedSB{VRAM}~\scalebox{0.88}[0.95]{0.8GB} \\ tLP$\downarrow$ 9.26};
  \node[lightblack_box_white_text, anchor=north east, font=\scriptsize, align=right, inner sep=1pt] at ($(L4.north east)+(-3pt,3pt)$) {\faClockO~223s, \condensedSB{VRAM}~\scalebox{0.88}[0.95]{0.8GB} \\ tLP$\downarrow$ 1.11};
\end{pgfonlayer}

\end{tikzpicture}
\caption{\textbf{Anti-flicker via freezing unchanged regions.}
We visualize inter-frame change with an error map
$E = 5\!\cdot\!(I_{1}-I_{17}) + 127$, where $I_{1}$ is the \textit{first} frame and $I_{17}$ the \textit{last} frame.
Mid-gray ($\approx127$) indicates no change; brighter/darker values denote larger differences.
If all primitives are freely optimized, spurious updates appear even in static regions, producing widespread flicker, as seen in \textcolor{cvprblue}{b}.
Our remedy \textcolor{cvprblue}{c} freezes primitives, concentrating updates only where the video truly changes.}
\label{Dynamic:FreezUnch}
\end{figure}

\begin{algorithm}[h]
\caption{Dynamic \OurMethod{} for Videos}
\label{alg:adaptive_control}
\small
\begin{algorithmic}[1]
\Require Primitives $\{\Theta_i = (x_i, y_i, s_i, \theta_i, \nu_i, \vc_i)\}_{i=1}^N$, spatial grid size $(n_h, n_w)$, inter-frame difference mask $D \in \{0,1\}^{H \times W}$, thresholds $\{\tau_{\text{scale}}, \tau_{\alpha}, \zeta\}$, per-region budget $K$, decay factor $\eta$
\Ensure Updated opacity logits $\{\nu_i^\star\}_{i=1}^N$

\State \textbf{Step 1: Compute freeze masks}
\For{$i = 1$ to $N$}
    \State Compute bounding box $\mathcal{B}_i$ from $(x_i, y_i, s_i, \theta_i)$
    \State $\text{freeze}_i \gets \mathbbm{1}(\mathcal{B}_i \cap \text{support}(D) = \emptyset)$ \Comment{freeze if no overlap with change}
\EndFor

\State \textbf{Step 2: Partition canvas into $n_h \times n_w$ spatial regions}
\For{each region $\mathcal{R}_{j}$ in $\{1, \ldots, n_h \times n_w\}$}
    \State $\mathcal{P}_j \gets \{i \mid (x_i, y_i) \in \mathcal{R}_j\}$ \Comment{primitives in region $j$}
    
    \State \textbf{Step 3: Score primitives by visibility and stuck criteria}
    \For{each primitive $i \in \mathcal{P}_j$}
        \State $\alpha_i \gets \alpha_{\max} \cdot \sigma(\nu_i)$
        \State Evaluate stuck criteria:
        \State \quad $c_1(i) \gets \mathbbm{1}(s_i \ge \tau_{\text{scale}} \cdot W)$ \Comment{large scale}
        \State \quad $c_2(i) \gets \mathbbm{1}(\alpha_i \ge \tau_{\alpha})$ \Comment{high opacity}
        \State \quad $c_3(i) \gets \mathbbm{1}(\text{depth-rank}_i \ge \zeta \cdot |\mathcal{P}_j|)$ \Comment{front z-order}
        \State $\text{is\_stuck}_i \gets (c_1(i) + c_2(i) + c_3(i) = 3) \land (\neg \text{freeze}_i)$
        \State $\text{score}_i \gets s_i \cdot \alpha_i \cdot \text{is\_stuck}_i$
    \EndFor
    
    \State \textbf{Step 4: Select and decay top-$K$ stuck primitives}
    \State $\mathcal{S}_j \gets \text{top-}K\{\text{score}_i \mid i \in \mathcal{P}_j\}$ \Comment{by descending score}
    \For{each $i \in \mathcal{S}_j$}
        \State $\nu_i^\star \gets \eta \cdot \nu_i$ \Comment{reduce opacity}
    \EndFor
\EndFor
\State \Return $\{\nu_i^\star\}_{i=1}^N$
\end{algorithmic}
\end{algorithm}

\subsubsection{Rendering with Spatial Constraint}
\paragraph{Re-initialization Mechanism}
When enabled, Primitives with $\sigma(\nu_i) <$ $\mathtt{prune\_threshold}$ (typically 0.3) are re-initialized. This occurs every $\mathtt{prune\_iterations}$ (50), excluding an initial warmup period and final iterations (optional) to prevent destabilization. Instead of pruning the transparent primitives as in \cite{rogge2025OC2DGS}, pruned primitives are re-initialized randomly, following our Structure-aware initialization (Sec. 3.2.2).

\section{Experimental Configuration Details}\label{sec:exp_configs}

This section provides a comprehensive description of the experimental settings used to generate all figures in the main paper. All configuration files are available in the $\mathtt{configs/}$ directory of our repository. Parameters not explicitly specified in individual configs default to values defined in $\mathtt{pydiffbmp/util/constants.py}$.

\subsection{Common Hyperparameters}

\Cref{tab:common_hyperparams} summarizes hyperparameters and settings that are consistently used across all or most experiments. These form the foundation of our optimization pipeline.

\begin{table*}[h]
\centering
\caption{\textbf{Common hyperparameters used across experiments.} These settings are applied to all experiments unless explicitly overridden in figure-specific configurations. Values marked with $^\dagger$ come from $\mathtt{constants.py}$ when not specified in config files.}
\label{tab:common_hyperparams}
\small
\setlength{\tabcolsep}{4pt}
\begin{tabular}{@{}llp{10cm}@{}}
\toprule
\textbf{Category} & \textbf{Parameter} & \textbf{Value / Description} \\
\midrule
\multirow{6}{*}{\textbf{Initialization}} 
& $\mathtt{initializer}$ & $\mathtt{structure\_aware}$ (default for most; $\mathtt{random}$ for Fig.~8) \\
& $\mathtt{v\_init\_bias}$ & $-4.0$ (yields $\sigma(-4) \approx 1.8\%$ initial opacity) \\
& $\mathtt{std\_c\_init}^\dagger$ & $0.02$ (color initialization noise std) \\
& $\mathtt{variance\_window\_size}^\dagger$ & $7$ (local variance computation window) \\
& $\mathtt{variance\_base\_prob}^\dagger$ & $0.1$ (base sampling probability for low-variance areas) \\
& $\mathtt{max\_prims\_per\_pixel}^\dagger$ & $100$ (200 for spatially constrained cases) \\
\midrule
\multirow{9}{*}{\textbf{Optimization}}
& $\mathtt{num\_iterations}$ & 100--500 (task-dependent; see \cref{tab:figure_specific}) \\
& $\mathtt{learning\_rate.default}$ & $0.1$ (base LR; scaled by gains below) \\
& $\mathtt{lr\_gain\_x}^\dagger$ & $10.0$ \\
& $\mathtt{lr\_gain\_y}^\dagger$ & $10.0$ \\
& $\mathtt{lr\_gain\_r}^\dagger$ & $10.0$ (scale parameter) \\
& $\mathtt{lr\_gain\_v}^\dagger$ & $1.5$ (5.0 for spatially constrained; see \cref{tab:figure_specific}) \\
& $\mathtt{lr\_gain\_theta}^\dagger$ & $1.0$ \\
& $\mathtt{lr\_gain\_c}^\dagger$ & $1.0$ \\
& $\mathtt{do\_decay}$ & $\mathtt{true}$ (exponential LR decay) \\
\midrule
\multirow{3}{*}{\textbf{Rendering}}
& $\mathtt{do\_gaussian\_blur}$ & $\mathtt{true}$ (soft rasterization, Sec. 3.2.1) \\
& $\mathtt{blur\_sigma}^\dagger$ & $1.0$ \\
& $\mathtt{alpha\_upper\_bound}$ & $1.0$ (0.7 for spatially constrained) \\
\midrule
\multirow{2}{*}{\textbf{Loss}}
& $\mathtt{loss\_config.type}$ & $\mathtt{mse}$ or $\mathtt{combined}$ (see \cref{tab:figure_specific}) \\
& $\mathtt{bg\_color}^\dagger$ & $\mathtt{white}$ (default); $\mathtt{random}$ for no-bg cases \\
\midrule
\multirow{2}{*}{\textbf{Postprocessing}}
& $\mathtt{psd\_scale\_factor}$ & $2.0$ or $4.0$ (export resolution multiplier) \\
& $\mathtt{compute\_psnr}$ & $\mathtt{true}$ (for quantitative evaluation) \\
\bottomrule
\end{tabular}
\end{table*}

\subsection{Figure-Specific Configurations}

\Cref{tab:figure_specific} details the key parameters that vary across different figures in the main paper. Each row corresponds to a specific subfigure or experiment. For more detail and specs, please check the code provided in the supplementary material. 

\begin{table*}[h]
\centering
\caption{\textbf{Figure-specific experimental configurations.} This table shows parameters that differ across experiments, including primitive details, number of primitives ($N$), image resolution, scale range, iterations, and special settings.}
\label{tab:figure_specific}
\small
\setlength{\tabcolsep}{3pt}
\begin{tabular}{@{}lp{3.2cm}ccccp{5cm}@{}}
\toprule
\textbf{Figure} & \textbf{Primitive(s)} & \textbf{$N$} & \textbf{Resolution} & \textbf{Scale Range} & \textbf{Iter.} & \textbf{Special Settings} \\
\midrule
\multicolumn{7}{@{}l}{\textit{Figure 1: Teaser examples}} \\
\quad 1(a)-fingerprint & fingerprint.jpg & 2000 & 1024 & [2, 8] & 300 & $\mathtt{radial\_transparency{=}true}$, $\mathtt{c\_blend{=}0.0}$ \\
\quad 1(a)-autograph & autograph\_seurat.png & 2000 & 1024 & [2, 10] & 300 & $\mathtt{c\_blend{=}0.0}$ \\
\quad 1(b) & logos/*.png (300 logos) & 300 & 512 & [4, 20] & 100 & $\mathtt{c\_blend{=}1.0}$, grayscale L1 + MSE loss \\
\quad 1(c)-flowers & 5 flower types & 1000 & 512 & [2, 31] & 300 & $\mathtt{exist\_bg{=}false}$, mask, pruning \\
\quad 1(c)-girl & 2 flower types & 1000 & 512 & [2, 20] & 300+100 & $\mathtt{exist\_bg{=}false}$, initial+sequential (17 frames), \cref{alg:adaptive_control}, freeze unchanged \\
\midrule
\multicolumn{7}{@{}l}{\textit{Figure 2: An illustration of the algorithm flow of \OurMethod{}}} \\
\quad 2 & 4 flower types & 500 & 512 & [2, 20] & 10 & $\mathtt{exist\_bg{=}false}$, 4 masks, re-initialization \\
\midrule
\multicolumn{7}{@{}l}{\textit{Figure 4: Heuristics for Dynamic \OurMethod{}}} \\
\quad 4(b) & cane.png, hat.png & 1000 & 256 & [8, 48] & 100 & sequential (8 frames), $\Theta^{f-1\star}$ init, MSE+perceptual loss \\
\quad 4(c) & cane.png, hat.png & 1000 & 256 & [8, 48] & 100 & sequential (8 frames), $\Theta^{f-1\star}$ init, \cref{alg:adaptive_control}, MSE+perceptual loss \\
\midrule
\multicolumn{7}{@{}l}{\textit{Figure 5: DiffVG vs ours}} \\
\quad 5(a) & square.svg & 2000 & 512 & [2, 10] & 100 & $\mathtt{use\_fp16{=}false}$ \\
\quad 5(b) & grass.svg & 2000 & 512 & [2, 10] & 100 & $\mathtt{use\_fp16{=}false}$ \\
\quad 5(c) & paw.jpg & 2000 & 512 & [2, 10] & 100 & $\mathtt{use\_fp16{=}false}$ \\
\midrule
\multicolumn{7}{@{}l}{\textit{Figure 6: Noisy Canvas Ablation}} \\
\quad 6(b) & Lisc\_lipy.jpg & 1000 & 312 & [2, 50] & 100 & $\mathtt{bg\_color{=}white}$ \\
\quad 6(c) & Lisc\_lipy.jpg & 1000 & 312 & [2, 50] & 100 & $\mathtt{bg\_color{=}random}$ \\
\midrule
\multicolumn{7}{@{}l}{\textit{Figure 7: Alpha Loss and Re Initialization Ablation}} \\
\quad 7(a) & paw\_complicated.png & 3000 & 512 & [2, 64] & 300 & $\mathtt{exist\_bg{=}false}$, $\mathtt{lr\_gain\_v{=}5.0}$, MSE only w/o mask \\
\quad 7(b) & paw\_complicated.png & 3000 & 512 & [2, 64] & 300 & $\mathtt{exist\_bg{=}false}$, $\mathtt{lr\_gain\_v{=}5.0}$, MSE+alpha loss, w/o re-initialization \\
\quad 7(c) & paw\_complicated.png & 3000 & 512 & [2, 64] & 300 & $\mathtt{exist\_bg{=}false}$, $\mathtt{lr\_gain\_v{=}5.0}$, MSE+alpha loss, w/ re-initialization\\
\midrule
\multicolumn{7}{@{}l}{\textit{Figure 8: CLIP-guided generation}} \\
\quad 8-amazon & flowers/*.png & 500 & 224 & [2, 20] & 500 & CLIP loss, random init, custom LR \\
\quad 8-galaxy & maxwell\_eq*.png & 1000 & 224 & [19, 20] & 500 & CLIP loss, random init, custom LR \\
\quad 8-witch & bat.png & 500 & 224 & [19, 20] & 500 & CLIP loss, random init, custom LR \\
\bottomrule
\end{tabular}
\end{table*}

\subsection{Detailed Descriptions of Main Figures}

\paragraph{Figure 1: Teaser Examples}
The teaser presents various applications of \OurMethod{}. \textbf{Seurat Painting Composition (1a)} uses 2000 fingerprints and 2000 autographs on separate image regions at 1024px resolution. The fingerprint primitives use $\mathtt{radial\_transparency{=}true}$ to create smooth fading effects. \textbf{The Marilyn Monroe assemblage (1b)} employs 300 brand logos~\cite{koustubhk_popular_2025} with $\mathtt{c\_blend{=}1.0}$ to preserve the original logo colors, using a combined loss with grayscale L1 (weight 1.0) and MSE (weight 0.2) to maintain luminance structure. \textbf{The flower video composition (1c)} consists of two parts. First, background flower primitives are optimized using a circle mask. Second, for the foreground video of the girl, an initial frame is optimized with a spatially constrained image (face region). The subsequent frames use warm-start initialization from the previous frame ($\Theta^{f-1\star}$), combined with our dynamic heuristics: removing stuck primitives and freezing unchanged regions, as detailed in Sec. 3.3.1 of the main paper.

\paragraph{Figure 2: An illustration of the algorithm flow of \OurMethod{}}
The target image $I^\text{target}$ is spatially constrained using four semantic masks (hair, skin, neck, cloth) from the CelebAMask-HQ dataset~\cite{lee2020maskgan}. The image $I$ depicts the intermediate output after 10 iterations, which continues for a total of 100 iterations. 

\paragraph{Figure 4: Heuristics for Dynamic DiffBMP}
The example uses an 8-frame clip from \emph{The Gold Rush} (Charlie Chaplin). 
Warm-starting the current frame from $\Theta^{f-1\star}$ can trap the optimizer in a local minimum—(b) shows an over-dominant primitive stuck across the face that washes out detail. 
In (c) we apply \emph{Removing Stuck Primitives}: adaptively decaying the opacity of large, front-ordered, high-opacity strokes so finer primitives take over, restoring facial detail and improving per-frame fidelity.

\paragraph{Figure 5: DiffVG vs ours}
These experiments demonstrate \OurMethod{}'s superior performance and versatility, particularly with complex bitmap or vector primitives where DiffVG struggles. All experiments use 2000 primitives at 512px resolution and are optimized for 100 iterations.

\paragraph{Figure 6: Noisy Canvas Ablation}
This ablation demonstrates the effect of canvas background on primitive coverage (Sec. 3.2.3). Configuration (b) uses $\mathtt{bg\_color{=}white}$, while (c) uses $\mathtt{bg\_color{=}random}$, encouraging primitives to fill all regions by blending with uniform noise per iteration.

\paragraph{Figure 7: Alpha Loss and Re Initialization Ablation}
These experiments ablate the alpha loss and re-initialization mechanisms for spatially constrained rendering(Sec. 3.3.2). Configuration (a) uses MSE loss only (but it follows the spatially constrained initialization), (b) adds the opacity loss component (weight 0.3) without re-initialization, and (c) enables both. We re-initialize primitives with opacity below 0.3 every 50 iterations, with a warmup period of 199 iterations. All use $\mathtt{gain\_v{=}5.0}$ to accelerate opacity optimization.

\paragraph{Figure 8: CLIP-Guided Generation}
These experiments use text-prompt guidance via CLIP loss instead of target images. The initialization is $\mathtt{random}$ rather than structure-aware as no target image structure is available. All configurations use ViT-B/32 CLIP model with 16 augmentations, normalized CLIP embeddings. The Amazon rainforest uses primitives with small scale [2, 20], while galaxy and witch use larger primitives [19, 20] to create distinct visual styles. Additionally, we apply the negative prompt ``blurry'' with a weight of 0.1 to mitigate unwanted artifacts.

\paragraph{Spatial Constraint Implementation}
Experiments with $\mathtt{exist\_bg{=}false}$ (Figures 1c, 2, 6, 7) optimize foreground-only rendering using Eq. 9 in the main paper. The alpha loss weight is typically 0.3, and $\mathtt{gain\_v}$ is increased to 5.0 to facilitate rapid opacity adjustments. The $\mathtt{max\_prims\_per\_pixel}$ is increased to 200 for these cases to ensure adequate coverage in complex regions.

\paragraph{Dynamic DiffBMP Implementation}
For dynamic \OurMethod{} (Figures 1c and 4), we optimize sequential frames by warm-starting from the previous frame's optimized parameters ($\Theta^{f-1\star}$). To prevent stuck primitives and flickering, we employ two heuristics: (1) \textbf{Removing Stuck Primitives} adaptively reduces opacity of over-dominant primitives that satisfy all three criteria simultaneously (large scale $s_i \ge \tau_{\text{scale}} \cdot W$, high opacity $\alpha_i \ge \tau_{\alpha}$, front z-order exceeding the $\zeta$ percentile). We partition the canvas into an $(n_h{=}4) \times (n_w{=}4)$ spatial grid and select up to $K{=}4$ problematic primitives per region, reducing their $\nu_i$ by a factor $\eta$. This occurs at specific epochs ([20, 45, 70] for Fig.~1c; [20, 40, 60, 80] for Fig.~4c). (2) \textbf{Freezing Unchanged Regions} computes an inter-frame difference mask $D$ and freezes primitives whose bounding boxes do not intersect $D$, preventing spurious updates in static regions. Hyperparameters: $\tau_{\text{scale}}{=}0.1$, $\tau_{\alpha}{=}0.7$, $\zeta{=}0.7$, $\eta{=}0.3$ (Fig.~1c) or $0.1$ (Fig.~4c).

\paragraph{FP16 Precision}
Most experiments use $\mathtt{use\_fp16{=}true}$ for memory efficiency. The FP16 implementation maintains accuracy through packed $\mathtt{\_\_half2}$ atomic operations as described in Sec. 3.1.2.

\section{Naive PyTorch Baseline Comparison}

To demonstrate the efficiency of our CUDA implementation, we have compared against a naive PyTorch-based renderer that uses standard tensor operations without tile-based culling or specialized kernels, in Sec. 4.1.

\subsection{Naive PyTorch Implementation}

The naive PyTorch approach processes all primitives for every pixel without spatial optimization, as shown in Algorithm~\ref{alg:pytorch_naive}.

\begin{algorithm}
\caption{Naive PyTorch Baseline (Sequential)}
\label{alg:pytorch_naive}
\small
\begin{algorithmic}[1]
\Require $N$ primitives $\{(x_i, y_i, s_i, \theta_i, \nu_i, \vc_i)\}_{i=1}^N$
\Ensure Rendered image $I$

\State $X, Y \gets \mathtt{meshgrid}()$ over canvas \Comment{Full $(H_{canvas}, W_{canvas})$ grids}
\State $I \gets \mathtt{zeros}()$; $T \gets \mathtt{ones}()$

\For{$i = 0$ \textbf{to} $N-1$} \Comment{Sequential CPU loop, no parallelism}
    \State $\Delta x \gets X - x_i$; $\Delta y \gets Y - y_i$ \Comment{$(H, W)$ tensors per primitive}
    \State $u, v \gets$ rotate/scale $(\Delta x, \Delta y)$ by $\theta_i, s_i$
    \State $mask \gets \mathtt{torch.grid\_sample}(M_i, (u,v))$ \Comment{Generic interpolation}
    \State $\alpha_i \gets \alpha_{max} \cdot \sigma(\nu_i) \cdot mask$; $\vc_i \gets \sigma(\vc_i)$
    \State $I \gets I + T \cdot \alpha_i \cdot \vc_i$; $T \gets T \cdot (1 - \alpha_i)$
\EndFor
\State \Return $I$
\end{algorithmic}
\end{algorithm}

\subsection{Key Inefficiencies and Performance Analysis}

Table~\ref{tab:pytorch_vs_cuda_supp} summarizes the architectural differences between the naive PyTorch baseline and our optimized CUDA implementation. The naive PyTorch baseline suffers from several critical bottlenecks:

\begin{table}[h]
\centering
\caption{Comparison of naive PyTorch vs. our CUDA implementation}
\label{tab:pytorch_vs_cuda_supp}
\setlength{\tabcolsep}{4pt}
\resizebox{\linewidth}{!}{%
\begin{tabular}{lcc}
\toprule
\textbf{Aspect} & \textbf{Naive PyTorch} & \textbf{Our CUDA} \\
\midrule
Spatial culling & None & Tile-based binning \\
Primitive loop & Sequential ($N$) & Parallel per-pixel \\
Memory per iteration & $O(N \times HW)$ & $O(k \times HW)$, $k \ll N$ \\
Precision & FP32 only & FP16 + FP32 mixed \\
Intermediate tensors & $(H, W)$ per primitive & Small per-pixel cache \\
Atomic operations & Many (unoptimized) & Packed $\mathtt{\_\_half2}$ \\
Gradient accumulation & Global sync & Per-pixel atomic \\
\midrule
\textbf{Speedup} & \textbf{1$\times$ (baseline)} & \textbf{30-50$\times$} \\
\bottomrule
\end{tabular}
}
\end{table}

\paragraph{Major Bottlenecks}
(1) \textbf{No spatial culling}---every primitive is evaluated at all $H \times W$ pixels, yielding $O(N \times HW)$ work even when most primitives contribute nothing. For 1000 primitives on $512^2$ canvas, this is 262M wasted evaluations. (2) \textbf{Massive memory allocation}---each primitive creates full-canvas tensors for $\Delta x, \Delta y, u, v, mask$ ($\sim$5 GB at FP32 for our example). (3) \textbf{Sequential loop}---the Python $\mathtt{for}$ loop prevents primitive-level parallelism and forces repeated kernel launches. (4) \textbf{Generic operations}---$\mathtt{torch.grid\_sample()}$ handles general cases rather than exploiting our structure.

Our CUDA kernel achieves 30-50$\times$ speedup via tile-based culling (10-100$\times$ work reduction), per-pixel parallelism, compact caches storing only $k \ll N$ affecting primitives, and FP16 optimization with tensor cores.

\section{More Results}
%Here, we provide more \OurMethod{} experimental results. 

\paragraph{Qualitative Results for Table 3}
Across our experiments, \OurMethod{} handles a wide variety of primitive libraries and target images within a single optimization framework, yet the optimization behavior is consistent. It first recovers the global layout and dominant color structure of a target and then progressively sharpens local details as optimization proceeds. This behavior is clearly visible in \cref{fig:opt_progress}, where we visualize the same configurations used in Table~3 of the main paper. Early iterations already place large primitives that capture silhouettes and large color regions, while later iterations refine edges, textures, and small features as overlapping primitives are repurposed and reweighted. The three rows in \cref{fig:opt_progress} cover different combinations of primitives and targets including photographic images, portraits, and more stylized graphics, which shows that this coarse to fine optimization pattern holds across diverse content without task specific tuning.

\paragraph{The Number of Primitives $N$}
In \cref{fig:diverse_example}, we vary the primitive count $N$ from 1000 to 4000 for several pairings of primitive collections and target images. Larger values of $N$ mainly improve sharpness, fine texture, and small details, resulting in PSNR gains.

\paragraph{Scaling with primitive footprint and tile density.}
To better understand how the rendering workload scales with the primitive count $N$, we analyze three complementary descriptors: the normalized primitive footprint $A_i/(HW)$, the primitive tile-hit degree $\tau_i$, and the per-tile primitive density $d_t$. We use the same primitive families as in the Seurat painting composition of  \cref{fig:teaser}(a), namely fingerprint and autograph primitives. To expose a broader primitive size distribution, we widen the allowed scale range to $[2,50]$ in this analysis instead of using the original  \cref{fig:teaser}(a) ranges. As in  \cref{fig:teaser}(a), radial transparency is enabled for the fingerprint primitives. Fig.~\ref{fig:supp_scaling_summary} summarizes the resulting distributions of $A_i/(HW)$ and $\tau_i$, together with the measured runtime, as $N$ increases. Here, $A_i$ denotes the area of the padded screen-space bounding box of primitive $i$, normalized by the image area $HW$. $\tau_i$ denotes the number of tiles intersected by that primitive. $d_t$ denotes the number of primitives assigned to tile $t$ during CPU binning.

For both primitive families, increasing $N$ shifts the footprint CDF toward smaller values and slightly lightens the $\tau_i$ distribution, indicating that individual primitives become smaller and intersect fewer tiles on average. However, runtime still increases monotonically with $N$. Fig.~\ref{fig:supp_tile_density_heatmaps} further shows that, despite the smaller average footprint of individual primitives, visually active regions accumulate progressively denser tile-local primitive lists as $N$ grows. Since the total primitive--tile interaction count satisfies $\sum_i \tau_i = \sum_t d_t$, the observed scaling depends not only on primitive count, but also on how primitive overlap is spatially distributed across tiles.

\begin{figure*}[t]
    \centering
    \includegraphics[width=0.195\linewidth]{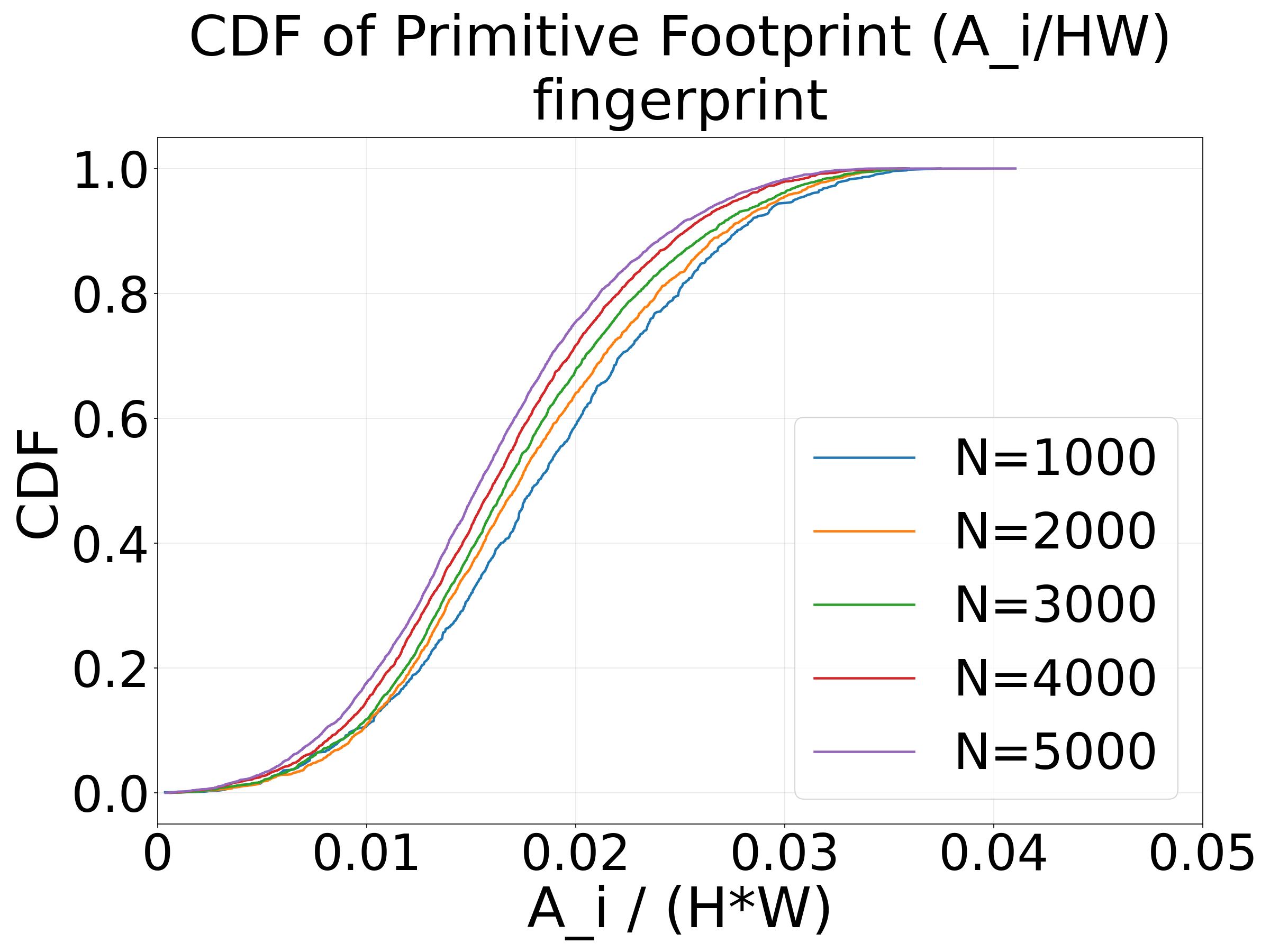}\hfill
    \includegraphics[width=0.195\linewidth]{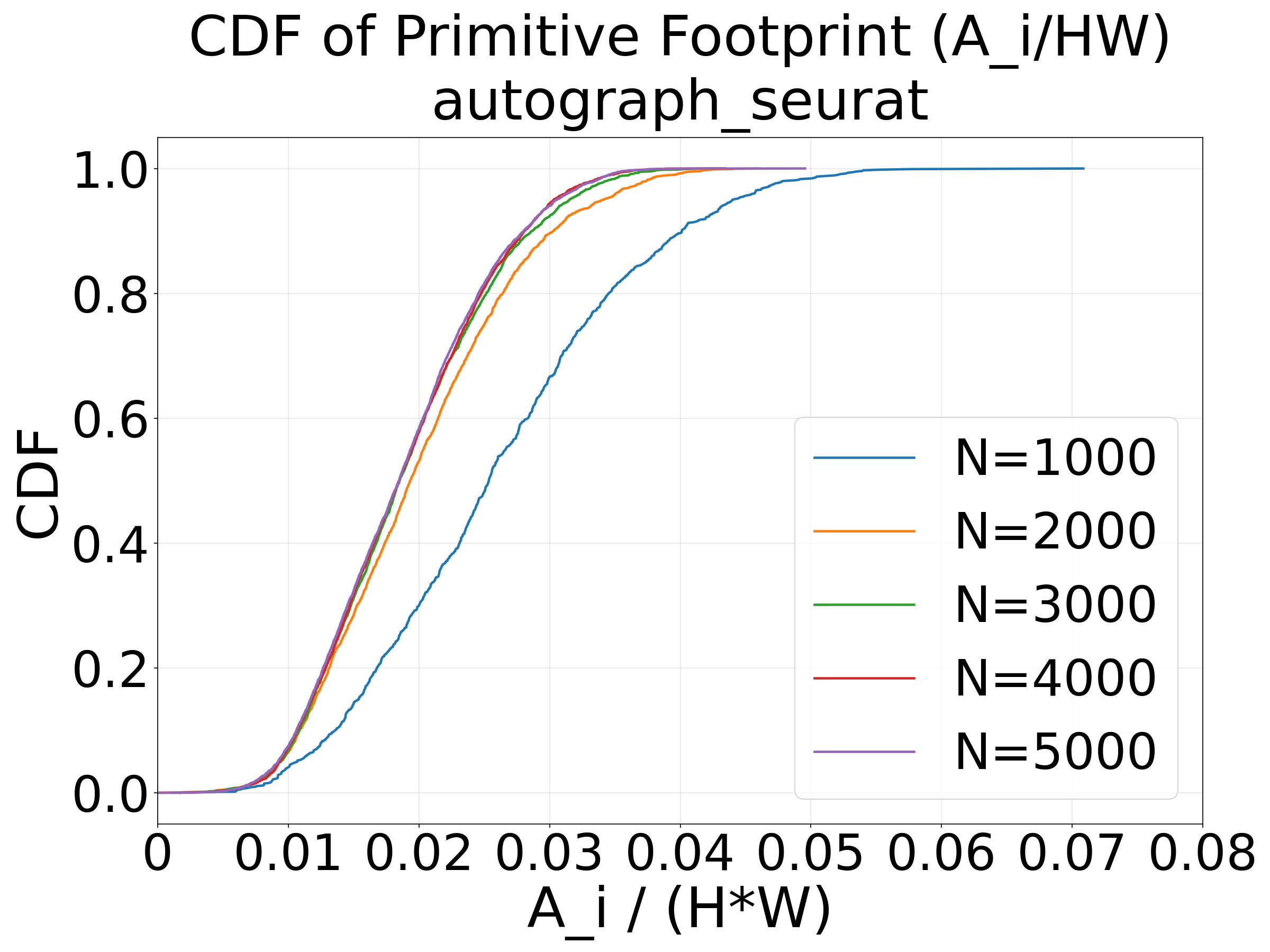}\hfill
    \includegraphics[width=0.195\linewidth]{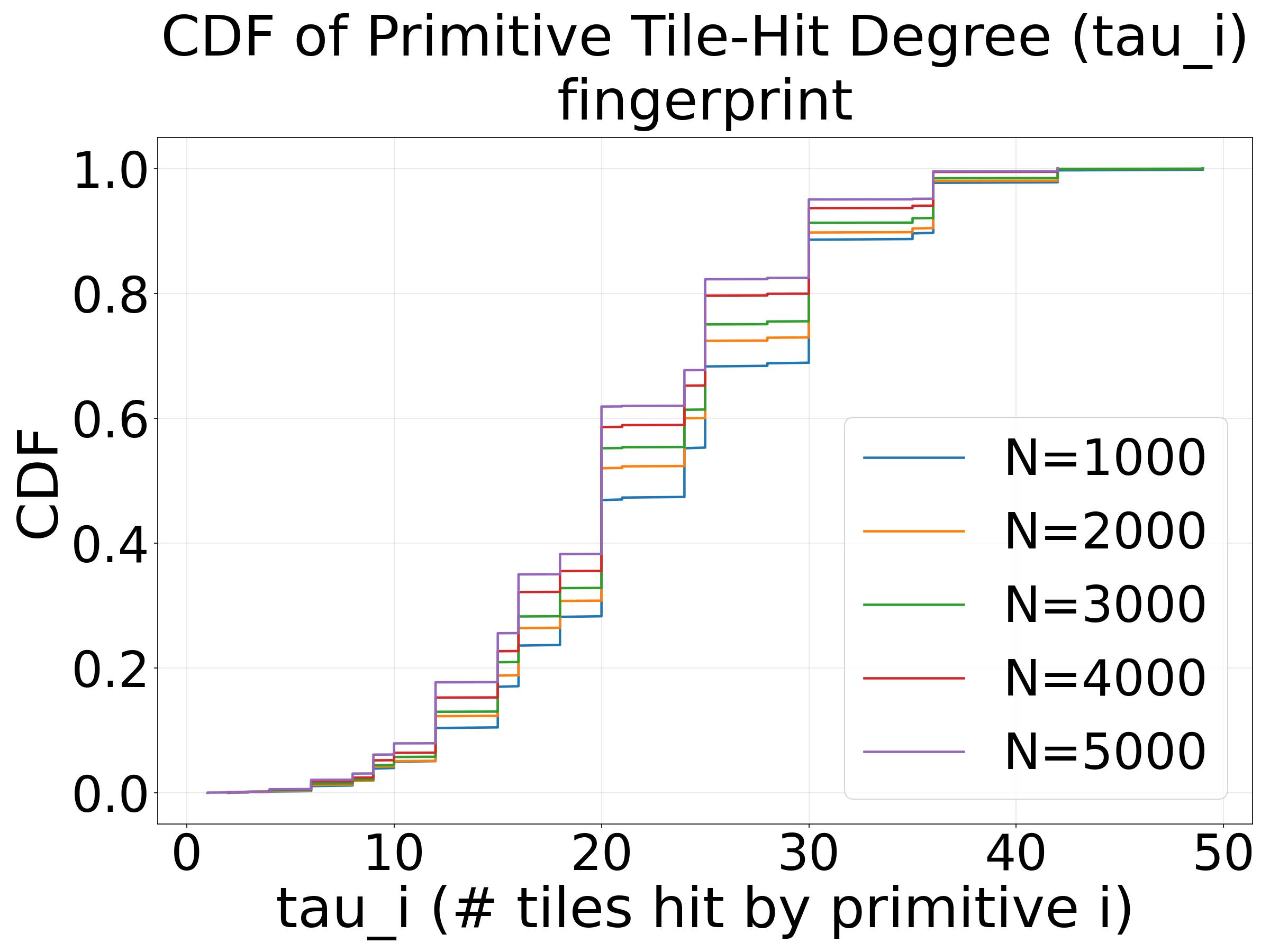}\hfill
    \includegraphics[width=0.195\linewidth]{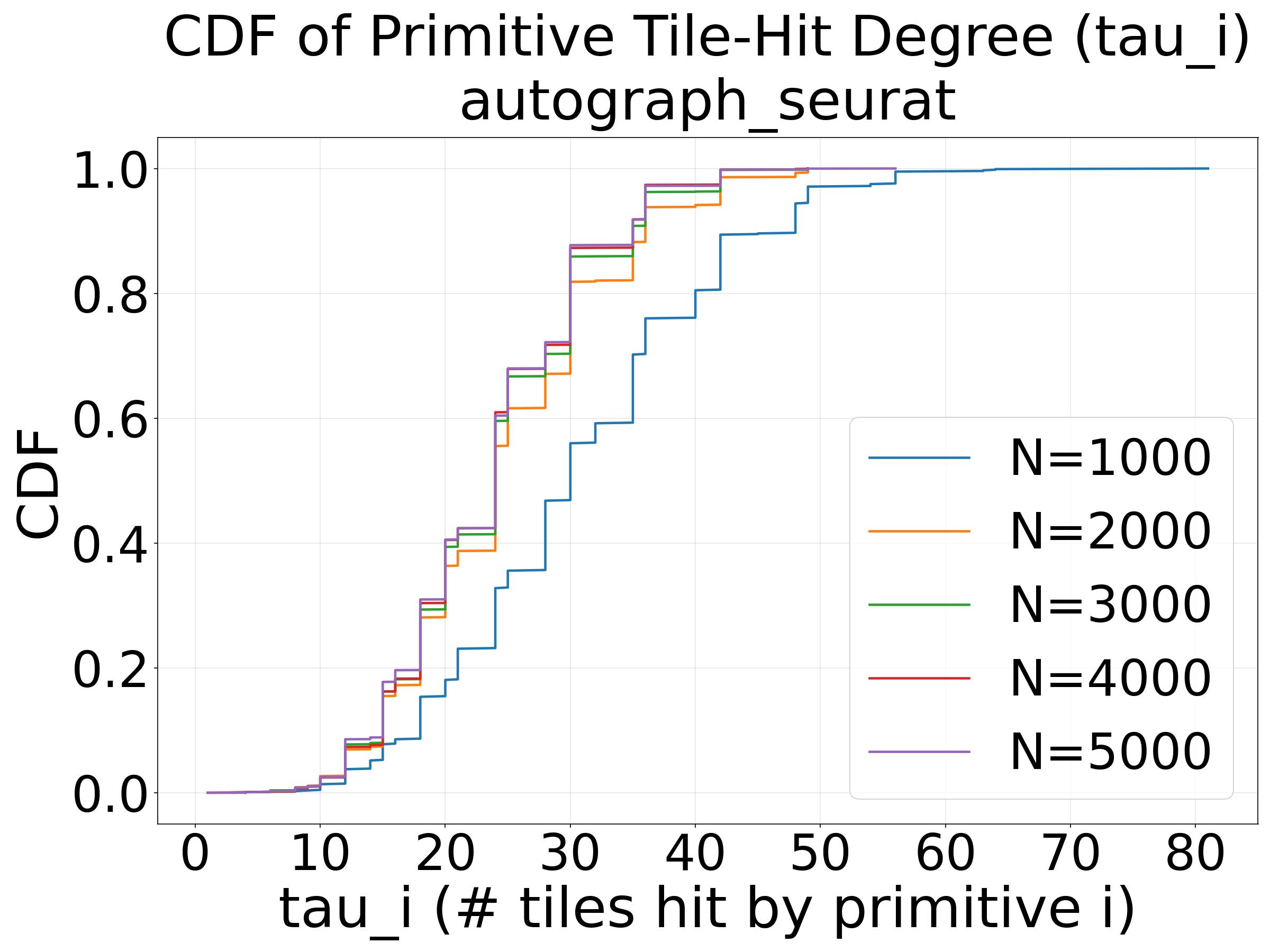}\hfill
    \includegraphics[width=0.195\linewidth]{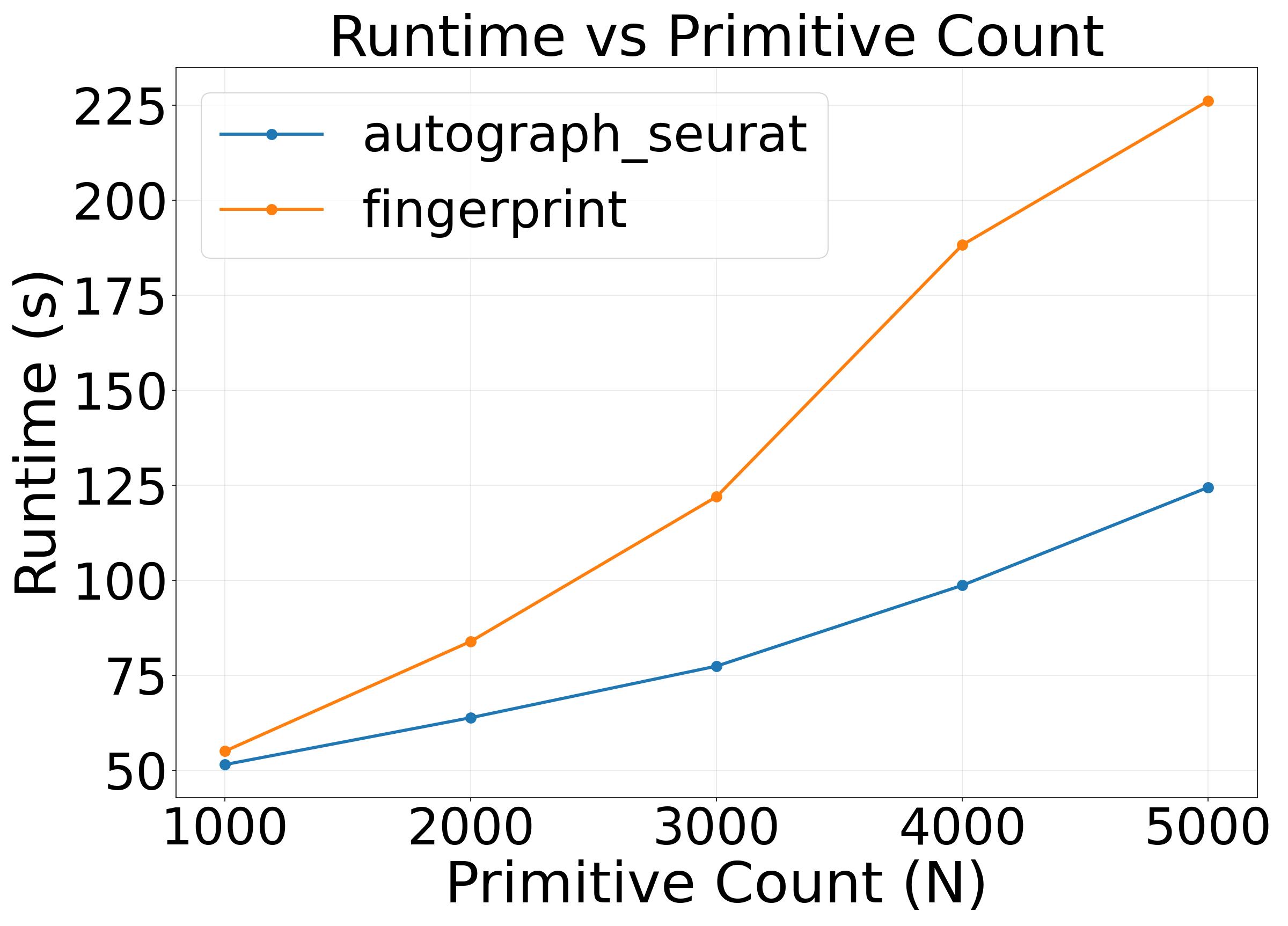}
    \caption{
    \textbf{Additional scaling statistics for the Seurat composition in \cref{fig:teaser}(a).} To expose a broader primitive size distribution, this analysis widens the allowed scale range to $[2,50]$ while sweeping the primitive count $N$. Fingerprint primitives retain radial transparency as in \cref{fig:teaser}(a). From left to right, the panels show the CDF of the normalized primitive footprint $A_i/(HW)$ for \textit{fingerprint}, the CDF of $A_i/(HW)$ for \textit{autograph\_seurat}, the CDF of the primitive tile-hit degree $\tau_i$ for \textit{fingerprint}, the CDF of $\tau_i$ for \textit{autograph\_seurat}, and the measured runtime versus $N$. Here, $A_i$ denotes the area of the padded screen-space bounding box of primitive $i$, normalized by the image area $HW$. $\tau_i$ denotes the number of tiles intersected by that primitive.}
    \label{fig:supp_scaling_summary}
\end{figure*}

\begin{figure*}[t]
    \centering
    \includegraphics[width=0.195\linewidth]{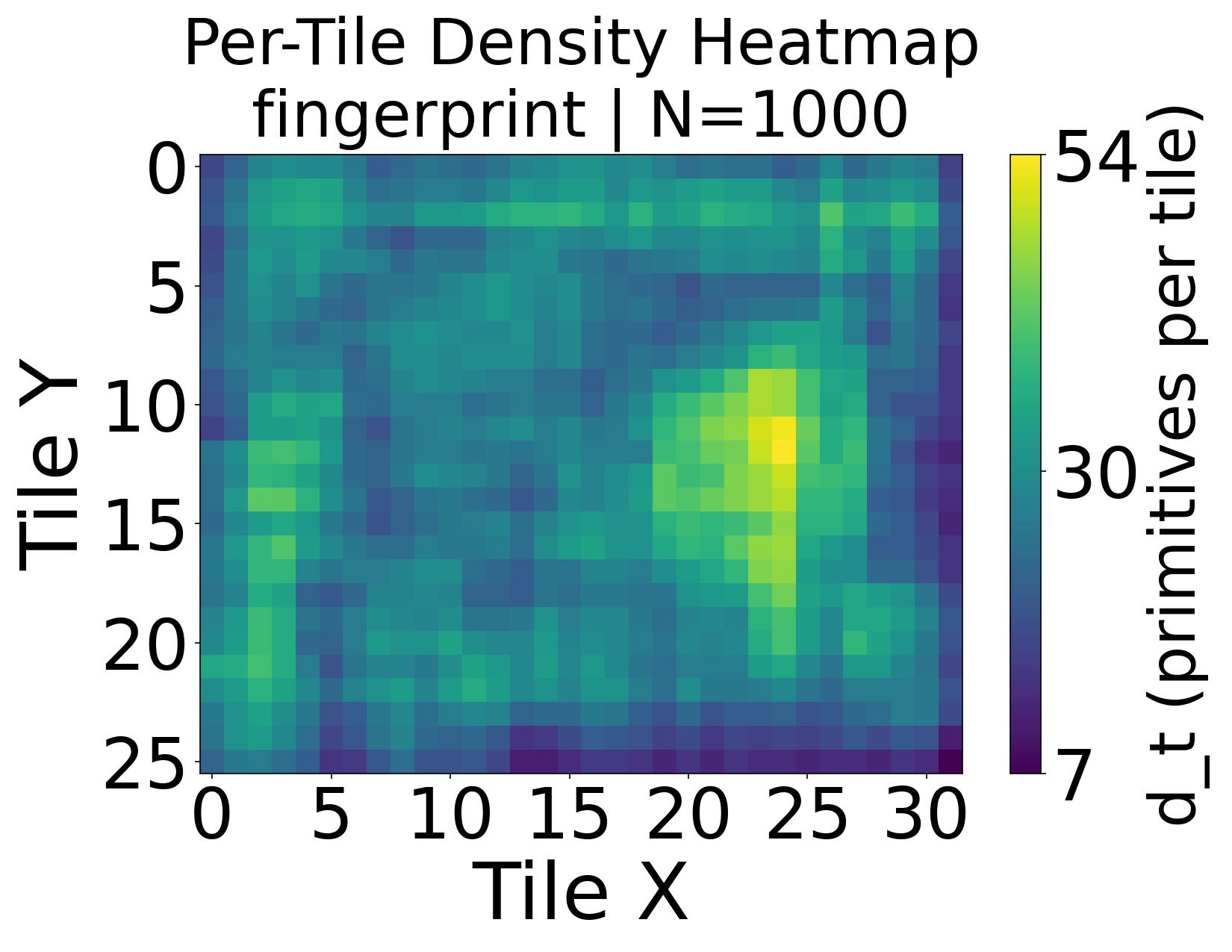}\hfill
    \includegraphics[width=0.195\linewidth]{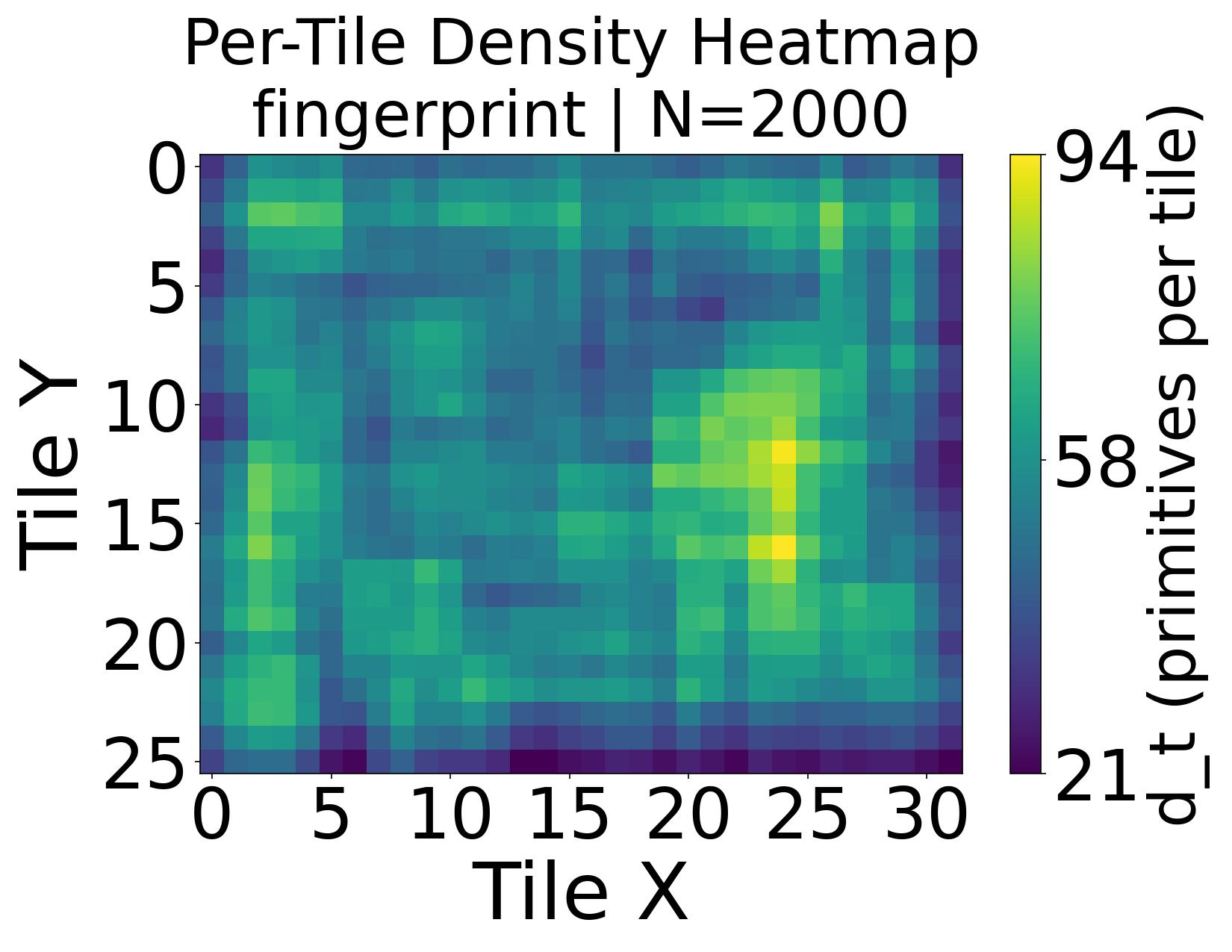}\hfill
    \includegraphics[width=0.195\linewidth]{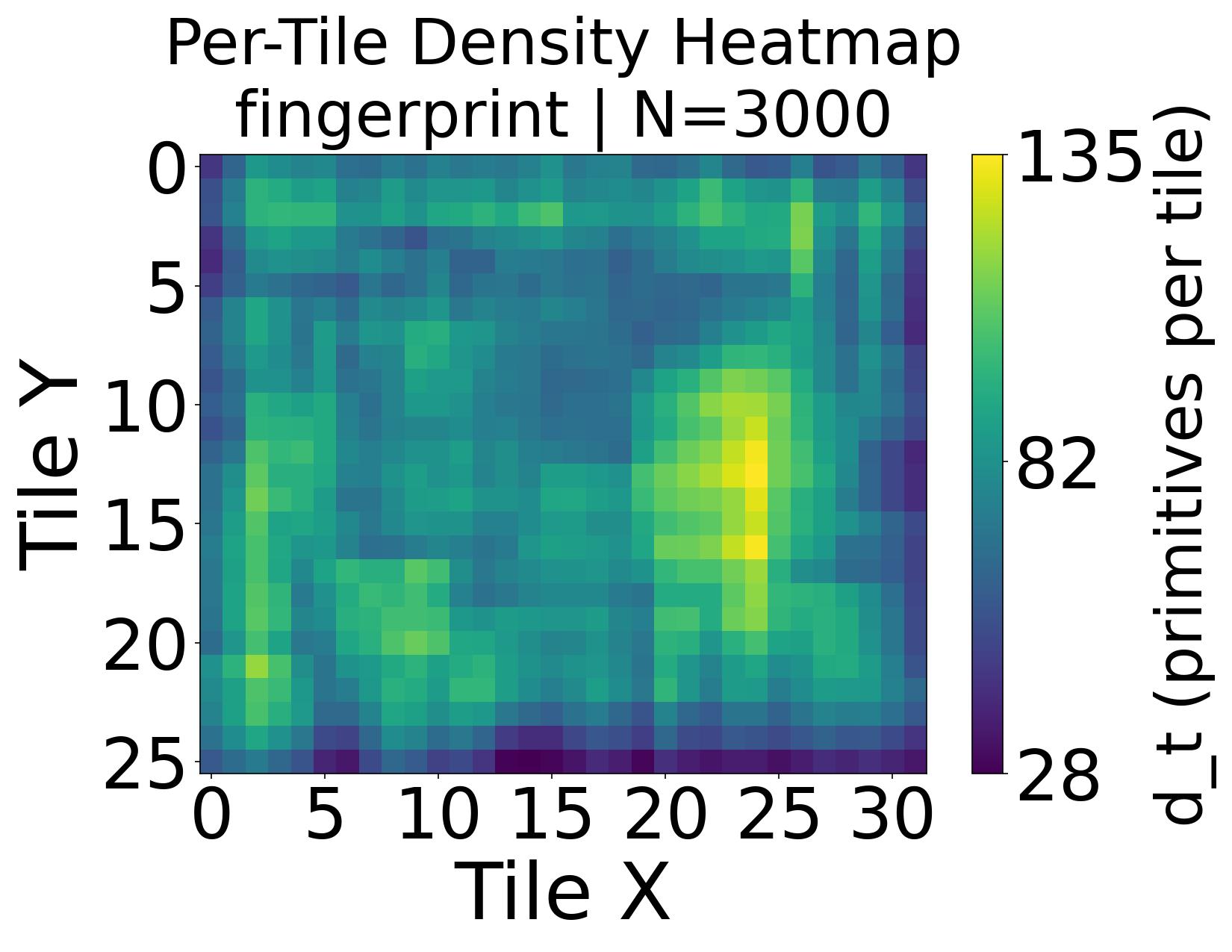}\hfill
    \includegraphics[width=0.195\linewidth]{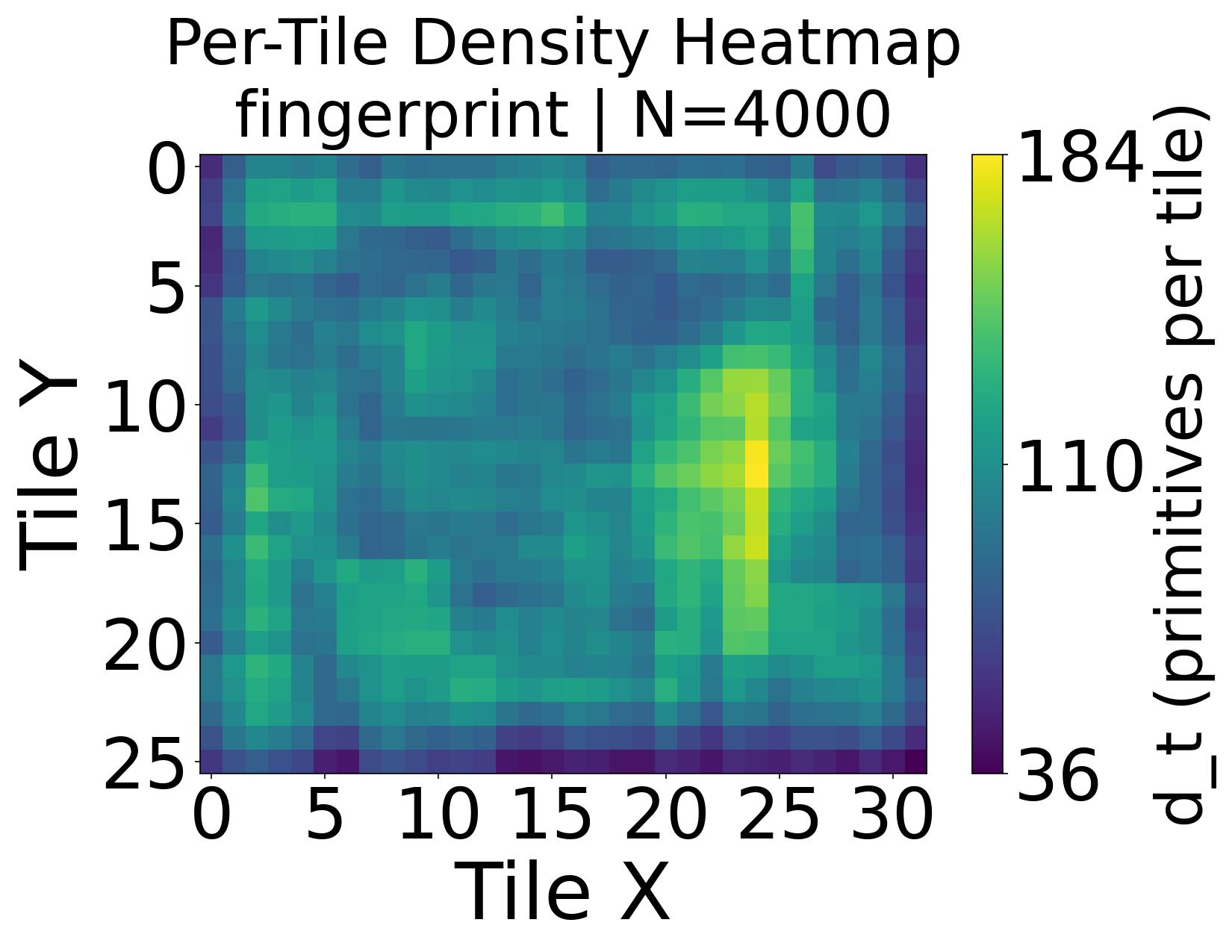}\hfill
    \includegraphics[width=0.195\linewidth]{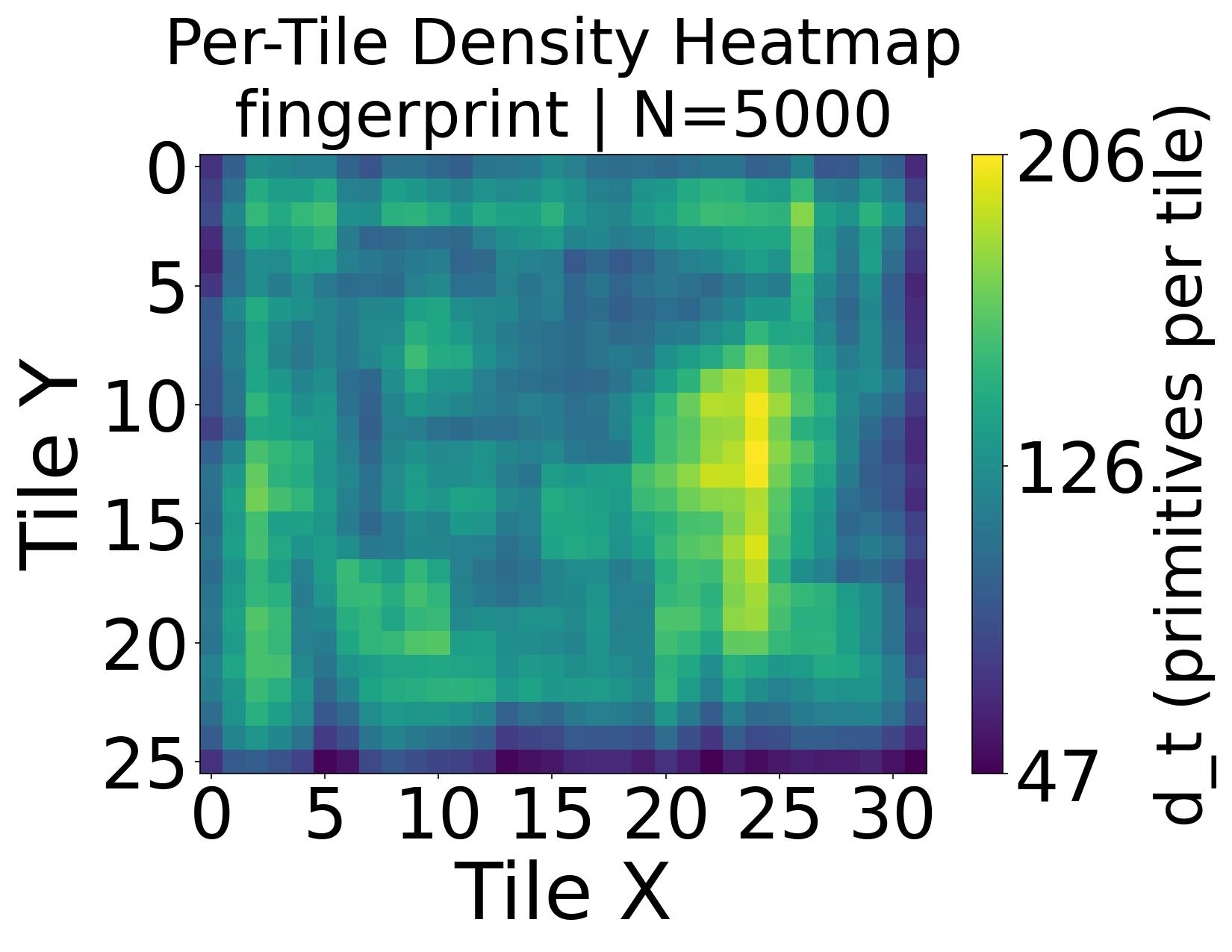}

    \vspace{1mm}

    \includegraphics[width=0.195\linewidth]{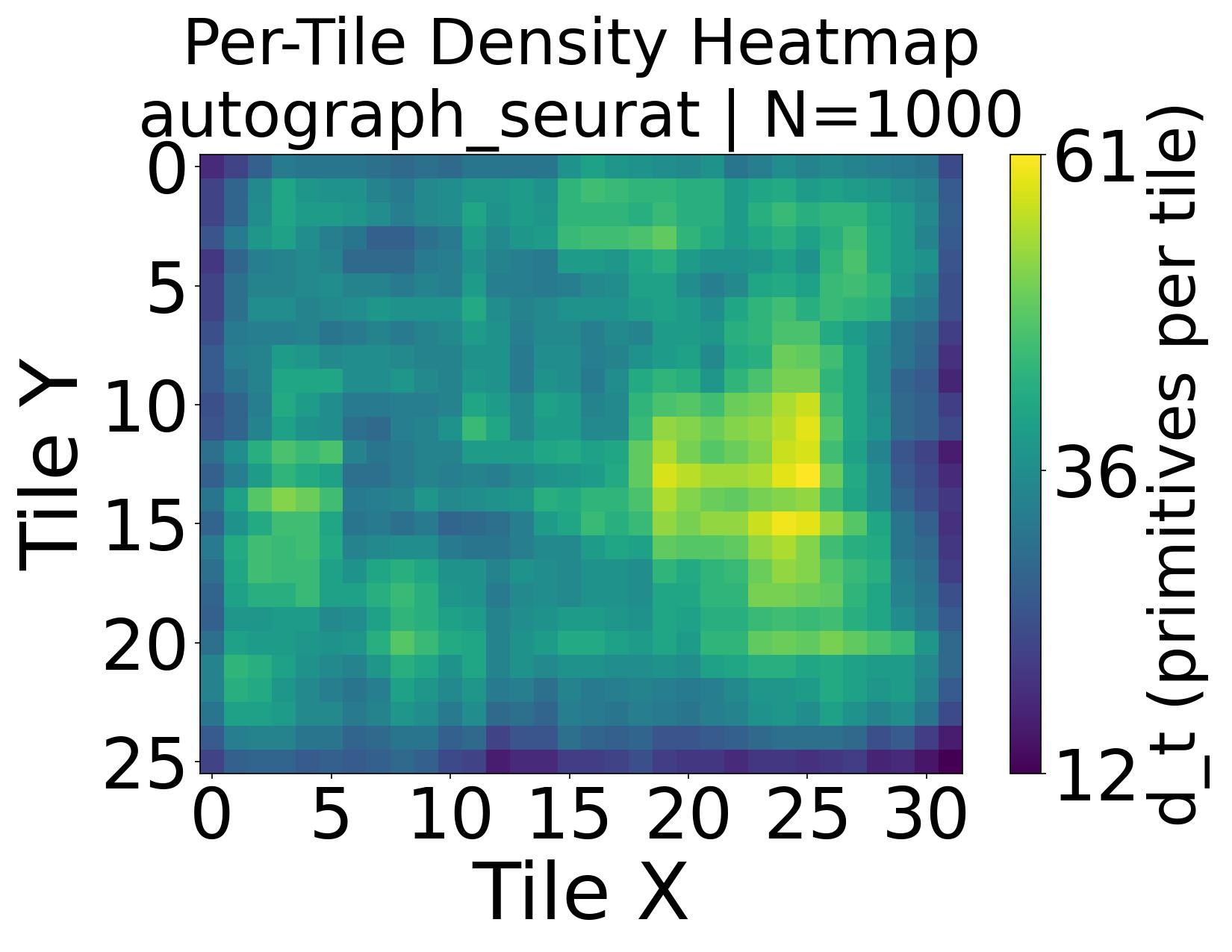}\hfill
    \includegraphics[width=0.195\linewidth]{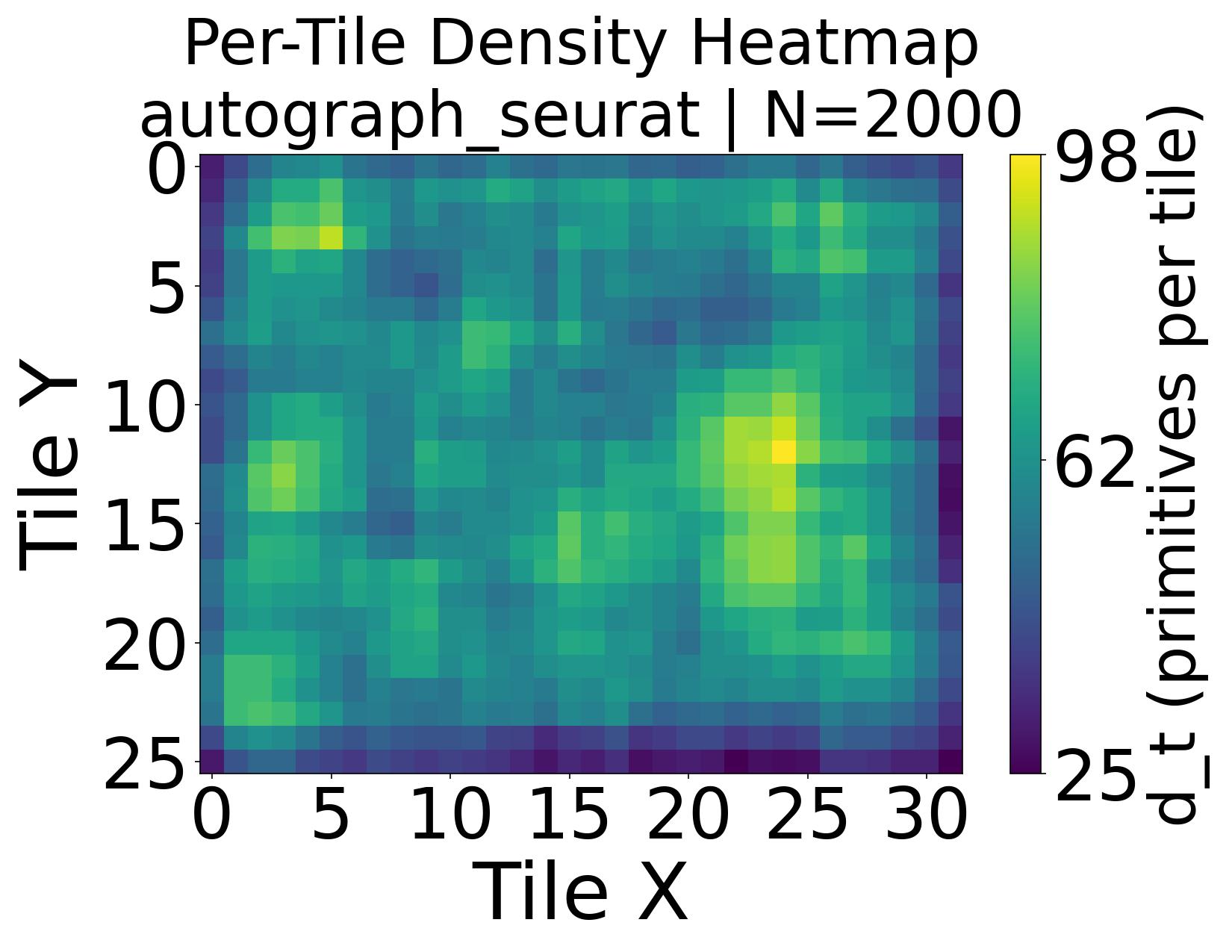}\hfill
    \includegraphics[width=0.195\linewidth]{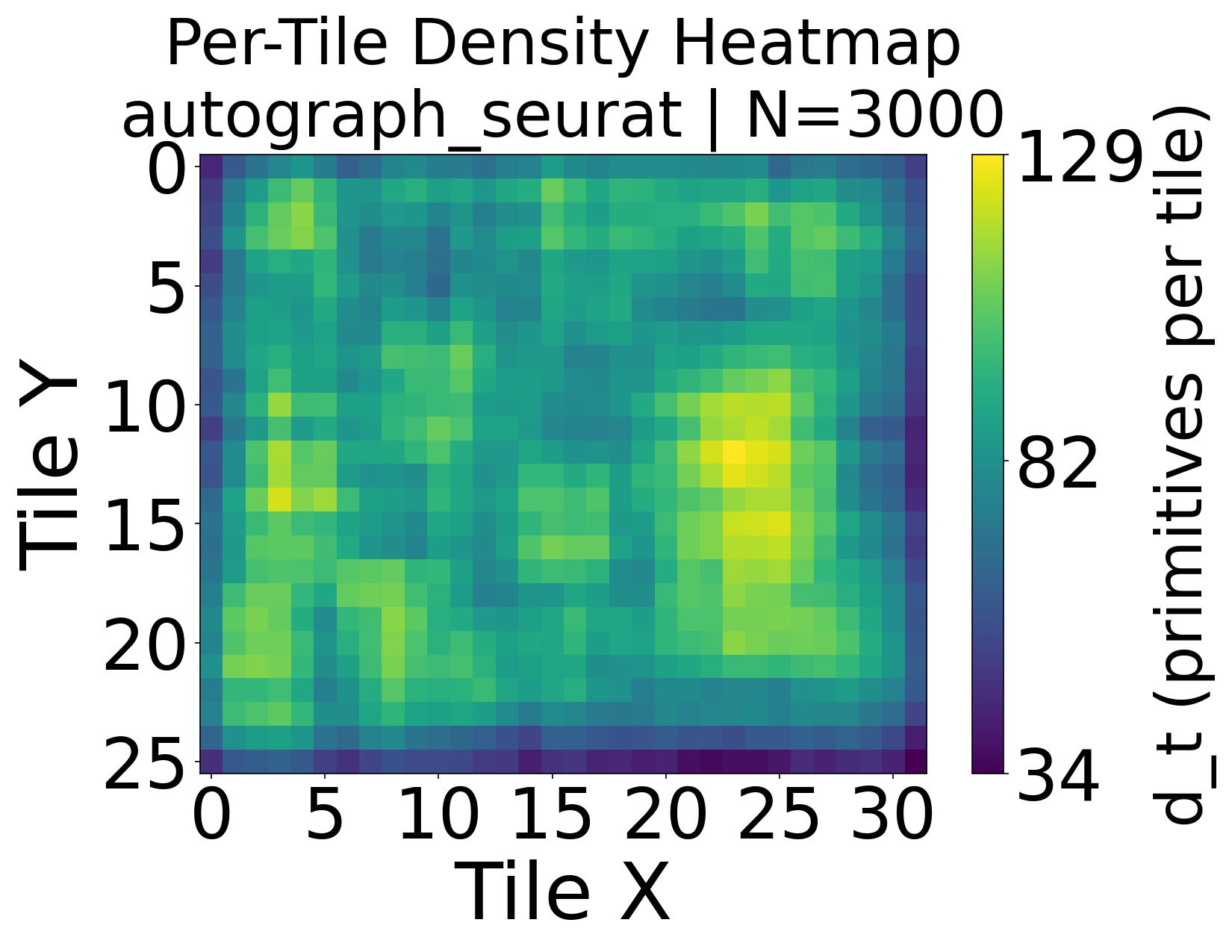}\hfill
    \includegraphics[width=0.195\linewidth]{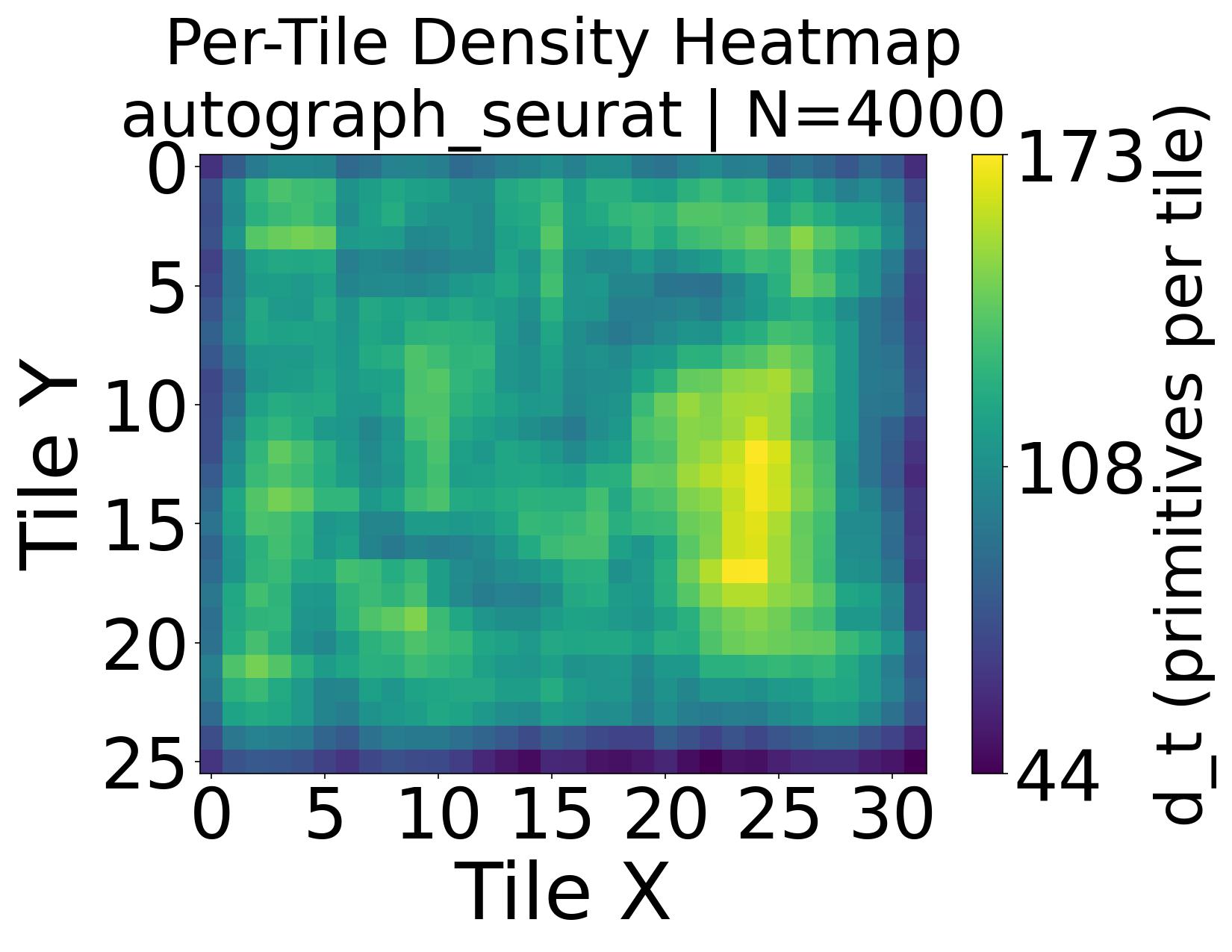}\hfill
    \includegraphics[width=0.195\linewidth]{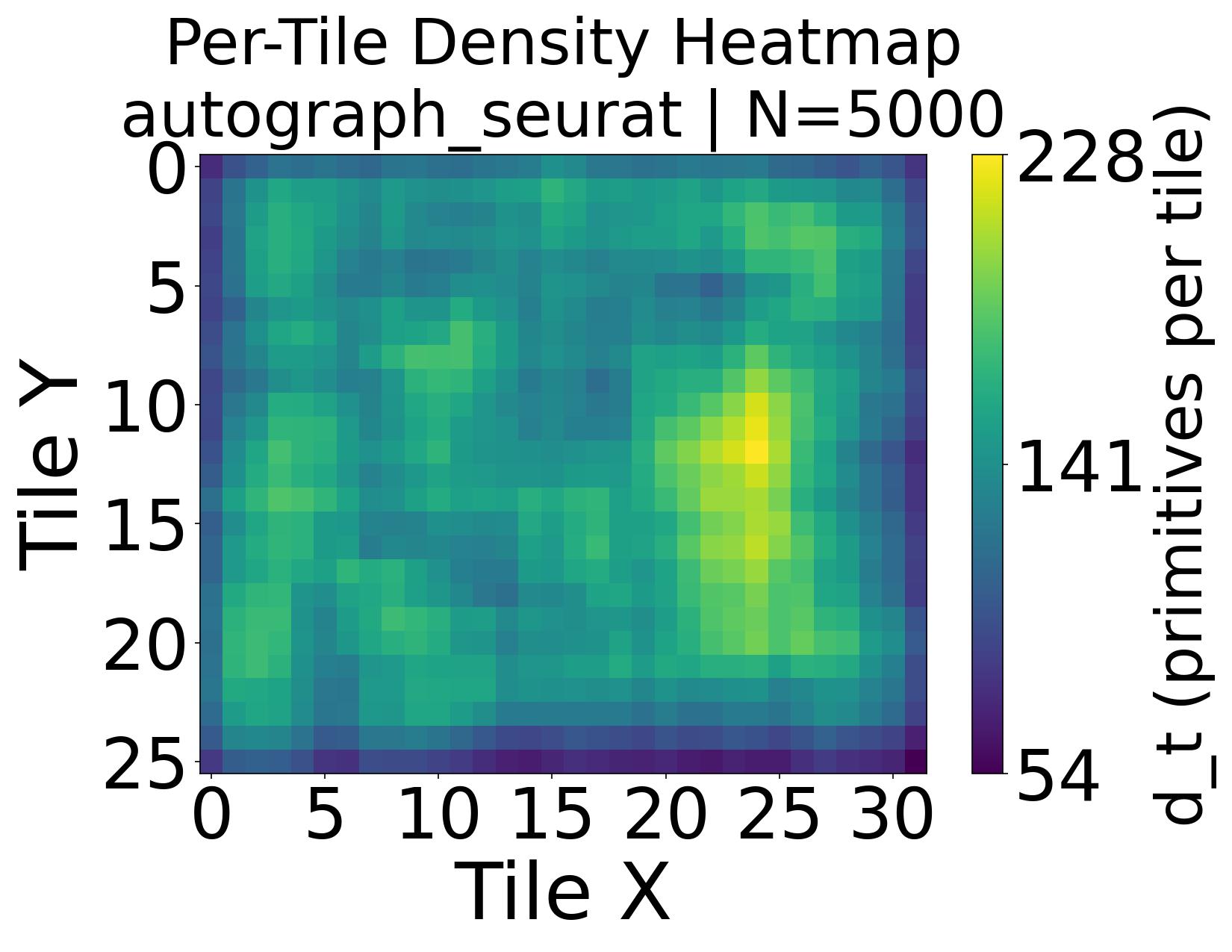}
    \caption{
    Spatial distribution of per-tile primitive density $d_t$ for the same scaling analysis setting as Fig.~\ref{fig:supp_scaling_summary}, with scale range $[2,50]$ and increasing primitive count $N$. Each heatmap visualizes the number of primitives assigned to tile $t$ during CPU binning. The top row shows \textit{fingerprint}, and the bottom row shows \textit{autograph\_seurat}. In each row, $N \in \{1000, 2000, 3000, 4000, 5000\}$ increases from left to right. Each panel uses its own color scale for readability, so absolute density values should be interpreted using the corresponding colorbar.}
    \label{fig:supp_tile_density_heatmaps}
\end{figure*}

\begin{figure*}[t!]
  \centering
  \begin{minipage}{1.0\textwidth}
  \centering
  \small
  % 컬럼 간 여백 조절 (필요에 따라 값 조절)
  \setlength{\tabcolsep}{4pt}
  \begin{tabular}{@{}*{6}{c}@{}}
    % ── 컬럼 헤더 ───────────────────────────────
    \textbf{Primitive} & \textbf{iter. 1} & \textbf{iter. 10}
      & \textbf{iter. 20} & \textbf{iter. 100} & \textbf{Ground~Truth} \\
    %\addlinespace[1ex]
    % ── 첫 번째 예시 (artwork) ─────────────────
    \begin{tikzpicture} [baseline={([yshift=-10.5ex]IMG)}]
      \node[inner sep=0pt] (IMG) {
        \includegraphics[width=0.13\textwidth]{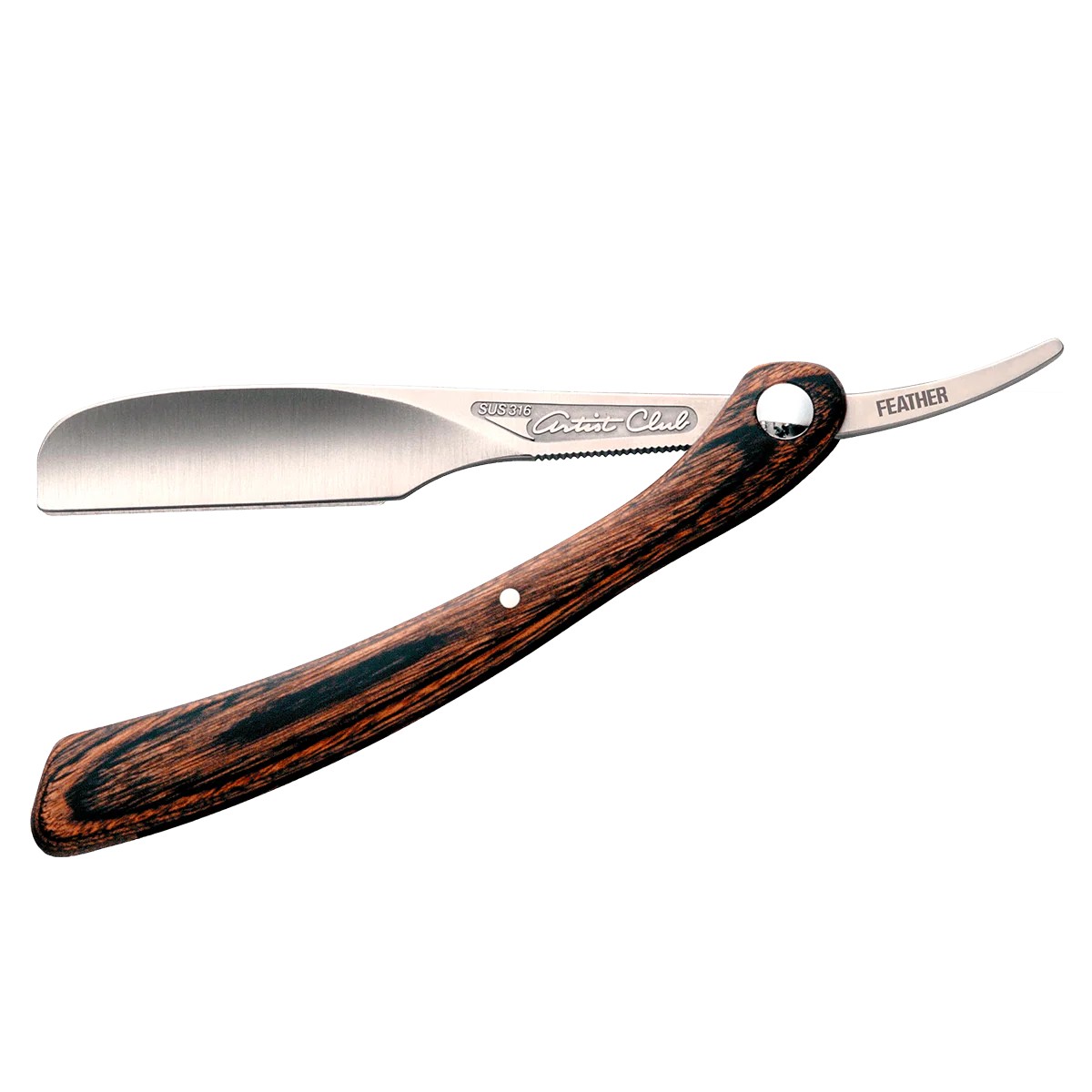}
      };
      \node[graybox,
            anchor=south east,
            align=right,
            fill opacity=0.6,
            inner sep=2pt,
            font=\footnotesize]
        at ([xshift=-0.3em,yshift=-0.8em]IMG.south east)
        {$N=2000$\\ 25.7 dB\\ \faClockO~20s};
    \end{tikzpicture} &
    \includegraphics[width=0.14\textwidth]{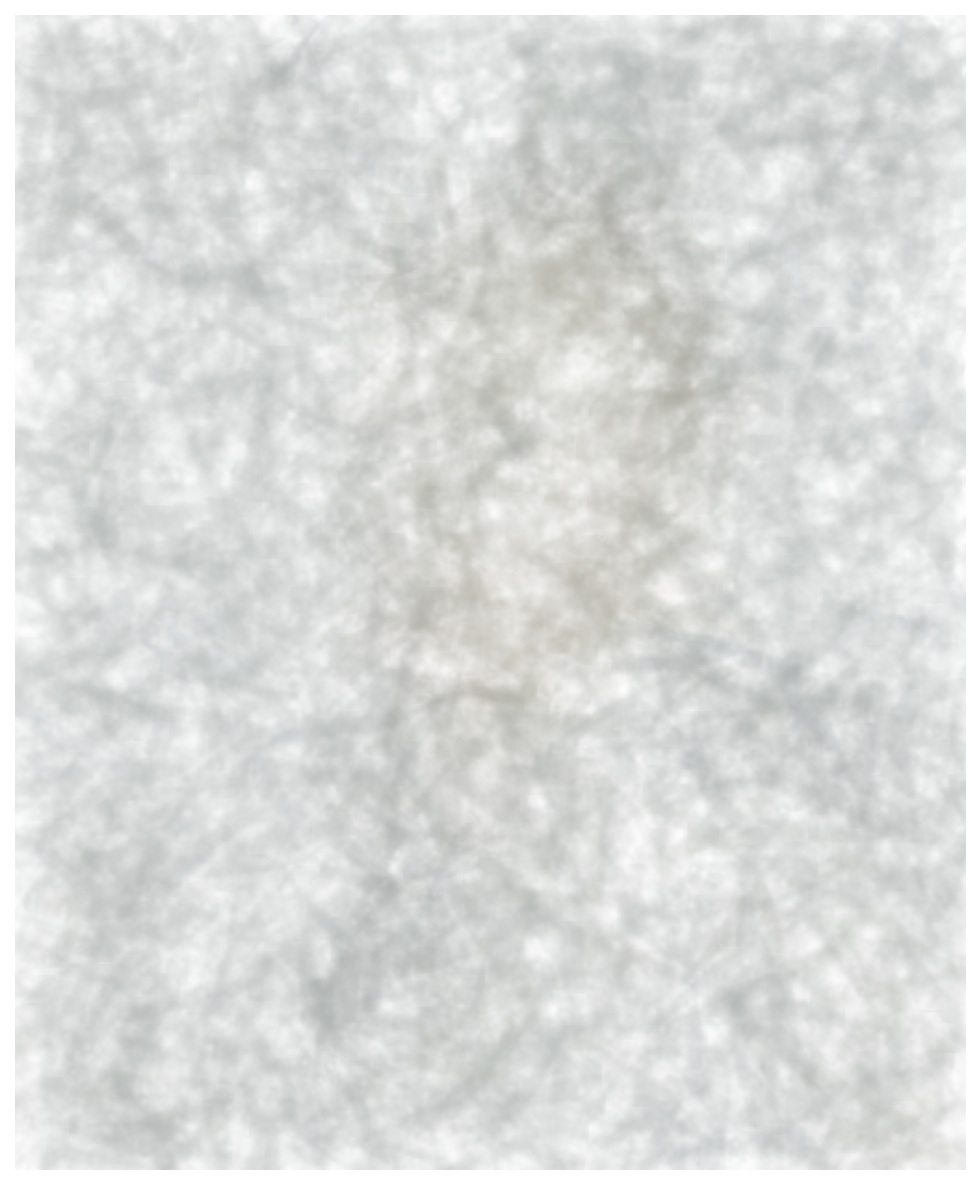} &
    \includegraphics[width=0.14\textwidth]{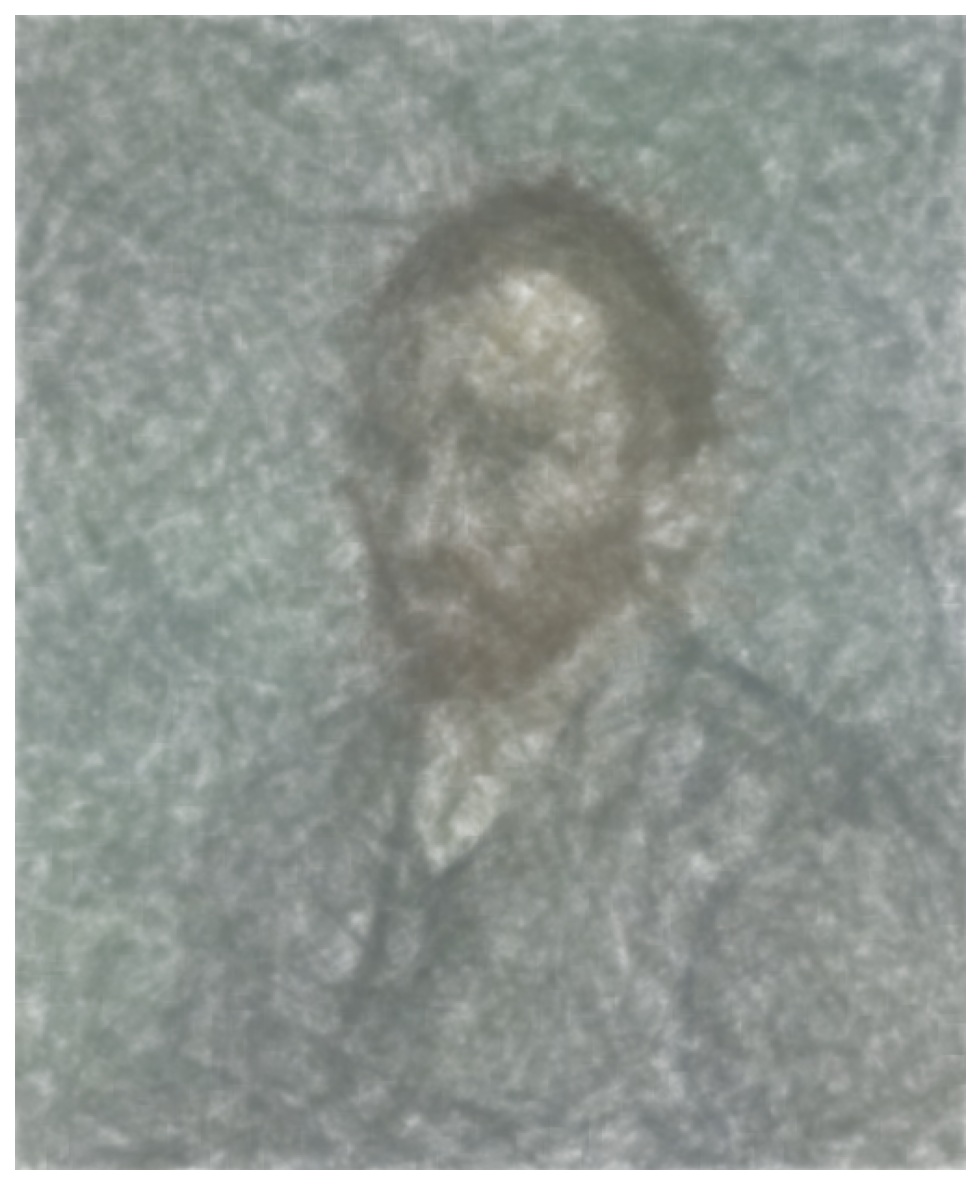} &
    \includegraphics[width=0.14\textwidth]{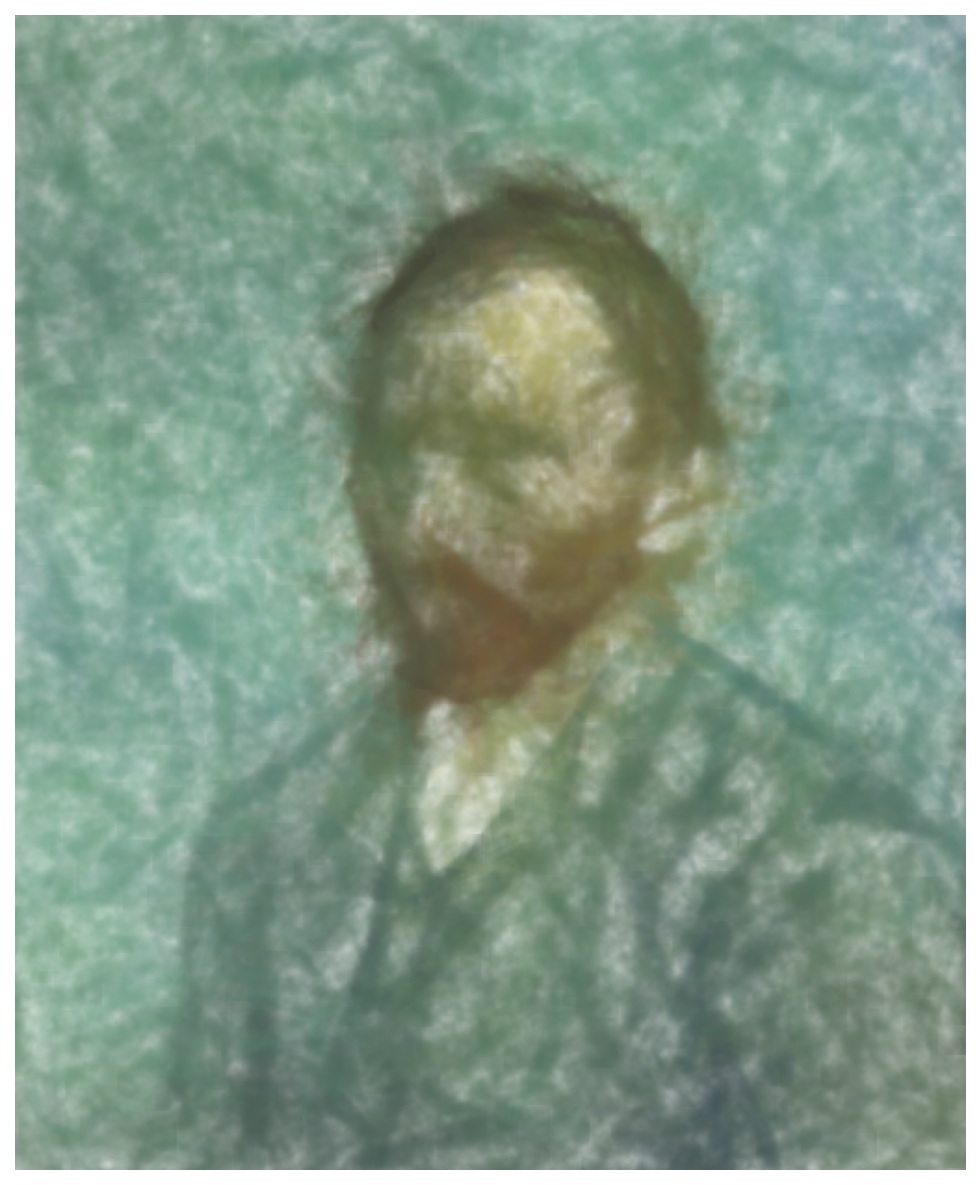} &
    \includegraphics[width=0.14\textwidth]{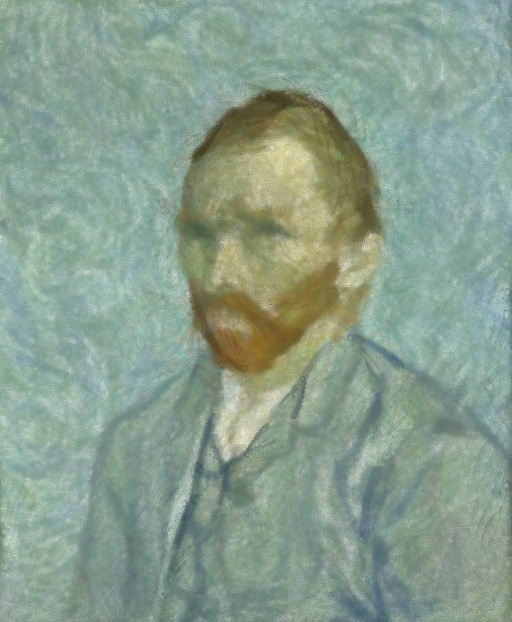} &
    \includegraphics[width=0.14\textwidth]{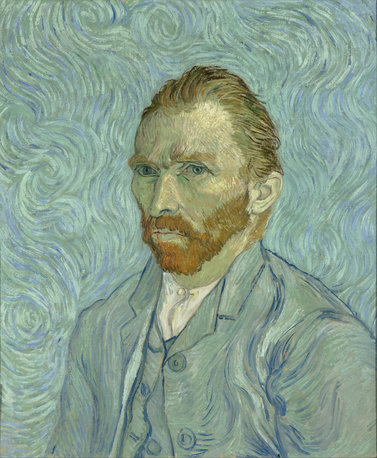} \\
    %\addlinespace[1ex]
    % ── 두 번째 예시 ────────────────
    \begin{tikzpicture} [baseline={([yshift=-12.0ex]IMG)}]
      \node[inner sep=0pt] (IMG) {
        \includegraphics[width=0.13\textwidth]{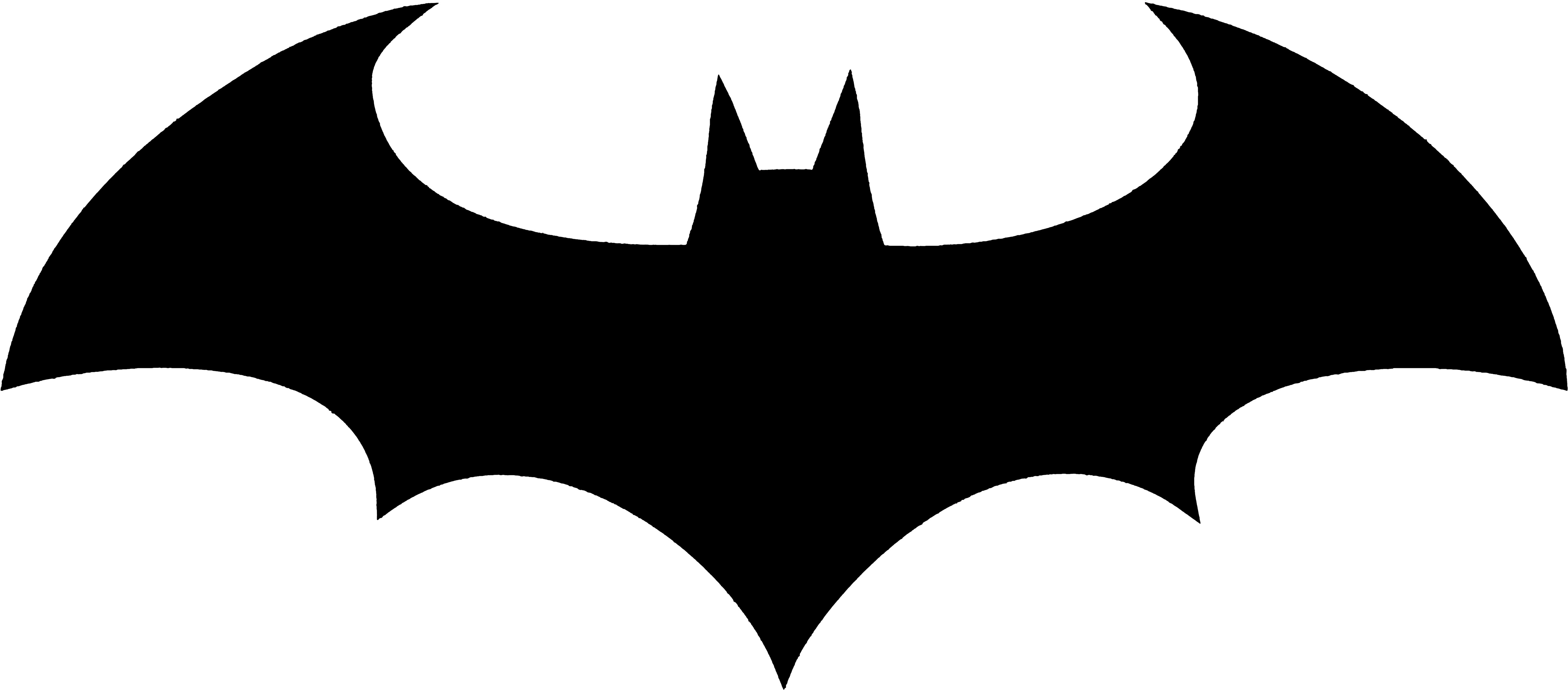}
      };
      \node[graybox,
            anchor=south east,
            align=right,
            fill opacity=0.6,
            inner sep=2pt,
            font=\footnotesize]
        at ([xshift=-0.3em,yshift=-3.5em]IMG.south east)
        {$N=2000$\\ 21.7 dB\\ \faClockO~22s};
    \end{tikzpicture} &
    \includegraphics[width=0.14\textwidth]{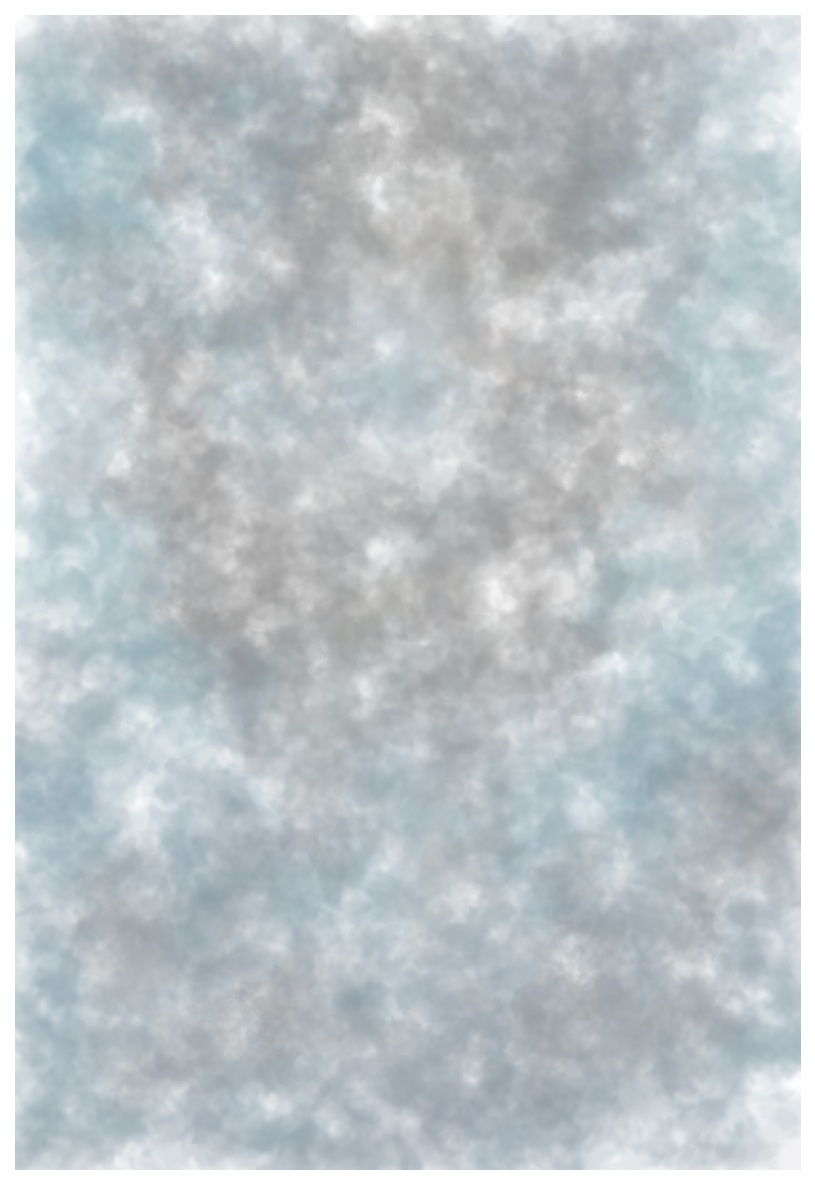} &
    \includegraphics[width=0.14\textwidth]{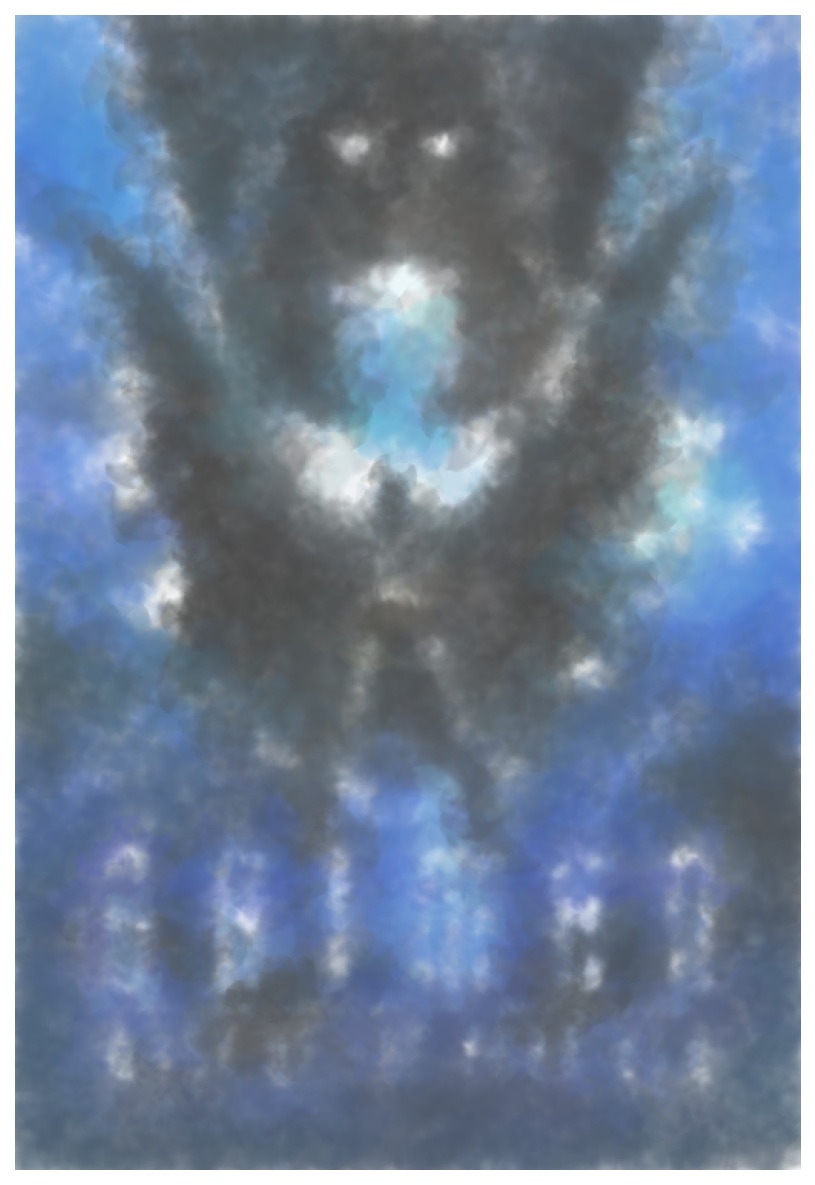} &
    \includegraphics[width=0.14\textwidth]{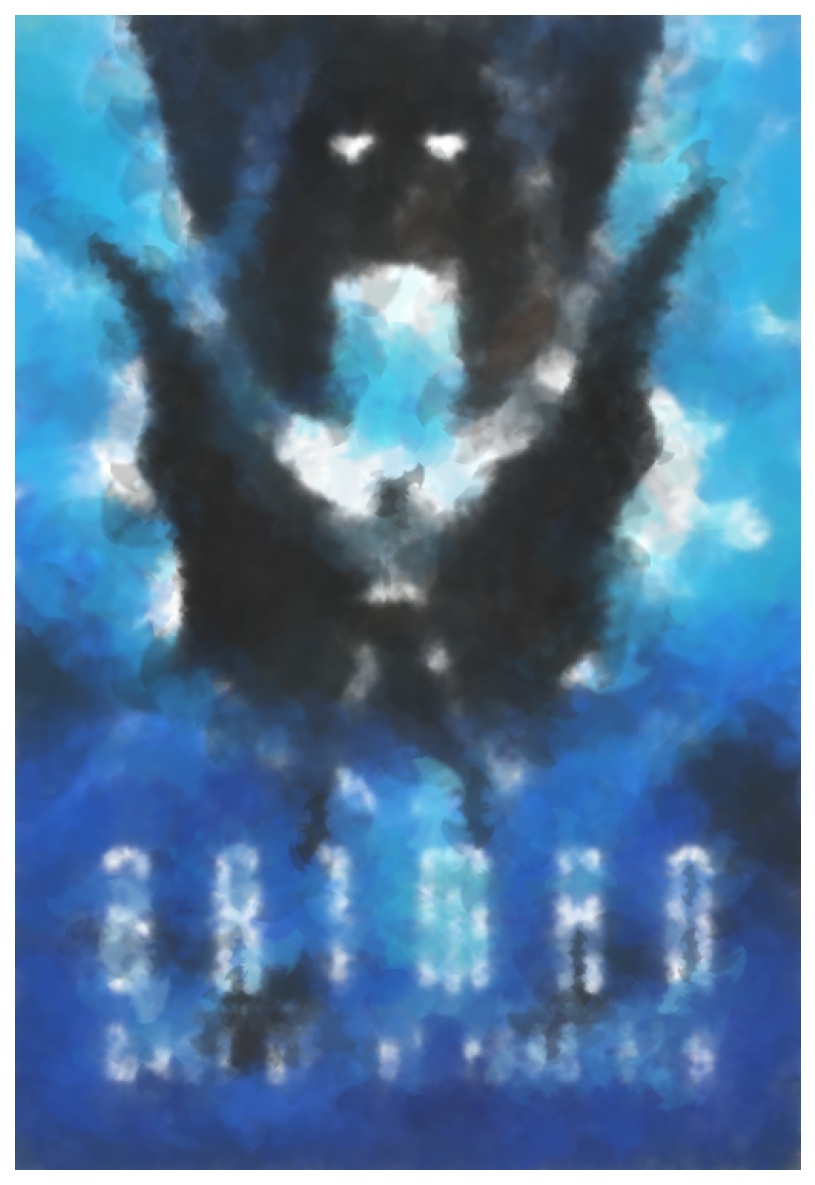} &
    \includegraphics[width=0.14\textwidth]{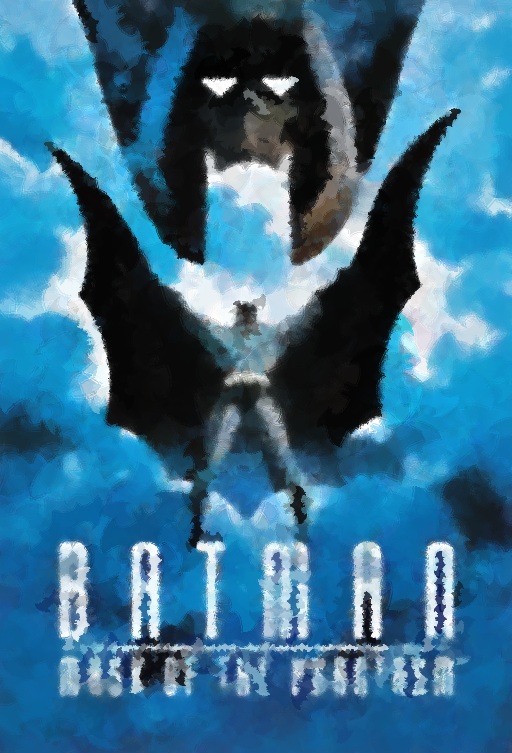} &
    \includegraphics[width=0.14\textwidth]{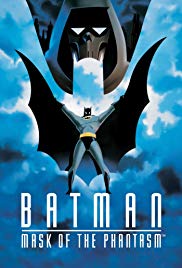} \\
    % ── 세 번째 예시 ────────────────
    \begin{tikzpicture} [baseline={([yshift=-12.0ex]IMG)}]
      \node[inner sep=0pt] (IMG) {
        \includegraphics[width=0.13\textwidth]{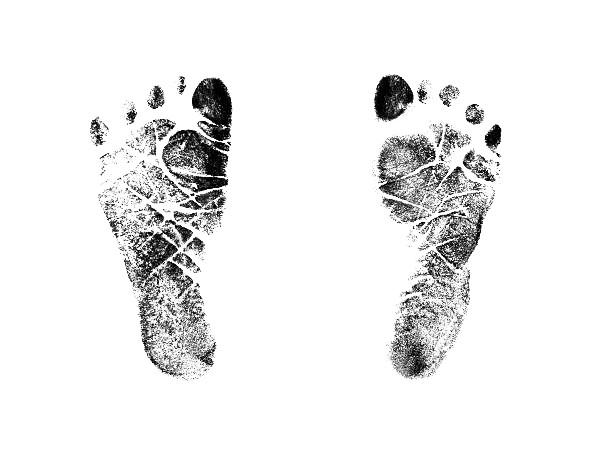}
      };
      \node[graybox,
            anchor=south east,
            align=right,
            fill opacity=0.6,
            inner sep=2pt,
            font=\footnotesize]
        at ([xshift=-0.3em,yshift=-2.5em]IMG.south east)
        {$N=2000$\\ 27.4 dB\\ \faClockO~30s};
    \end{tikzpicture} &
    \includegraphics[width=0.14\textwidth]{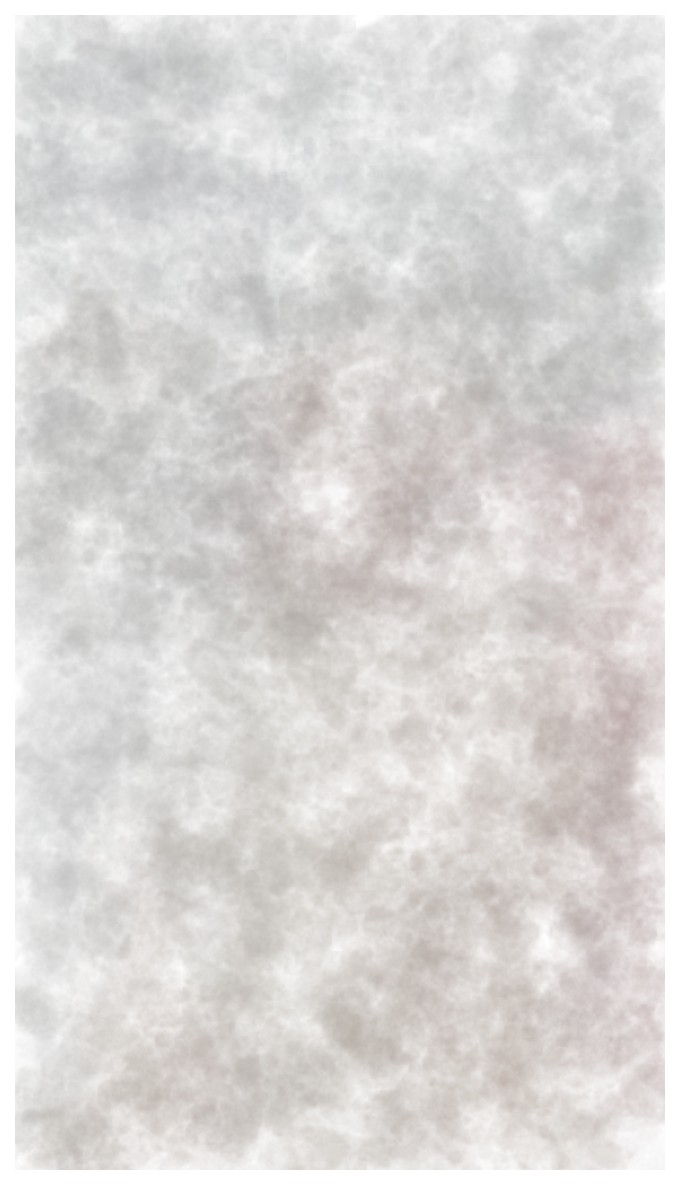} &
    \includegraphics[width=0.14\textwidth]{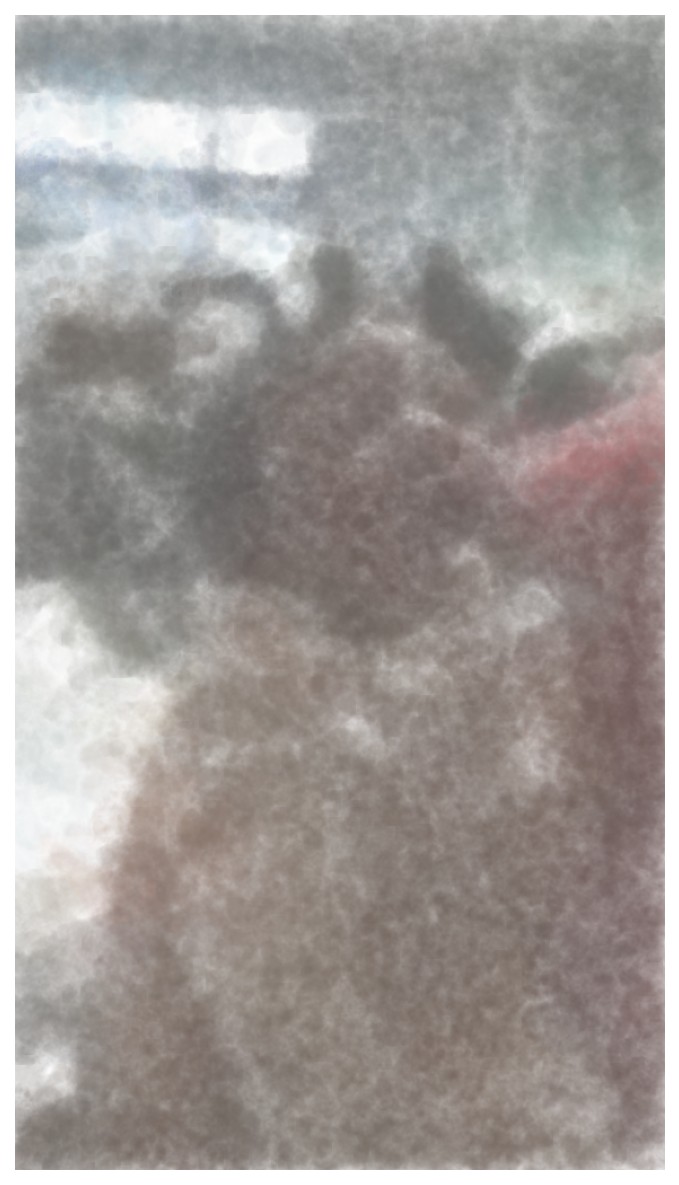} &
    \includegraphics[width=0.14\textwidth]{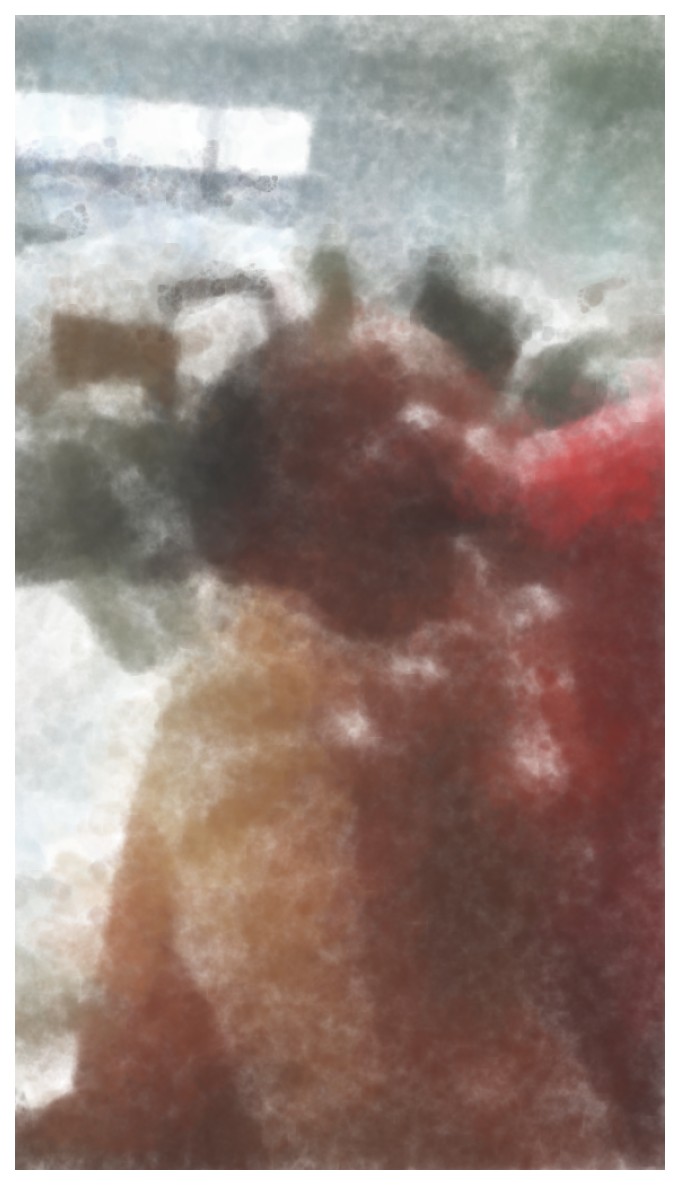} &
    \includegraphics[width=0.14\textwidth]{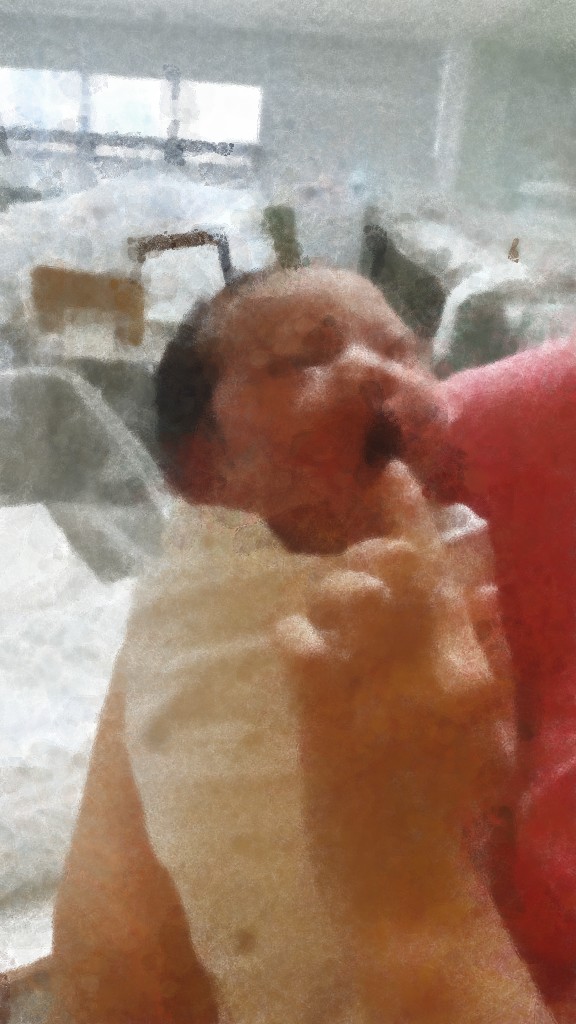} &
    \includegraphics[width=0.14\textwidth]{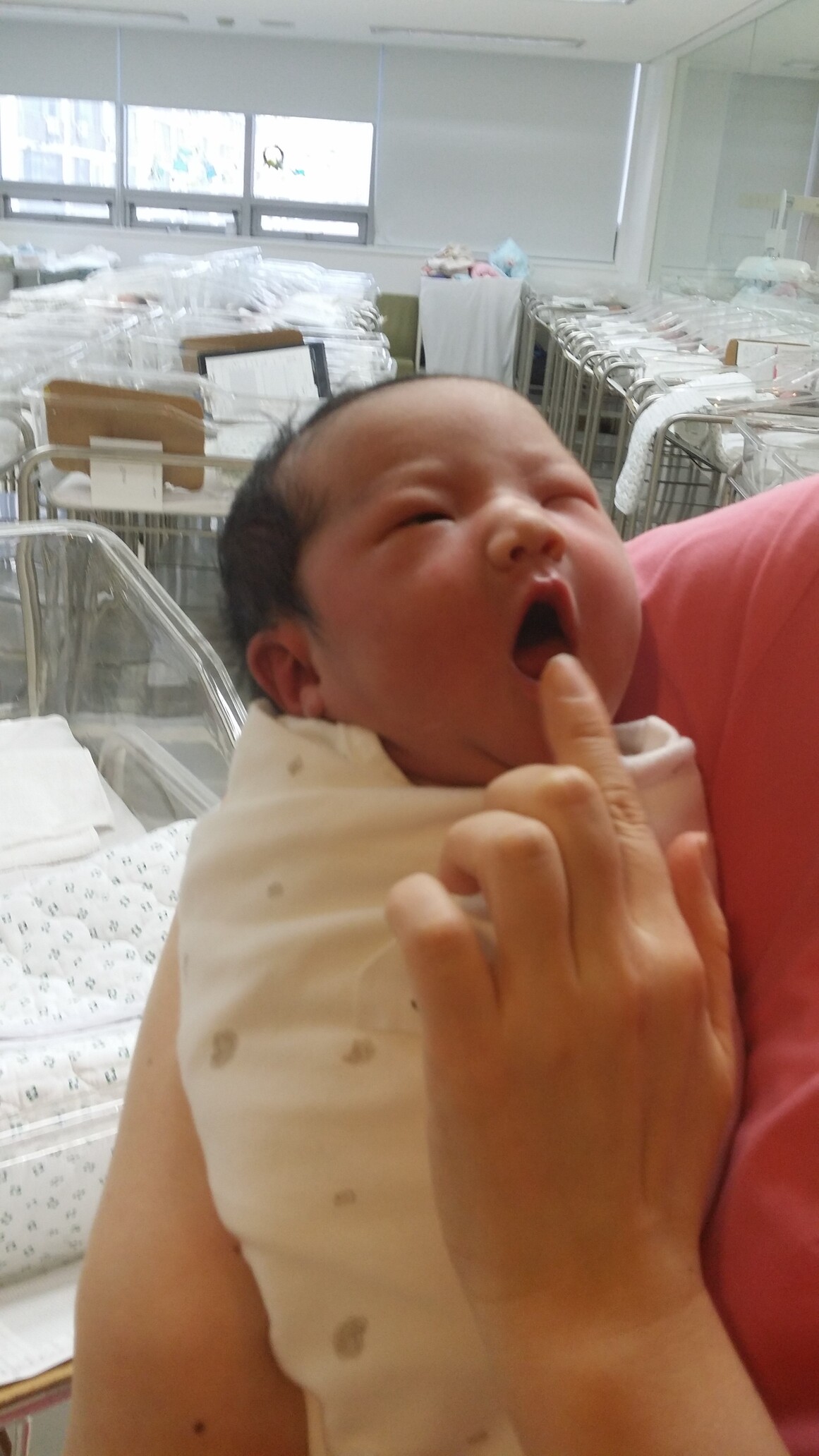} \\
  \end{tabular}
  \caption{%
    \textbf{Optimization progress across iterations for three different examples.} Each row shows the evolution from iteration~1 (left) to iteration~100 (right), illustrating how coarse structure emerges early and fine details are refined in later iterations. All three rows reuse the same primitive sets and target images as the ablation study in Table~3 of the main paper, enabling direct visual comparison with the quantitative results.
  }
  \label{fig:opt_progress}
  \end{minipage}
\end{figure*}

\begin{figure*}[t!]
  \centering
  \begin{minipage}{1.0\textwidth}
  \centering
  \small
  % 컬럼 간 여백 조절 (필요에 따라 값 조절)
  \setlength{\tabcolsep}{4pt}
  \begin{tabular}{@{}*{6}{c}@{}}
    % ── 컬럼 헤더 ───────────────────────────────
    \textbf{Primitive} & \textbf{$N=1000$} & \textbf{$N=2000$}
      & \textbf{$N=3000$} & \textbf{$N=4000$} & \textbf{Ground~Truth} \\
    %\addlinespace[1ex]
    % ── 첫 번째 예시 (artwork) ─────────────────
    \raisebox{0.1\height}{
      \includegraphics[width=0.1\textwidth]{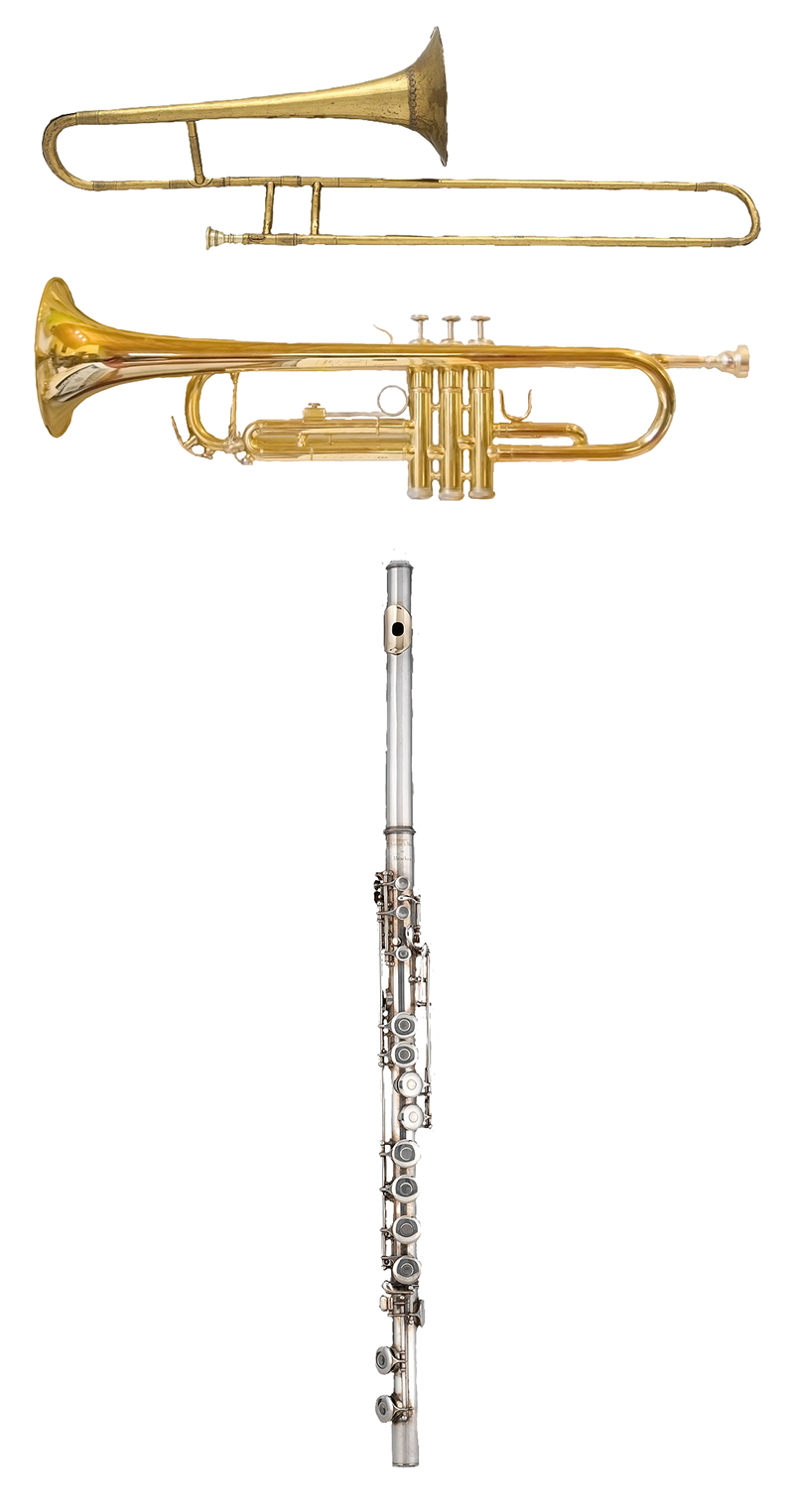}
    } &
    \begin{tikzpicture}
      \node[inner sep=0pt] (IMG) {
        \includegraphics[width=0.16\textwidth]{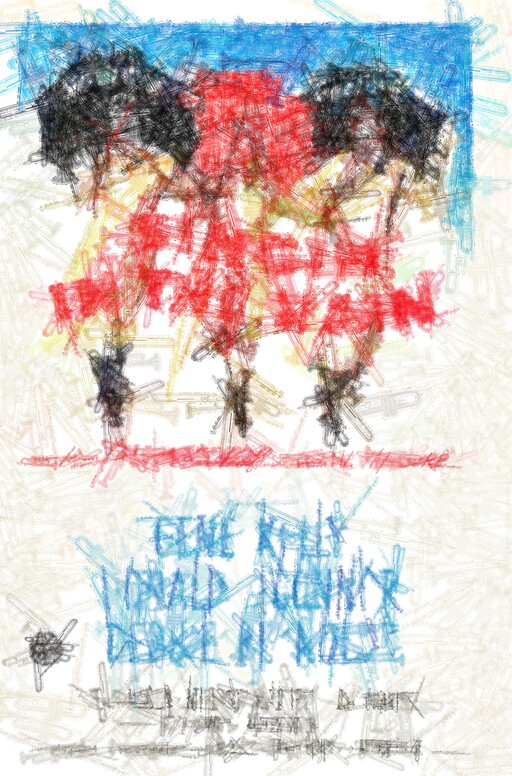}
      };
      \node[graybox,
            anchor=south east,
            align=right,
            fill opacity=0.6,
            inner sep=2pt,
            font=\footnotesize]
        at ([xshift=-0.3em,yshift=0.3em]IMG.south east)
        {15.4 dB\\ \faClockO~36s};
    \end{tikzpicture} &
    \begin{tikzpicture}
      \node[inner sep=0pt] (IMG) {
        \includegraphics[width=0.16\textwidth]{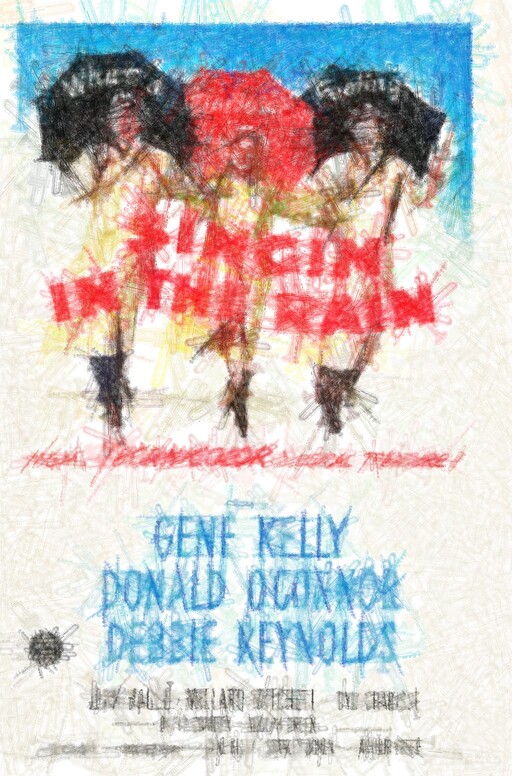}
      };
      \node[graybox,
            anchor=south east,
            align=right,
            fill opacity=0.6,
            inner sep=2pt,
            font=\footnotesize]
        at ([xshift=-0.3em,yshift=0.3em]IMG.south east)
        {16.9 dB\\ \faClockO~49s};
    \end{tikzpicture} &
    \begin{tikzpicture}
      \node[inner sep=0pt] (IMG) {
        \includegraphics[width=0.16\textwidth]{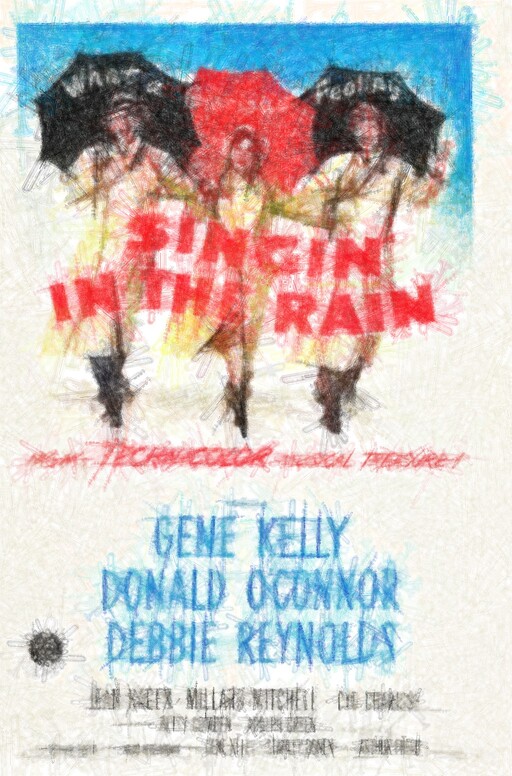}
      };
      \node[graybox,
            anchor=south east,
            align=right,
            fill opacity=0.6,
            inner sep=2pt,
            font=\footnotesize]
        at ([xshift=-0.3em,yshift=0.3em]IMG.south east)
        {17.5 dB\\ \faClockO~41s};
    \end{tikzpicture} &
    \begin{tikzpicture}
      \node[inner sep=0pt] (IMG) {
        \includegraphics[width=0.16\textwidth]{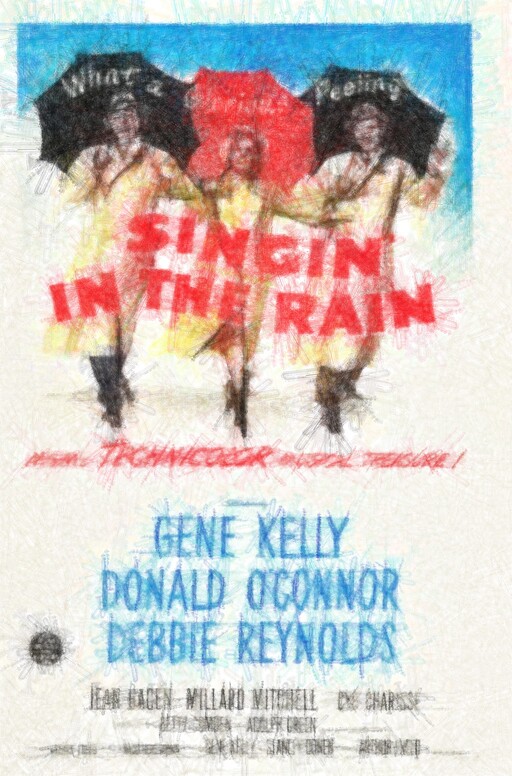}
      };
      \node[graybox,
            anchor=south east,
            align=right,
            fill opacity=0.6,
            inner sep=2pt,
            font=\footnotesize]
        at ([xshift=-0.3em,yshift=0.3em]IMG.south east)
        {17.8 dB\\ \faClockO~47s};
    \end{tikzpicture} &
    \includegraphics[width=0.16\textwidth]{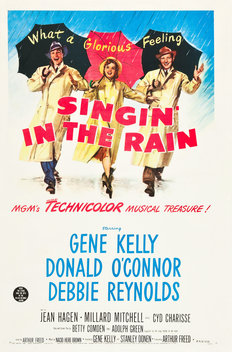} \\
    %\addlinespace[1ex]
    % ── 두 번째 예시 ────────────────
    \raisebox{0.7\height}{
      \includegraphics[width=0.1\textwidth]{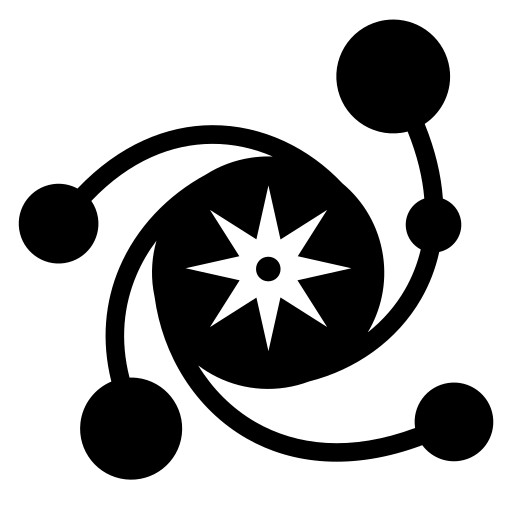}
    } &
    \begin{tikzpicture}
      \node[inner sep=0pt] (IMG) {
        \includegraphics[width=0.16\textwidth]{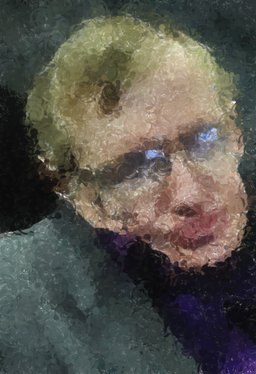}
      };
      \node[graybox,
            anchor=south east,
            align=right,
            fill opacity=0.6,
            inner sep=2pt,
            font=\footnotesize]
        at ([xshift=-0.3em,yshift=0.3em]IMG.south east)
        {22.5 dB\\ \faClockO~41s};
    \end{tikzpicture} &
    \begin{tikzpicture}
      \node[inner sep=0pt] (IMG) {
        \includegraphics[width=0.16\textwidth]{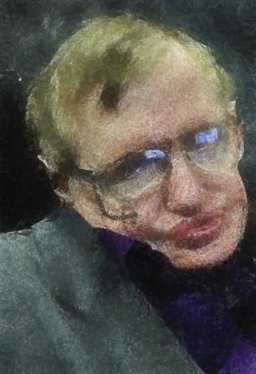}
      };
      \node[graybox,
            anchor=south east,
            align=right,
            fill opacity=0.6,
            inner sep=2pt,
            font=\footnotesize]
        at ([xshift=-0.3em,yshift=0.3em]IMG.south east)
        {24.1 dB\\ \faClockO~46s};
    \end{tikzpicture} &
    \begin{tikzpicture}
      \node[inner sep=0pt] (IMG) {
        \includegraphics[width=0.16\textwidth]{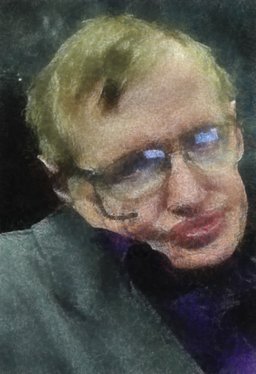}
      };
      \node[graybox,
            anchor=south east,
            align=right,
            fill opacity=0.6,
            inner sep=2pt,
            font=\footnotesize]
        at ([xshift=-0.3em,yshift=0.3em]IMG.south east)
        {24.8 dB\\ \faClockO~42s};
    \end{tikzpicture} &
    \begin{tikzpicture}
      \node[inner sep=0pt] (IMG) {
        \includegraphics[width=0.16\textwidth]{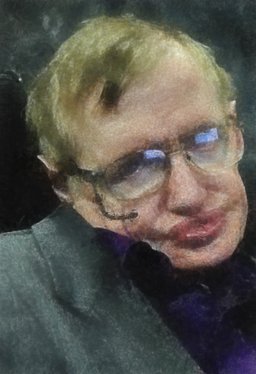}
      };
      \node[graybox,
            anchor=south east,
            align=right,
            fill opacity=0.6,
            inner sep=2pt,
            font=\footnotesize]
        at ([xshift=-0.3em,yshift=0.3em]IMG.south east)
        {25.4 dB\\ \faClockO~40s};
    \end{tikzpicture} &
    \includegraphics[width=0.16\textwidth]{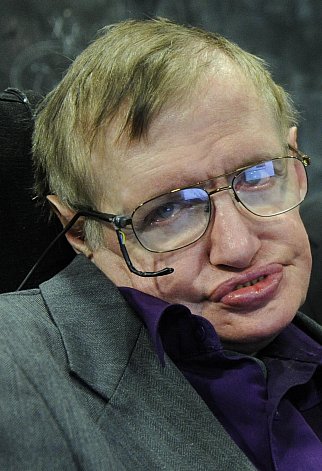} \\
    % ── 세 번째 예시 ────────────────
    \raisebox{0.3\height}{
      \includegraphics[width=0.1\textwidth]{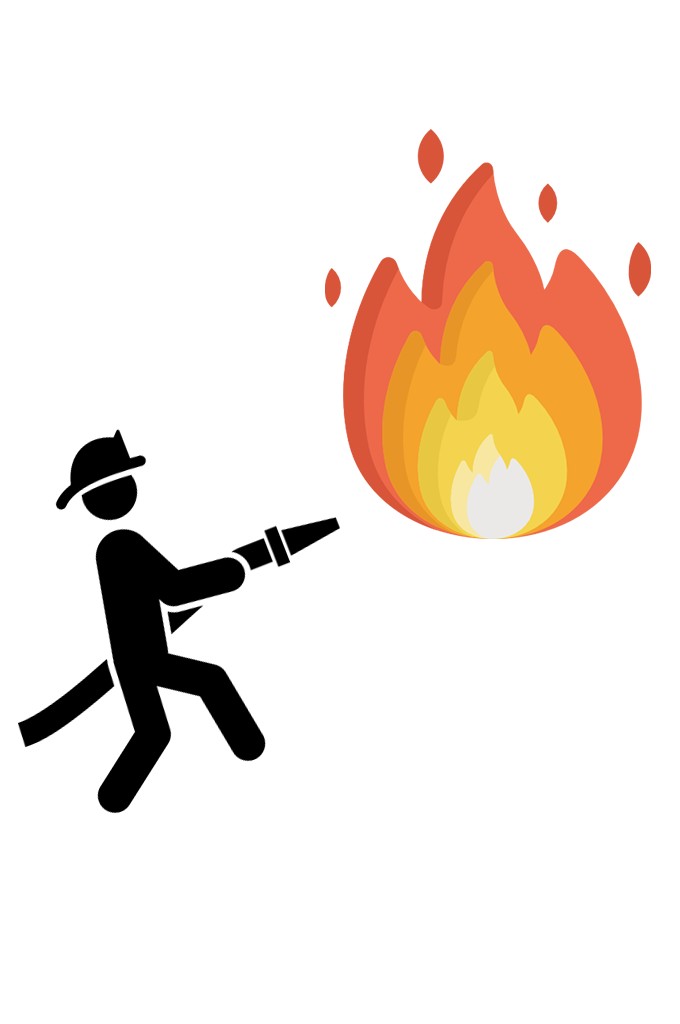}
    } &
    \begin{tikzpicture}
      \node[inner sep=0pt] (IMG) {
        \includegraphics[width=0.16\textwidth]{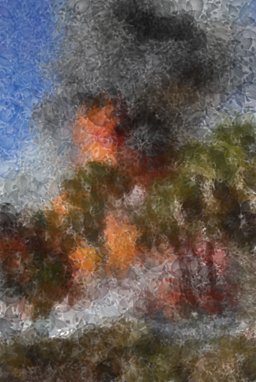}
      };
      \node[graybox,
            anchor=south east,
            align=right,
            fill opacity=0.6,
            inner sep=2pt,
            font=\footnotesize]
        at ([xshift=-0.3em,yshift=0.3em]IMG.south east)
        {19.6 dB\\ \faClockO~36s};
    \end{tikzpicture} &
    \begin{tikzpicture}
      \node[inner sep=0pt] (IMG) {
        \includegraphics[width=0.16\textwidth]{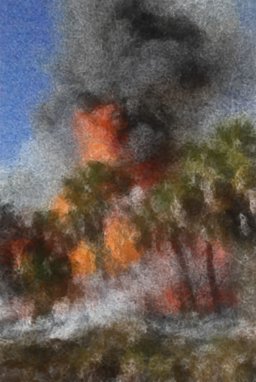}
      };
      \node[graybox,
            anchor=south east,
            align=right,
            fill opacity=0.6,
            inner sep=2pt,
            font=\footnotesize]
        at ([xshift=-0.3em,yshift=0.3em]IMG.south east)
        {20.6 dB\\ \faClockO~44s};
    \end{tikzpicture} &
    \begin{tikzpicture}
      \node[inner sep=0pt] (IMG) {
        \includegraphics[width=0.16\textwidth]{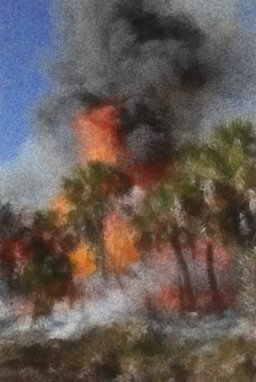}
      };
      \node[graybox,
            anchor=south east,
            align=right,
            fill opacity=0.6,
            inner sep=2pt,
            font=\footnotesize]
        at ([xshift=-0.3em,yshift=0.3em]IMG.south east)
        {21.1 dB\\ \faClockO~39s};
    \end{tikzpicture} &
    \begin{tikzpicture}
      \node[inner sep=0pt] (IMG) {
        \includegraphics[width=0.16\textwidth]{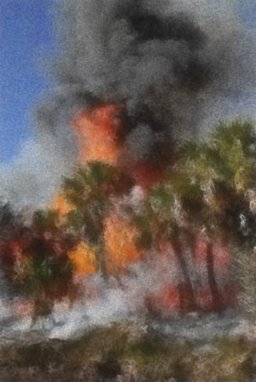}
      };
      \node[graybox,
            anchor=south east,
            align=right,
            fill opacity=0.6,
            inner sep=2pt,
            font=\footnotesize]
        at ([xshift=-0.3em,yshift=0.3em]IMG.south east)
        {21.3 dB\\ \faClockO~42s};
    \end{tikzpicture} &
    \includegraphics[width=0.16\textwidth]{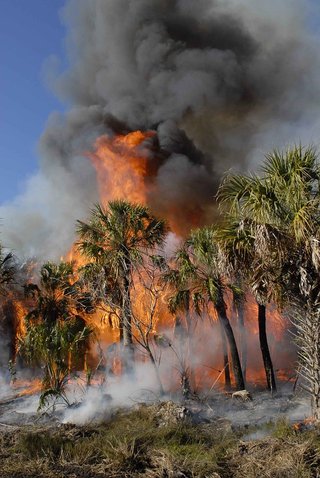} \\
    Avg. gain & 0.0 dB & + 1.4 dB & + 2.0 dB & + 2.3 dB & - \\
  \end{tabular}
  \caption{%
    \textbf{Effect of primitive count on reconstruction quality across diverse targets and primitive types.}
    Each row uses a different primitive set (left) to approximate a different target image (right), while columns sweep the number of primitives $N$ from 1000 to 4000. As $N$ increases, our optimizer consistently sharpens details and improves PSNR across movie posters, portraits, and natural scenes, demonstrating robust approximation capability over varied image content and primitive shapes.
  }
  \label{fig:diverse_example}
  \end{minipage}
\end{figure*}

\section{More discussion}
\paragraph{Initialization with autoregressive/RL methods}
Initializing our method with autoregressive/RL-based methods~\cite{hertzmann1998painterly,  huang2019learning, schaldenbrand2021content, ganin2018synthesizing,mellor2019unsupervised,jia2019paintbot,huang2025attention,liu2021paint, tang2024attentionpainter,  song2024processpainter,wang2023stroke} and then fine-tuning with \OurMethod{} could potentially yield high-quality images with fewer primitives. However, the primary goal of this work is to provide a general rendering engine usable for diverse objectives. Utilizing these existing methods, which often rely on primitives with pre-defined shapes, falls outside the scope of our current research. Nevertheless, given that we have created an easy-to-hack Python interface, we hope that creators will find it easy to integrate these techniques.

\paragraph{Diff-3D-Raster} Similar to recent work on non-Gaussian splatting in 3D \cite{chen2024linear, von2025linprim, Held20243DConvex}, our work could be extended to 3D for artistic expression, even if it does not offer computational advantages for real-time rendering. However, existing studies often use simple analytic primitives like polyhedra, and we anticipate that 3D raster primitives would present a challenge due to their significantly higher computational demands.
\fi

\end{document}